\documentclass[twocolumn,prb,groupedaddress,nofootinbib]{revtex4}

\usepackage[utf8]{inputenc}
\usepackage[english]{babel}
\usepackage[T1]{fontenc}
\usepackage{lmodern}
\usepackage{graphicx}
\usepackage{amsmath}
\usepackage{amsfonts}
\usepackage{amssymb}
\usepackage{microtype}
\usepackage{braket}
\usepackage{subfigure}
\usepackage[breaklinks=true,colorlinks, citecolor=blue]{hyperref}

\newcommand{\si}{\sigma}
\newcommand\be{\beta}
\newcommand\De{\Delta}

\newcommand\om{\omega}
\newcommand\ta{\theta}
\newcommand\dg{\dagger}

\newcommand\beq{\begin{equation}}
\newcommand\eeq{\end{equation}}
\newcommand\bea{\begin{eqnarray}}
\newcommand\eea{\end{eqnarray}}

\newcommand{\ig}{\includegraphics}
\newcommand{\ct}{\cite}
\newcommand{\la}{\lambda}
\newcommand{\bi}{\bibitem}
\newcommand{\al}{\alpha}

\newcommand{\ga}{\gamma}
\newcommand{\non}{\nonumber}
\newcommand{\de}{\delta}

\newcommand{\Tr}{{\rm Tr}}

\newcommand{\ttil}{\tilde{t}}
\newcommand{\Htil}{\widetilde{\mathcal{H}}}
\newcommand{\ktil}{\widetilde{\kappa}}
\newcommand{\Mtil}{\widetilde{M}}
\newcommand{\Ytil}{\widetilde{Y}}

\begin{document}

\begin{@twocolumnfalse}
	
\title{Driven quantum many-body systems and out-of-equilibrium topology}
\author{Souvik Bandyopadhyay$^{(1)}$, Sourav Bhattacharjee$^{(1)}$ and 
Diptiman Sen$^{(2)}$} 
\affiliation{$^{(1)}$Department of Physics, Indian Institute of Technology 
Kanpur, Kanpur 208016, India \\
$^{(2)}$Center for High Energy Physics and Department of Physics,
Indian Institute of Science, Bengaluru 560012, India}

\begin{abstract}
In this review we present some of the work done in India in the area of
driven and out-of-equilibrium systems with topological phases. After 
presenting some well-known examples of topological systems in one and
two dimensions, we discuss the effects of periodic driving in some of them.
We discuss the unitary as well as the non-unitary dynamical preparation of topologically non-trivial states in one and two dimensional systems. We then discuss the
effects of Majorana end modes on transport through a Kitaev chain and 
a junction of three Kitaev chains. Following this, transport through the surface states
of a three-dimensional topological insulator has also been reviewed. The effects
of hybridization between the top and bottom surfaces in such systems and the application
of electromagnetic radiation on a strip-like region on the top surface
are described. Two unusual topological systems are mentioned briefly,
namely, a spin system on a kagome lattice and a Josephson junction of 
three superconducting wires. We have also included a pedagogical discussion on topology and topological invariants in the appendices, where the connection between topological properties and the intrinsic geometry of many-body quantum states is also elucidated.
\end{abstract}

\maketitle
\end{@twocolumnfalse}

{\hypersetup{linkcolor = blue}
\tableofcontents
}

\section{Introduction}

Topological phases of quantum matter have been investigated extensively for
many years~\cite{hasan10,qi11}. Typically, systems in such phases have bulk
states which are gapped and states at the boundaries which are gapless; the
boundary states lie within the bulk gap and they contribute to electronic
transport at low temperatures. In addition, the number of species of boundary
states (for, say, a given value of the boundary momentum) is
given by a topological invariant which can be calculated from the bulk
properties of the system. The properties of the boundary states (such
as their energies) are often robust against small amounts of disorder. 
The form of the topological invariant depends on the dimensionality of the
system and the symmetries of its Hamiltonian; for example, it may be
a winding number, a Chern number, a Bott invariant,
and so on. The transport properties of the boundary states are of great
interest as they often have universal features, such as quantized values
of the differential conductance or peaks in the differential conductance
at special values of the voltage bias.

The dynamics of quantum systems is another subject of considerable
interest from the point of view of topology. For instance, periodic driving
or the quenching
of some parameters in the Hamiltonian can sometimes turn a non-topological
system which has no boundary states into a topological system which has
boundary states, and vice versa. Further, the value of the topological
invariant for such systems can be controlled by parameters such as the
amplitude and frequency of the periodic driving or the choice of initial
and final Hamiltonians of the quenching protocol. The effects of dissipation
on the dynamics is also of great interest.

In this review, we plan to provide a glimpse of various systems in which
topology and dynamics combine to play an important role. The plan of this
review is as
follows. In Sec. II, we introduce some well-studied topological systems
such as the Su-Schrieffer-Heeger (SSH) model and the Kitaev chain in one dimension,
and the Kitaev honeycomb model and the Haldane model of graphene in two
dimensions, and the different phases that these models have depending on the
system parameters. In Sec. III, we discuss the rich structure of phases which
can arise when the Kitaev chain and the Kitaev honeycomb model are driven
periodically in time. In Sec. IV, we discuss some aspects of one-dimensional
topological systems which are out-of-equilibrium, in particular, their
topological classification and their non-unitary dynamics in the presence
of dissipation. We have primarily focused on the dynamical preparation of topological states which also exhibit the dynamical emergence of a topological bulk-boundary correspondence.
In Sec. V, we study out-of-equilibrium Chern topological
systems in two dimensions, and the possibility of using dynamics and
dissipation to experimentally produce a desired topological phase.
In Sec. VI, we study the Kitaev chain and the
effect that the boundary states (which are zero energy Majorana modes)
of such a system have on the transport through such a system. Transport
through a junction of three Kitaev chains is also discussed. 

Apart from one-dimensional symmetry protected topological insulators and Chern insulators in two dimensions, there also exist various three-dimensional systems which exhibit an intricate connection between topology and the discrete symmetries, such as time-reversal symmetric three-dimensional topological insulators \ct{hasan10,qi11}. Although the equilibrium topological properties of three-dimensional systems are beyond the scope of this review, we will nevertheless briefly discuss a few recent studies into the non-equilibrium transport through topological surface states in such systems in Sec.~\ref{sec:trti}. For example, we discuss the effects of hybridization between the opposite
surfaces of a thin topological insulator and the effects of
periodic driving (which can be produced by applying electromagnetic radiation)
on a strip-like region of the top surface. In Sec. VIII,
two unusual examples of topological systems are presented. These include
a spin system on a kagome lattice in two dimensions where the magnon bands
can have Berry curvature, and a Josephson junction of three superconducting
wires with different pairing phases in which the bands of Andreev bound state
at the junction have a Berry curvature as a function of the pairing phases.
We summarize our review in Sec. IX. We end with some Appendices. Appendix
A presents a pedagogical review of the geometry of topological systems
and the various topological invariants. Appendix B discusses a perturbative
expansion for the edge correlations which arise in the dissipative
dynamics of a Kitaev chain. 


\section{Topology of fermionic systems in equilibrium}\label{sec:models}

In this section we introduce a few minimalistic systems which manifest topologically non-trivial phases of matter, particularly topological insulators (we 
refer to Appendix~\ref{sec:adiabatic} for a mathematical interlude). In this section, we introduce various quantum many-body topological systems and also discuss the physical manifestations of topology in their properties. We will see that non-trivial topological states of thermodynamically large systems hold robust non-local signatures which are protected against local perturbations and can therefore find extensive use in processing of quantum information. Depending on the dimension and symmetries respected by the respective Hamiltonians, {every topological system can be classified into a generic periodic table of topological insulators.} To begin with, we consider the simplest topological systems in one dimension and then move on to discuss more complex models in two dimensions. 

\subsection{Su-Schrieffer-Heeger Model}

\begin{figure}[ht]
\ig[width=\columnwidth]{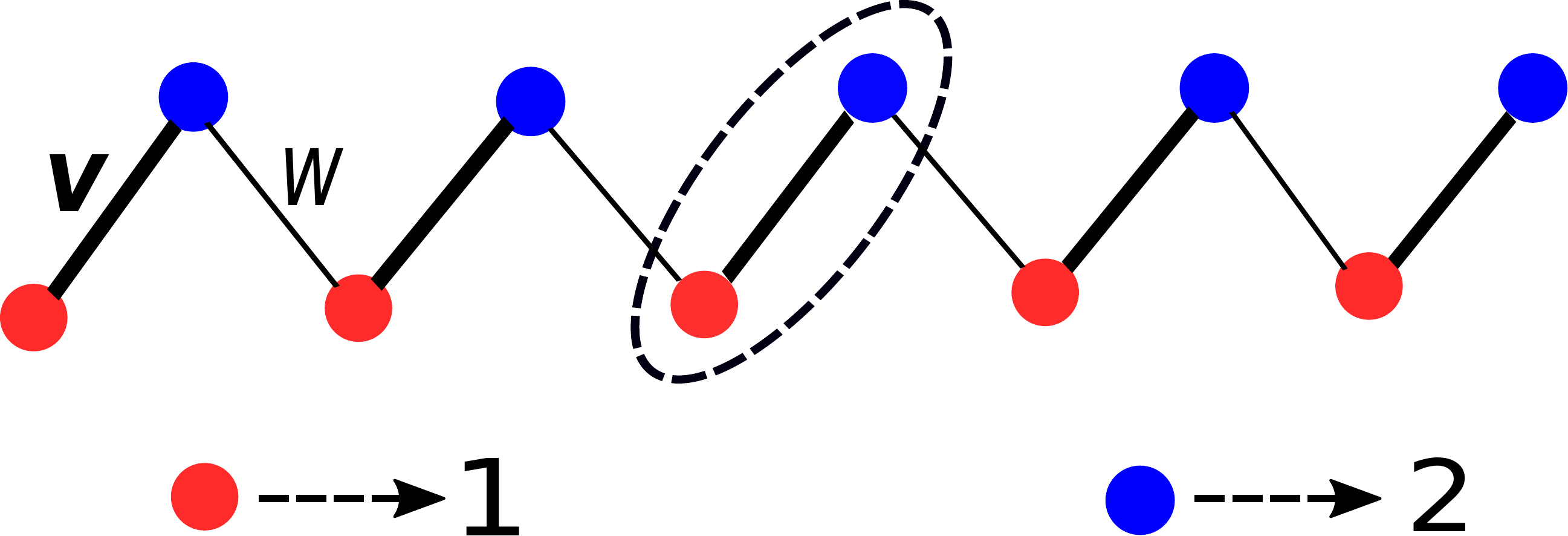}	
\caption[]{The SSH chain with two sublattices (shown in red and blue and 
labeled as sublattices 1 and 2 respectively). A typical unit cell containing 
a site from each sublattice has been circled with dotted lines.} \label{SSH} 
\end{figure} 

The SSH model\cite{shen12,bernevig13} belongs to the BDI symmetry class of topological insulators and is the simplest one-dimensional model exhibiting an 
underlying topological structure and end states\ct{asboth16}. Physically, it describes a lattice with a two atom sublattice structure in which the {intra-cell} hopping amplitude is in general different from the {inter-cell} hopping amplitude. The Hamiltonian for the SSH model can be written in terms of the (spin polarized) fermion creation and annihilation operators as,
\begin{equation}\label{eq:H}
H=\sum_{n=1}^{\mathcal{N}}(vc^{\dagger}_{n,1}c^{}_{n,2}+wc^{\dagger}_{n,2}c^{}_{n+1,1}+{\rm H.c.}),
\end{equation}
where H.c. denotes the Hermitian conjugate, $v$ and $w$ are the intra-cell (see Fig.~\ref{SSH}) and inter-cell hopping amplitudes respectively and $\mathcal{N}$ is the total number of unit cells in the chain. The complex fermionic operator $c^{\dagger}_{n,i}$ ($c_{n,i}$) creates (destroys) a fermion in the sublattice position $i~(i=1,2)$ of the $n^{th}$ unit cell and satisfies the fermionic anti-commutation rules,
\begin{equation}\label{eq:anticom}
\{c^{\dagger}_p,c_q\}=\delta_{pq} \text{~~~and~~~} 
\{c_p,c_q\}=\{c^{\dagger}_p,c^{\dagger}_q\}=0. \end{equation}

Note that with periodic boundary conditions, the lattice has 
a discrete translational symmetry for a translation by a complete unit cell, 
i.e., the system remains invariant under a translation by 
two consecutive lattice points. Utilizing this translational invariance, 
one can rewrite the Hamiltonian in Fourier space in terms of the spinors 
$\Psi_k$, defined as
\begin{equation}
\Psi_k=\left(
\begin{array}{c}
 c_{k,1} \\
 c_{k,2} \\
\end{array} \right)=\frac{1}{\mathcal{\sqrt{N}}}\sum\limits_{l=1}^{\mathcal{N}}e^{-ikl}\left(
\begin{array}{c}
c_{l,1} \\
c_{l,2} \\
\end{array} \right),
\end{equation}
such that the Hamiltonian in Eq.~\eqref{eq:H} is decoupled for each conserved quasimomentum mode $k$ as,
\begin{equation}\label{eq:Hk}
\begin{split}
H=\bigoplus_{k}\mathcal{H}_k=\bigoplus_{k}\Psi^\dagger_kH(k)\Psi_k,\\ 
H(k)=\vec{h}(k) \cdot \vec{\sigma},
\end{split}
\end{equation}
where $\vec \sigma$ is the vector of Pauli matrices denoting pseudo-spin operators and $k=2\pi m/\mathcal{N}$, where $m=0,1,\dots,\mathcal{N}$. For the SSH model, the functions $\vec{h}(k)$ and hence, the single-particle Hamiltonian $H(k)$ are found to be
\begin{eqnarray}\label{eq:hk}
\nonumber h_{x}(k)&=& Re(v)+|w|\cos(k+arg(w)), \\
\nonumber h_{y}(k)&=& -Im(v)+|w|\sin(k+arg(w)), \\
\noindent h_{z}(k)&=& 0, \end{eqnarray}
with the lattice parameter set equal to unity.

The single-particle Hamiltonian $H(k)$ has the following eigenvalue spectrum,
\begin{equation}\label{eq:Ek}
E(k)=\pm|\vec{h(k)}|,
\end{equation}
and the respective eigenvectors are
\begin{equation}\label{eq:eg}
|\pm\rangle = \frac{1}{\sqrt{2}} \left(
\begin{array}{c}
\pm e^{-i \phi(k) } \\
1 \\
\end{array} \right), \end{equation}
where $\phi=\tan^{-1}\left(\frac{h_{y}}{h_{x}}\right)$.
It is clear from the Eq.~\eqref{eq:hk}, that $\vec{h}(k)$ is periodic in $k$, {$0\leq k\leq 2\pi$ (the Brillouin zone (BZ) has the geometry of $\mathcal{S}^1$). Sweeping over the BZ is thus equivalent to tracing out a closed loop in the parameter space, with $k$ parametrizing the curve. Further, it follows from Eq.~\eqref{eq:Ek} that the system is in the ground state at half-filling, i.e., when only the lowest band is completely occupied. The different phases of the SSH model are therefore classified} through the bulk topological winding number $\nu$, which is given as (see Eq.~\eqref{Berry} of Appendix \ref{sec_app_A}),
\begin{equation}\label{eq:winding}
\nu=\frac{i}{2\pi }\oint_{BZ}\langle\psi_{0}^{k}|\partial_{k}|\psi_{0}^{k}\rangle dk=\frac{i}{2\pi}\int^{\pi}_{-\pi}dk \frac{d}{dk}\ln(h_{x}+ih_{y}).
\end{equation}
{The second equality in the above equation can be understood as follows. We note that $\vec{h}(k)$ traces out a closed curve over a plane as $k$ varies over the first BZ. The winding number is proportional to the change in the argument of $\vec{h}(k)$ as $k$ varies over the first BZ and is thus integer quantized.} If the curve does not enclose the origin, $\nu$ is zero (i.e., if $|v| > |w|$). On the other hand,
$\nu$ is one if the curve encloses the origin (i.e., if $|v| < |w|$). It is also evident from Eq.~\eqref{eq:Ek} that the energy gap between the two bands 
vanishes at $|v| = |w|$ and the winding number becomes undefined. Thus if the gap is not closed, $\nu$ is well-defined and is robust to external changes in the Hamiltonian and hence is a topological invariant clearly demarcating the topologically trivial and non-trivial phases.\\

It is also known that a topological non-triviality of the bulk system induces {conducting boundary states} under open boundary conditions. This {\it bulk-boundary correspondence}\ct{shen12} is simple to understand in the particular situation of the SSH Hamiltonian when $v,w\in\mathbb{R}$. Consider the two different topological phases characterized by the sign of the quantity $m=v-w$ separated by the QCP at $v=w$ ($m=0$). At the QCP, the energy gap vanishes at $k^*=\pi$. We therefore focus on the low energy spectrum of the SSH model near the QCP ($m\sim 0$) such that $k=k^*+\delta k$, where $\delta k<<1$. Plugging in these approximations in the Hamiltonian in Eq.~\eqref{eq:hk} one sees that
\begin{equation}\label{eq:jr}
H_c(k)\simeq-\delta k\sigma^y+m\sigma^x.
\end{equation}
The linear dispersion at the QCP at $m=0$ is why it is called the Dirac point. Also, the low-energy Hamiltonian $H_c$ in real space translates to the one-dimensional Dirac Hamiltonian with $m$ being the mass,
\begin{equation} H_c=i\partial_x\sigma^y+m\sigma^x. \end{equation}
This equation can then be utilized to model a junction of two systems along the x-axis such that there is a mass discontinuity. Specifically, we take the mass to be of different signs on the positive and negative x-axis respectively signifying that the two systems across the junction are topologically inequivalent. It is straightforward to check that under such a mass discontinuity, the Dirac Hamiltonian hosts zero energy eigenstates which are exponentially localized at the junction. This exemplifies the existence of {edge states} localized at the boundary of topologically non-trivial systems. These edge states are protected by the bulk topology of the thermodynamically large system. In other words, they are robust against all such local perturbations which are incapable of changing the bulk topology. The thermodynamic limit is necessary as otherwise the exponentially localized states at the opposite edges might hybridize to produce finite tunneling into the bulk.\\

The topological classification of non-interacting many-body quantum systems is performed by considering three different discrete symmetries viz., the time-reversal symmetry ($\mathcal{T}$), the particle-hole symmetry ($\mathcal{P}$) and the sublattice (chiral) symmetry ($\mathcal{S}$). The constraints imposed upon the Hamiltonian of a system possessing the above symmetries in the quasi momentum basis are expressed as,
\begin{equation}\label{eq:symm}
\begin{split}
\mathcal{T}^{-1} H(k)\mathcal{T}=H(-k),\\
\mathcal{P}^{-1}H(k)\mathcal{P}=-H(-k),\\
\mathcal{S}^{-1}H(k)\mathcal{S}=-H(k),
\end{split}
\end{equation}
where $\mathcal{T}$ and $\mathcal{P}$ are anti-unitary operators such that $\mathcal{T}^2=\pm\mathbb{I}$ and $\mathcal{P}^2=\pm\mathbb{I}$, whereas $\mathcal{S}$ is an unitary operator satisfying $\mathcal{S}^2=\mathbb{I}$ and $\mathbb{I}$ is the $2 \times 2$ identity operator. We also note that the sublattice symmetry is a combined effect of the time-reversal symmetry and the particle hole symmetry as,
\begin{equation}\label{eq:sublattice}
\mathcal{S}=\mathcal{T}\mathcal{P}.
\end{equation}
It is now evident from the Hamiltonian of the SSH model in Eqs.~\eqref{eq:Hk} and~\eqref{eq:hk} and the symmetry transformations in Eq.~\eqref{eq:symm} that the SSH model is symmetric under the sublattice transformation $\mathcal{S}=\bigotimes_{k}\sigma^z$ which results in the vanishing of $h_z(k)$. Also, if the hopping coefficients $v$ and $w$ are real, the Hamiltonian possesses time-reversal symmetry $\mathcal{T}=\bigotimes_k\mathcal{K}$, $\mathcal{K}$ being the complex conjugation operator. Hence, it is clear from Eq.~\eqref{eq:sublattice} that the system is also symmetric under the particle-hole/charge conjugation operation with $\mathcal{P}=\bigotimes_k\mathcal{K}\sigma^z$. As such the SSH model is classified in the BDI class of Hamiltonians within the topological classification scheme. It must also be noted that the Hamiltonian $H(k)$ spans a $\mathcal{S}^1$ on a plane because $h_z(k)=0$ is satisfied for all $k$. This in turn is guaranteed when the system lies in the BDI class or, in particular, respects the chiral/sublattice symmetry $\mathcal{S}$. In other words, the chiral symmetry protects the topological phases of the system and hence, these phases are also known as {\it symmetry protected topological} (SPT) phases.\\

\subsection{Kitaev chain}
\label{sec:kitaev_chain}

Similar to the SSH chain, the Kitaev chain is a one dimensional lattice system of spin-polarized fermions that represents a prototype model of $p$-wave superconductors (Since the system is spin-polarized we will not write the spin label below). As we will see below, it has topological phases in which a long system has a pair of zero energy Majorana modes, with one localized at each end~\cite{kitaev01,lutchyn10,oreg10,gott1,gott2}. For a chain of linear dimension $N$, with the lattice parameter once again set to unity, the many-body Hamiltonian is given by
\bea H &=& \sum_{n=1}^{N-1} [ \ga ( c_n^\dag c_{n+1} + c_{n+1}^\dag c_n ) 
+ \De ( c_n c_{n+1} + c_{n+1}^\dag c_n^\dag )] \non \\
&& - \sum_{n=1}^N \mu (2c_n^\dag c_n - 1), \label{ham1} \eea
where $\ga$ and $\De$ are the hopping and superconducting pairing amplitudes between neighboring sites, respectively, while $\mu$ is the on-site chemical potential. All of these parameters are considered to be real and one may assume that $\ga > 0$ without loss of generality.

The topological features of the Kitaev chain, particularly the existence of localized end states, are more explicitly manifested in the Majorana representation. Defining the Majorana operators, 
\beq a_{2n-1} = c_n + c_n^\dag ~~~~{\rm and}~~~~ a_{2n} = i (c_n - c_n^\dag),
\label{majo} \eeq
where $n=1,2,\cdots,N$, Eq.~\eqref{ham1} assumes the form
\bea H &=& i ~ \sum_{n=1}^{N-1} ~[~ J_x a_{2n} a_{2n+1} ~-~ J_y a_{2n-1} 
a_{2n+2}] \non \\
& & +~ i ~ \sum_{n=1}^N ~\mu a_{2n-1} a_{2n}, \non \\
J_x &=& \frac{1}{2} (\ga - \De) ~~~~{\rm and}~~~~ J_y = \frac{1}{2} 
(\ga + \De). \label{ham2} \eea
The Majorana operators are Hermitian operators satisfying $\{ a_m, a_n \} = 2 \de_{mn}$. 
A careful look into the above Hamiltonian uncovers an interesting connection with the SSH Hamiltonian. To be precise, the Kitaev chain in the Majorana representation is exactly the SSH chain (see Eq.~\eqref{eq:jr}) with Majorana fermions replacing Dirac fermionic operators. 

Furthermore, a Jordan-Wigner transformation of the Majorana operators ~\cite{lieb}, given as,
\bea a_{2n-1} &=& \left( \prod_{j=1}^{n-1} \si_j^z \right) ~\si_n^x, \non \\
a_{2n} &=& \left( \prod_{j=1}^{n-1} \si_j^z \right) ~\si_n^y, \label{jw} \eea
maps the above system to a spin-1/2 $XY$ chain placed in a magnetic field pointing in the $\hat z$ direction, with the Hamiltonian modified as,
\beq H ~=~ - \sum_{n=1}^{N-1} ~[~ J_x \si_n^x \si_{n+1}^x ~+~ J_y \si_n^y
\si_{n+1}^y] ~- \sum_{n=1}^N ~\mu \si_n^z, \label{ham3} \eeq
where the $\si_n^a$ denote the Pauli matrices at site $n$.
In all the numerical calculations that follow, we set $\ga = - \De$ which implies $J_y = 0$ and $J_x = \ga$; the system is then equivalent to an Ising model (with interaction $J_x$) in a transverse magnetic field $\mu$.

\begin{figure}
\ig[width=\columnwidth]{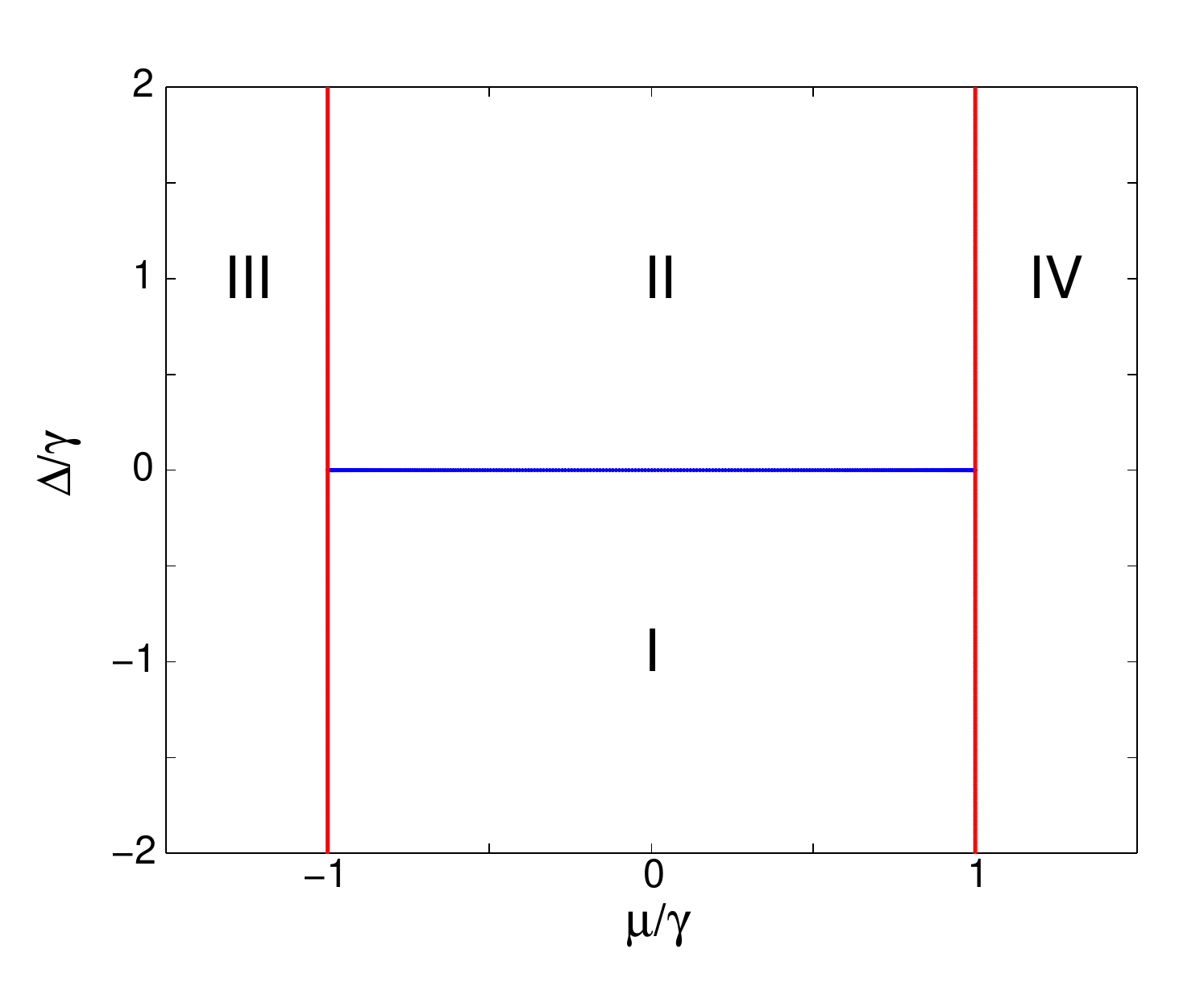}	
\caption[]{Phase diagram of the model in Eq.~\eqref{ham1} as a function of 
$\mu/\ga$ and $\De/\ga$. Phases I and II are topological while III and IV are
non-topological.} \label{flo0} \end{figure} 

To elucidate the bulk properties, let us first consider a chain with periodic boundary conditions. Substituting the Fourier transform 
$c_k = \frac{1}{\sqrt N}~ \sum_{n=1}^N c_n e^{ikn}$ in Eq.~\eqref{ham1}, we obtain,
\bea H &=& 2 (\ga - \mu) c_0^\dag c_0 ~+~ 2 (- \ga - \mu) c_\pi^\dag c_\pi
\non \\
&& +~ \sum_{0 < k < \pi} ~\left( \begin{array}{cc}
	c_k^\dag & c_{-k} \end{array} \right) ~h_k ~\left( \begin{array}{c}
	c_k \\
	c_{-k}^\dag \end{array} \right), \non \\
h_k &=& {\vec{h}}(k)\cdot {\vec{\tau}}
=2(\ga \cos k - \mu) ~\tau^z ~+~ 2 \De \sin k ~\tau^y, \label{hk} \eea
where ${\vec {h}}(k) = (h^x(k), h^y(k), h^z(k))=(0, 2 \De \sin k, 2(\ga \cos k - \mu))$ and ${\vec{\tau}}=(\tau^x, \tau^y,\tau^z)$ is the vector of Pauli matrices denoting pseudo-spin. Note that the effective BZ is halved ($0<k<\pi$) as the pairing term in Eq.~\eqref{ham1} couples the momenta $k$ and $-k$.
The dispersion relation follows from Eq.~\eqref{hk} and is given 
by~\cite{gott1,gott2}
\beq E_k ~=~ \pm \sqrt{4(\ga \cos k - \mu)^2 ~+~ 4 \De^2 \sin^2 k}.
\label{disp} \eeq

As shown in Fig.~\ref{flo0}, the Kitaev chain, or equivalently the $XY$ spin chain, can exist in four different phases~\cite{gott1,gott2}. In the spin chain representation, phases I and II correspond to long-range ferromagnetic orders of $\si^x$ and $\si^y$, respectively. On the other hand, phases III and IV correspond 
to paramagnetic phases with no long-range order. The four phases are 
separated from each other by quantum critical lines
where the energy $E_k$ vanishes for some values of $k$. The critical lines are given by $\mu/\ga = \pm 1$ for all values of $\De$, and $-1 \le \mu/\ga \le 1$ for $\De = 0$. Interestingly, these phases are also topologically distinct and can be characterized by a winding number.

To define the winding number, we proceed in a similar way as we did in the case of the SSH chain. First, note that the vector ${\vec {h}}(k)$ is confined to the $y-z$ plane. Let $\phi_k = \tan^{-1} (h^z(k)/h^y(k))$ denote the angle sustained by the vector $\vec {h}(k) $ with respect to the $\hat z$ axis. The winding number can be defined as in Eq.~\eqref{eq:winding} by following the change in $\phi_k$ as we go around the full BZ ($-\pi\leq k\leq\pi$) \cite{niu,tong}, i.e., 
\beq W ~=~\int_{-\pi}^{\pi} ~\frac{dk}{2\pi} ~\frac{d\phi_k}{dk}. 
\label{wind} \eeq
This can take any integer value and is a topological invariant, namely, it
does not change under small changes in $h_k$ unless $h_k$ happens to pass
through zero for some value of $k$ in which case the winding number becomes
ill-defined; this can only happen if the energy $E_k = 0$ at some value of 
$k$ which means that the bulk gap is zero. In a gapped phase, therefore,
Eq.~\eqref{wind} defines a $Z$-valued topological invariant. 

In the ferromagnetic phases, it is easy to check that $W = \pm1$; the system is said to exist in a topologically non-trivial phase. As in the case of the SSH chain, in the topologically non-trivial phase, a long and open Kitaev chain hosts a pair of zero energy Majorana modes at each end ~\cite{gott2} and correspondingly, the two-point correlation function of the Majorana end modes in the ground state of the Hamiltonian, defined as {$\bra{GS}\theta\ket{GS}$, where $\theta=i a_1a_{2N}$}, also remains finite and approaches unity as $\mu\to 0$. This Majorana end correlation in a thermodynamically large system is a remarkable manifestation of topologically protected long range correlations. This can be verified by considering the extreme case $J_x > 0$ and $J_y = \mu = 0$ in Eq.~\eqref{ham2}, in which case the Hamiltonian is independent of the terms $a_1$ and $a_{2N}$; hence the corresponding modes have zero energies. In the paramagnetic phase, one finds $W = 0$ and hence the system is topologically trivial. Also, no zero energy Majorana modes are present at either end of an open chain. The winding number thus exhibits a bulk-boundary correspondence.

It is important to analyze the role played by symmetry in the existence of the topological structure discussed above. The crucial ingredient is once again the confinement of the vector $\vec {h}(k)$ to a 2d parameter space, which results from the underlying chiral symmetry of the Kitaev chain. This can be verified from the symmetry transformation $\tau_x\vec{h}(k)\tau_x = -\vec{h}(k)$; which is satisfied in the absence of any $x-$ component of the vector $\vec{h}(k)$. The Kitaev chain also possesses the particle-hole, $\mathcal{P}\vec{h}(k)\mathcal{P}^\dagger = -\vec{h}(-k)$, and  the time-reversal symmetries, $\mathcal{T} \vec{h}(k)\mathcal{T}^\dagger=\vec{h}(-k)$. The symmetry operators can be identified as $\mathcal{P}=\mathcal{K}\tau_x$ and $\mathcal{T}=\mathcal{K}$, where $\mathcal{K}$ is the complex conjugation operator. In general, the presence of any two of these symmetries guarantees that the third must also be present. This establishes that the completely symmetric Kitaev chain also lies in the BDI symmetry class.

\subsection{Kitaev honeycomb model}\label{subsec_kitaev_honey}

The Kitaev model is a two-dimensional model of spin-$1/2$'s placed on the sites of a honeycomb lattice, with a Hamiltonian of the form 
\beq H~=~\sum_{j+l=even}^{} (J_1 \si^x_{j,l} \si^x_{j+1,l} + J_2 \si^y_{j-1,l}
\si^y_{j,l} + J_3 \si^z_{j,l} \si^z_{j,l+1}), \label{hsig} \eeq 
where $j,l$ are the column and row indices, respectively. For simplicity, we assume that $J_1, J_2, J_3 \ge 0$.

\begin{figure} 
\ig[width=0.95\columnwidth]{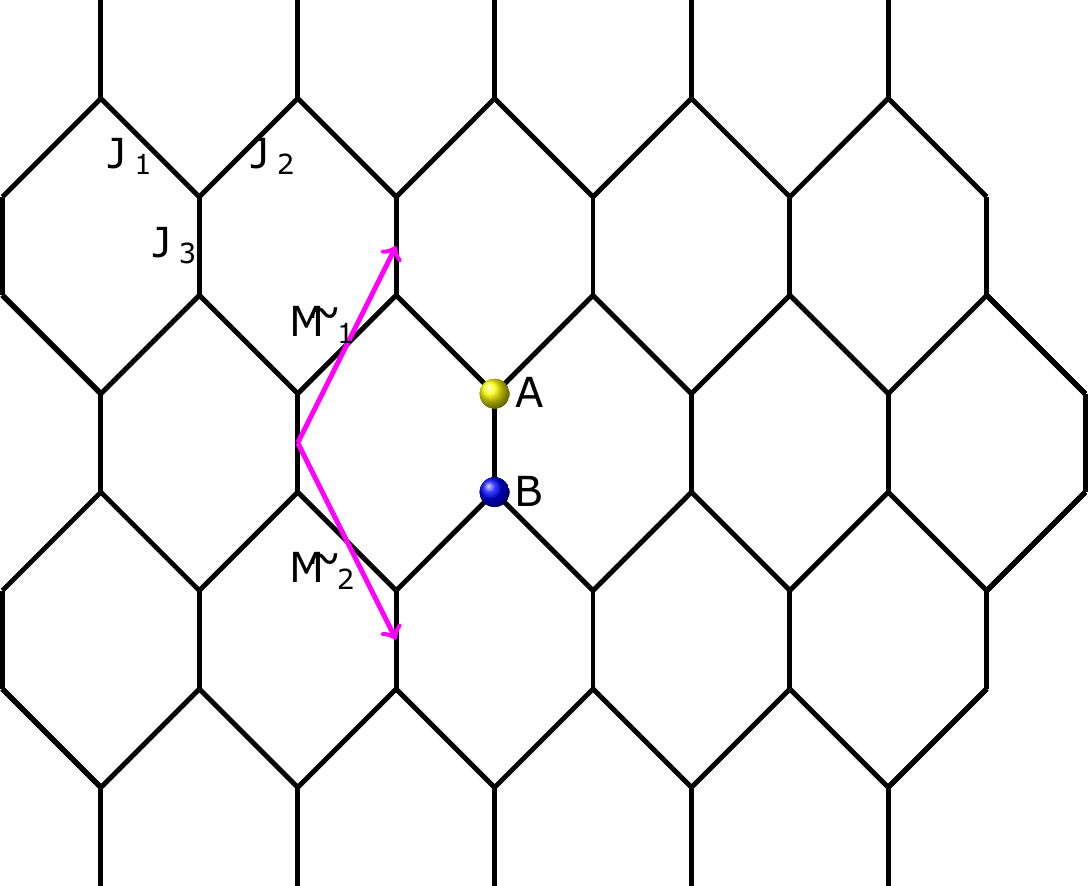}
\caption[]{Kitaev model on the honeycomb lattice with $xx$ coupling $J_1$, $yy$ coupling $J_2$ and $zz$ coupling $J_3$. $\vec{M_1}$ and $\vec{M_2}$ are the spanning vectors of the lattice, and $A$ and $B$ denote the two sites of a unit cell. (See Ref.~\onlinecite{thakurathi14}).} \label{fig01} 
\end{figure}

Fig.~\ref{fig01} shows a schematic representation of the honeycomb lattice. The unit cells are chosen as the vertical bonds, with the sublattices labeled as $A$ and $B$; these have $j+l$ equal to odd and even integers, respectively. It is convenient to set the nearest-neighbor distance to be $1/\sqrt{3}$, which allows us to label each unit cell by the vector, $\vec{n}= \hat{i} n_1+(\frac{1}{2} \hat{i} + \frac{\sqrt{3}}{2} \hat{j}) n_2$, where $n_1, ~n_2$ are integers that can be calculated from the coordinates of the $B$ site in the corresponding unit cell as $n_1 = (j-l)/2$ and $n_2 = l$. Fig.~\ref{fig01} shows
the spanning vectors $\vec{M_1} =\frac{1}{2} \hat{i} + \frac{\sqrt{3}}{2}
\hat{j}$ and $\vec{M_2} =\frac{1}{2} \hat{i} - \frac{\sqrt{3}}{2} \hat{j}$
which join some neighboring unit cells.

As with the Kitaev chain, the topological aspects of the Kitaev model are also best understood in the Majorana representation. We define the Majorana 
operators~\cite{kitaev06, feng07, chen07, nussinov08, baskaran07, lee07}
\bea \hat a_{j,l} &=& \left( \prod_{i,k<l} \si^z_{i,k} \right)\left( \prod_{i<j} \si^z_{i,l} \right)
\si^y_{j,l}, \non \\
\hat{a}'_{j,l} &=& \left( \prod_{i,k<l} \si^z_{i,k} \right)\left( \prod_{i<j} \si^z_{i,l} \right)
\si^x_{j,l}~~ \text{for} ~j+l=\text{even},\label{eq_2dmajoa} \non \\
&& \\
\hat b_{j,l} &=& \left( \prod_{i,k<l} \si^z_{i,k} \right)\left( \prod_{i<j} \si^z_{i,l} \right)
\si^x_{j,l}, \non \\
\hat b'_{j,l} &=& \left( \prod_{i,k<l} \si^z_{i,k} \right)\left( \prod_{i<j} \si^z_{i,l} \right)
\si^y_{j,l} ~~ \text{for} ~j+l=\text{odd}, \non \\
&& \label{eq_2dmajob}\eea
These are Hermitian operators satisfying the anticommutation relations
$\{ \hat a_{m,n}, \hat a_{m',n'} \} = 2 \de_{mm'} \de_{nn'}$, $\{ \hat a'_{m,n}, \hat a'_{m',n'} \} = 2 \de_{mm'} \de_{nn'}$, $\{ \hat b_{m,n},
\hat b_{m',n'} \} = 2 \de_{mm'} \de_{nn'}$, $\{ \hat b'_{m,n},
\hat b'_{m',n'} \} = 2 \de_{mm'} \de_{nn'}$, and $\{ \hat a_{m,n}, 
\hat a'_{m',n'} \} = \{ \hat a_{m,n}, 
\hat b_{m',n'} \} = \{ \hat a_{m,n}, 
\hat b'_{m',n'} \} = \{ \hat a'_{m,n}, 
\hat b_{m',n'} \} = \{ \hat a'_{m,n}, 
\hat b'_{m',n'} \} = \{ \hat b_{m,n}, 
\hat b'_{m',n'} \} = 0$. The Hamiltonian then takes the form
\beq H~=~ i~\sum_{\vec{n}}^{} (J_1 \hat b_{\vec{n}} \hat a_{\vec{n}-\vec{M_1}}
+ J_2 \hat b_{\vec{n}} \hat a_{\vec{n}+\vec{M_2}} + J_3 \hat D_{\vec{n}} 
\hat b_{\vec{n}} \hat a_{\vec{n}}), \label{hmaj} \eeq
where $\hat{D}_{\vec{n}} = i\hat b'_{\vec{n}} \hat a'_{\vec{n}}$. It is straightforward to see that the operators $\hat{D}_{\vec{n}}$ commute with each other and also with the Hamiltonian; hence they represent conserved quantities. In particular, the eigenvalues of $\hat D_{\vec{n}}$ can assume the values $\pm 1$, independently, for each $\vec{n}$, thereby decomposing the
$2^N$-dimensional Hilbert space ($N$ being the total number of sites) into $2^{N/2}$ sectors. Further, it is known that the ground state of the model lies in the sector in which the eigenvalue of $\hat D_{\vec{n}}$ equals $1$ for all $\vec{n}$ and as such, we will restrict ourselves to this sector. We note that the operators $\hat{D}_{\vec{n}}$, when expressed in terms of spin operators using Eqs.~\eqref{eq_2dmajoa} and~\eqref{eq_2dmajob}, are highly non-local string operators.

Assuming periodic boundary conditions, the Majorana operators can be Fourier transformed as
\bea \hat a_{\vec{n}} &=&\sqrt{\frac{4}{N}} \sum_{\vec{k}\in \frac{1}{2}
	\text{BZ}}{}(\hat a_{\vec{k}} e^{i\vec{k} \cdot \vec{n}}+ \hat a_{\vec{k}}^\dg
e^{-i\vec{k} \cdot \vec{n}}), \non \\
\hat b_{\vec{n}} &=&\sqrt{\frac{4}{N}} \sum_{\vec{k}\in \frac{1}{2}
	\text{BZ}}{}(\hat b_{\vec{k}} e^{i\vec{k} \cdot \vec{n}}+ \hat b_{\vec{k}}^\dg
e^{-i\vec{k} \cdot \vec{n}}), \label{fourier} \eea
which satisfy the anticommutation relations $\{ \hat a_{\vec{k}}, 
\hat a_{\vec{k}'}^\dag \} = \{ \hat b_{\vec{k}}, \hat b_{\vec{k}'}^\dag \} = 
\de_{\vec{k},\vec{k}'}$. Note that the sums over $\vec k$ in 
Eq.~\eqref{fourier} only go over half the BZ;
a convenient choice of the BZ is given by a rhombus whose vertices lie
at $(k_x,k_y)= (\pm 2\pi, 0)$ and $(0,\pm 2\pi /\sqrt{3})$.
The Hamiltonian in Eq.~\eqref{hmaj} can then be written in a form similar to the Kitaev chain Hamiltonian in Eq.~\eqref{hk}
\bea H &=& \sum_{ \vec{k}\in \frac{1}{2}\text{BZ}} ~\left( \begin{array}{cc}
	\hat a_{\vec{k}}^\dg & \hat b_{\vec{k}}^\dg \end{array} \right) ~H_k ~\left(
\begin{array}{c}
	\hat a_{\vec{k}} \\
	\hat b_{\vec{k}} \end{array} \right), \non \\
H_{\vec{k}} &=& 2 [J_1 \sin(\vec{k}.\vec{M}_1)-J_2 \sin(\vec{k}.\vec{M}_2)]
\tau^x \non \\
&& + 2 [J_3+J_1 \cos(\vec{k}.\vec{M}_1)+J_2 \cos(\vec{k}.\vec{M}_2)] \tau^y.
\label{hmom} \eea
The dispersion relation can be derived from Eq.~\eqref{hmom}; it consists of two bands with energies
\bea E_{\vec{k}}^\pm &=& \pm 2[\{J_1 \sin(\vec{k}.\vec{M}_1)-J_2 \sin(\vec{k}.
\vec{M}_2)\}^2 \non \\
&& +\{J_3+J_1 \cos(\vec{k}.\vec{M}_1)+J_2 \cos(\vec{k}.\vec{M}_2)\}^2 ]^{1/2}.
\non \\
\label{disp1} \eea

\begin{figure*}
\subfigure[]{
\ig[width=0.6\columnwidth]{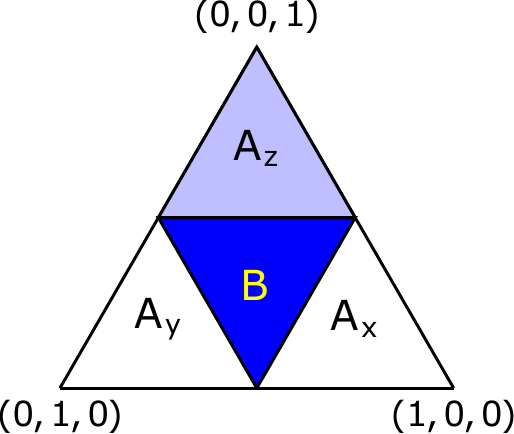}
\label{fig03}}
\subfigure[]{
\ig[width=0.6\columnwidth]{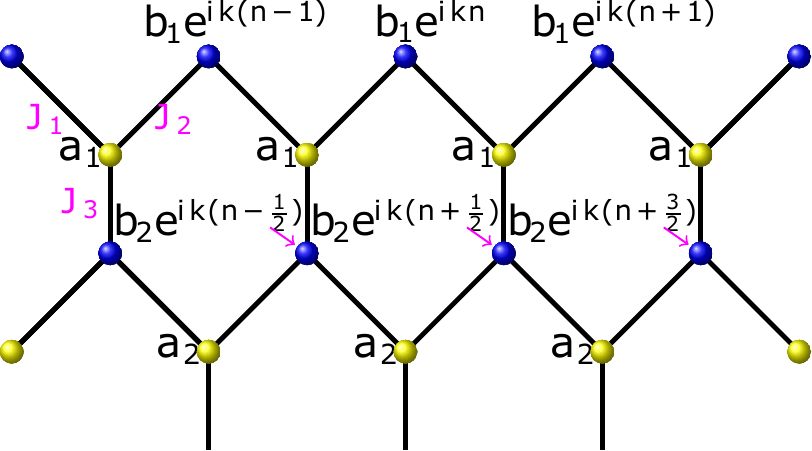}
\label{fig02}}\quad\quad%
\subfigure[]{
\ig[width=0.6\columnwidth]{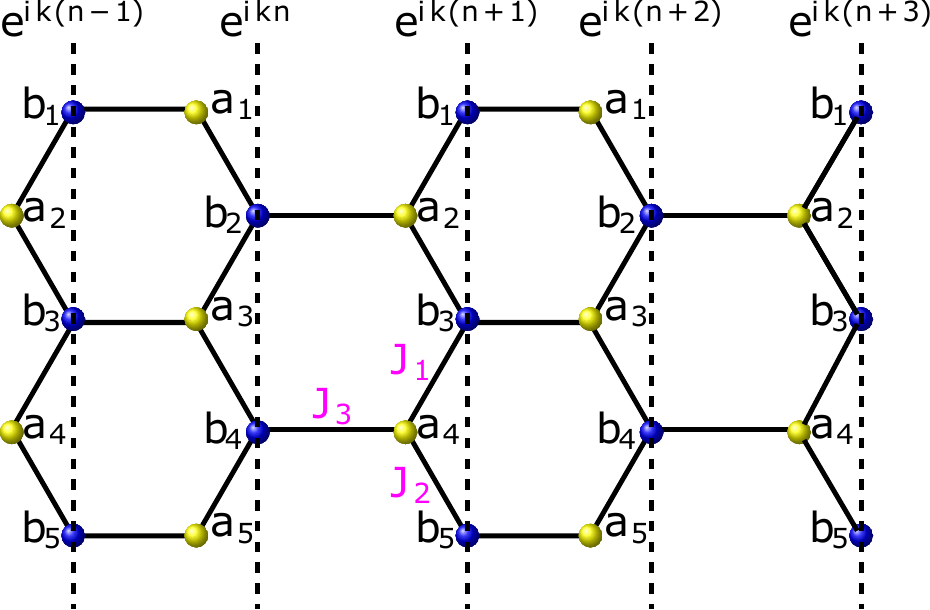}
\label{fig04}}
\caption{(a) Phase diagram for zigzag edge states in the triangle with $J_1+J_2+J_3=1$. Majorana modes exist in the regions $A_z$ and $B$. (b) Zigzag and (c) armchair edge configuration of the Kitaev model. Majorana fermions with a momentum $k$ along the edge are indicated. (See Ref.~\onlinecite{thakurathi14}).}
\end{figure*}

To construct a phase diagram of the model, it is convenient to impose the condition $J_1 + J_2 + J_3 = 1$, which describes points lying within (or on) an equilateral triangle. This triangle can be divided into four smaller equilateral triangles as shown in Fig.~\ref{fig03}, namely, $A_x$, $A_y$, $A_z$ and $B$. It turns out~\cite{kitaev06} that the system is gapped in the three $A$ phases, with $E_{\vec{k}}$ being non-zero for all $\vec k$, and is gapless in the $B$ phase, with $E_{\vec{k}} = 0$ for some value of $\vec k$ whose value depends on the location of the point in that phase. The four phases are separated from each other by quantum critical lines where one of the couplings is equal to the sum of the other two.

\begin{figure*}
\subfigure[]{
\ig[width=0.6\columnwidth]{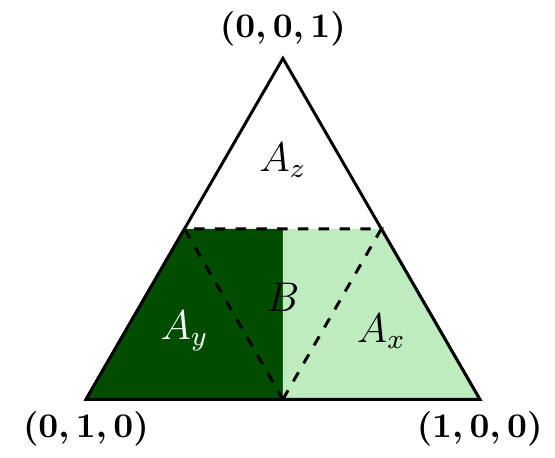}
\label{fig05}} \quad \quad%
\subfigure[]{
\ig[width=0.6\columnwidth]{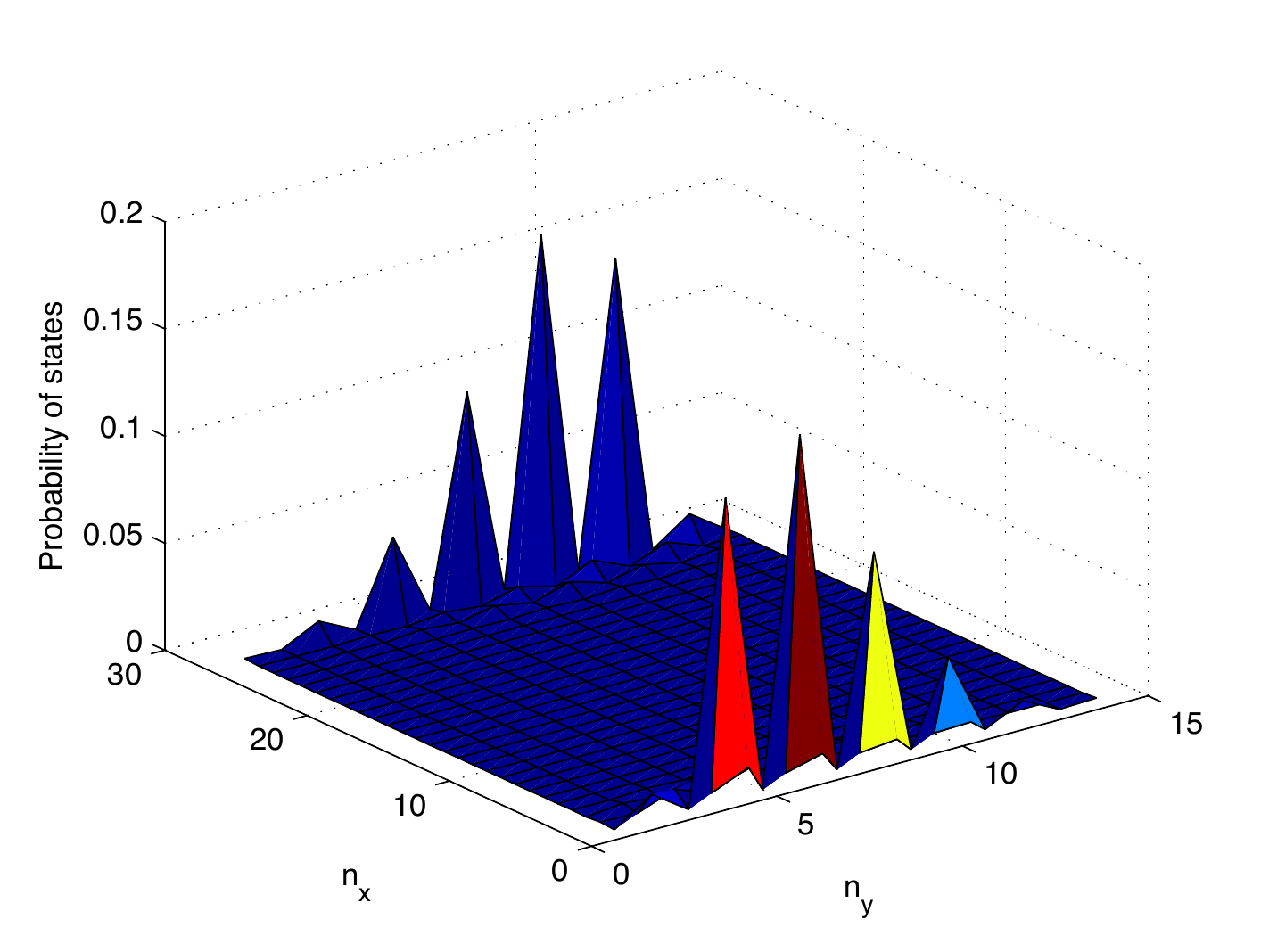}
\label{fig06}} \quad \quad%
\subfigure[]{
\ig[width=0.6\columnwidth]{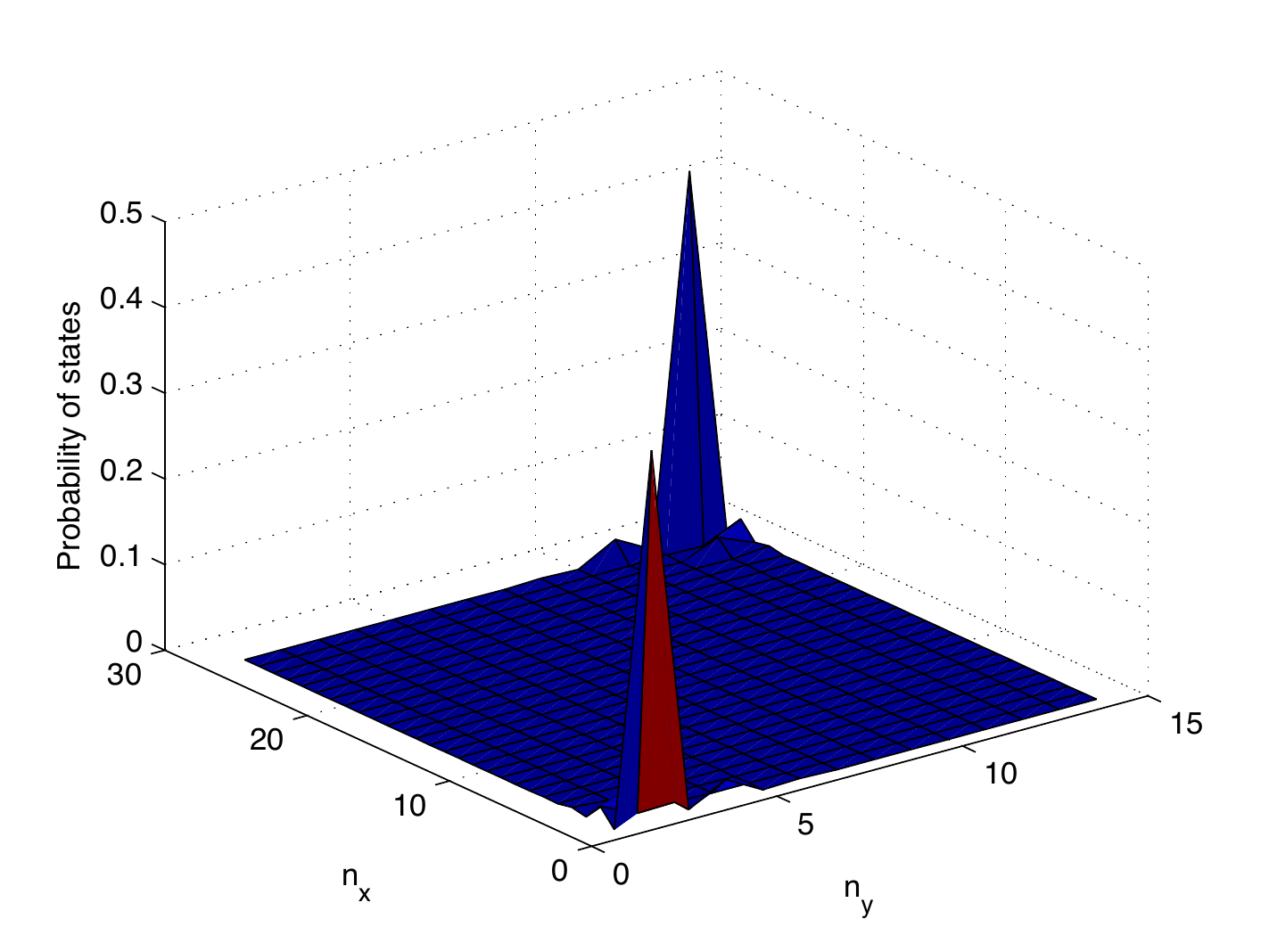}
\label{fig07}}	
\caption{(a) Phase diagram for armchair edge states in the triangle with $J_1+J_2+J_3=1$. Majorana modes of type $\hat b$ ($\hat a$) exist in the regions $A_y$ ($A_x$) and the left (right) half of $B$. These regions are indicated by dark (light) shades. (b) Armchair edge states and (c) corner states for a system with $N_x \times N_y = 27 \times 14$ sites, with $J_1=0.7, ~J_2=0.15$ and $J_3=0.15$. (See Ref.~\onlinecite{thakurathi14}).}
\end{figure*}

The nature of the edge modes which can appear in the Kitaev model is dependent on the geometry of the edge themselves. Depending on the orientation, the edges in a semi-infinite lattice can have different structures, two of such structures studied commonly are the zigzag
and armchair~\cite{nakada,kohmoto} edges. Assuming that the edges are infinitely long, translational
invariance permits us to label the edge states by their momentum
$k$. To find the edge states, we first write the Hamiltonian in
Eq.~\eqref{hmaj} in the form
\beq H ~=~ 2i ~\sum_{\al \be} ~\hat b_\be ~L_{\be \al} ~\hat a_\al, 
\label{hgen} \eeq
where $\al,~\be$ label the sites, and $L_{\al \be}$ is a real matrix.
We now use the Heisenberg equations of motion $d \hat a_\al / dt ~=~ i ~[H,
\hat a_\al]$ and similarly for $\hat b_\be$. We then obtain
\bea \frac{d\hat a_\al}{dt} &=& - 4~ \sum_{\be} ~\hat b_\be ~L_{\be \al}, 
\non \\
\frac{d\hat b_\be}{dt} &=& 4~ \sum_{\al} ~L_{\be \al} ~\hat a_\al. 
\label{eom1} \eea
Zero energy modes, if they exist, can only appear in the form of localized edge states and therefore can be found by equating the above Heisenberg equations to zero. We will henceforth denote wave functions by alphabet letters 
without hats (such as $a$ and $b$) to distinguish them from operators which 
are denoted by $\hat a$ and $\hat b$.

In Fig.~\ref{fig02}, the wave functions for the Majorana operators of type $\hat b$ are denoted by $b_{m,n}$, where $n$ goes from $-\infty$ to $\infty$ and increases towards the right along the edge, and $m=1,2,3,...$ increases as we go down away from the top edge and into the bulk of the system. Further, they can be decomposed into Fourier components as $b_{m,n} = b_m e^{ikn}$ or $b_m e^{ik(n+1/2)}$ depending on whether $m$ is odd or even. Similarly, the wave functions for Majorana operators of type $\hat a$ (not shown in Fig.~\ref{fig02}) are denoted by $a_{m,n} = a_m e^{ikn}$ or $a_m e^{ik(n+1/2)}$. Substituting in Eq.~\eqref{eom1} equating with zero reveals that there exists zero energy solutions with $a_m = 0$ for all $m$ and
\beq J_1 b_m e^{ikn}+ J_2 b_m e^{ik(n+1)}+ J_3 b_{m+1} e^{ik(n+\frac{1}{2})}
=0 \eeq
for all $m \ge 1$. From Fig.~\ref{fig02}, it is clear that the zero energy solution corresponds to localized edge states at the top where the Majorana operators are of type $\hat b$. Further, assuming $b_m=(\la_k)^m$ leads to
$\la_k = -(J_1 e^{-ik/2}+ J_2 e^{ik/2})/J_3$. 
For a normalizable edge state, we require $|\la_k| < 1$; this occurs if 
\beq \cos k ~<~ \frac{J_3^2-J_1^2-J_2^2}{2 J_1 J_2}. \label{zigcon} \eeq
The above condition holds true for all values of $k$ in region $A_z$ and for a finite range of values of $k$ in region $B$. On the contrary, no solution exist for Eq.~\eqref{zigcon} in the regions $A_x$ and $A_y$, which implies the absence of edge states in these regions. A similar analysis for zero energy solutions with $b_m = 0$ for all $m$ shows the presence of localized edge states at the bottom of the semi-infinite lattice (see Fig.~\ref{fig02}) where the Majorana operators are of type $\hat a$.

For armchair edges (see Fig.~\ref{fig04}), the $\hat a$ and $\hat b$ Majorana operators have wave functions given by $a_m e^{ikn}$ and $b_m e^{ikn}$, respectively. In this case, the Heisenberg equations of motion have zero energy solutions with $a_m = 0$ for all $m$, provided that 
\beq J_1 b_m + J_2 b_{m+2} + J_3 b_{m+1} e^{-ik} ~=~ 0 \label{arm1} \eeq 
for all $m \ge 1$, and 
\beq J_2 b_2 + J_3 b_1 e^{-ik} ~=~ 0. \label{arm2} \eeq
Assuming $b_m=(\la_k)^m$, Eq.~\eqref{arm1} has two roots given by
$\la_{k\pm} = (- J_3 e^{-ik} \pm \sqrt{J_3^2 e^{-2ik} - 4 J_1J_2})/(2J_2).$
Eqs.~(\ref{arm1}-\ref{arm2}) imply that a normalizable edge state
will exist if both $|\la_{k\pm}| < 1$. We find that this occurs if
$J_1 < J_2$ and $J_3 \le J_1 + J_2$. 
The corresponding region in the phase diagram is shown as a dark shaded region on the left side of Fig.~\ref{fig05}. A similar analysis shows that zero energy edge modes of type $\hat a$ (i.e., with $b_m = 0$ for all $m$) also exist, the corresponding region is shown as light shaded in Fig.~\ref{fig05}.

Finally, it is important to consider a finite system with armchair edges along one direction and zigzag edges along the other. Interestingly, as shown in Figs.~\ref{fig06} and~ \ref{fig07}, while some edge modes are found to be localized along the armchair edges, the rest appear only at the corners.

We remark here that although the Kitaev honeycomb model is capable of hosting Majorana edge modes, the latter do not arise as a consequence of an underlying bulk topological structure of the complete two-dimensional Hamiltonian. This can be argued from a symmetry analysis of the system. From Eq.~\eqref{hmom}, it is straightforward to see that the Hamiltonian possesses chiral, time-reversal as well as particle-hole symmetries, with the symmetry operators identified as 
$\mathcal{S} = \tau^z$, $\mathcal{T} = i\tau^y$ and $\mathcal{P} = \tau^x$, respectively. The model thus belongs to the BDI symmetry class which lacks any topological structure for two-dimensional systems. However, as shown in Ref.~\onlinecite{thakurathi14}, the two-dimensional semi-infinite Kitaev ribbon can be mapped to an aggregate of one-dimensional Kitaev chains for each quasimomentum along the infinitely long direction (see also Ref.~\onlinecite{chitra21}) which might host topological end Majoranas for some particular quasimomenta. It is these Majorana modes which appear at the edges of the Kitaev Honeycomb lattice.

\subsection{Haldane model of graphene} 

The Haldane model of graphene is a paradigmatic system which encompasses the basic physics of all Chern insulators and have tremendous application in the design of topologically protected quantum states. The bare Hamiltonian for the Haldane model \ct{haldane83} is obtained by breaking the time-reversal and sublattice symmetries of graphene,

\begin{figure*}
\subfigure[]{
\centering
\includegraphics[width=0.8\columnwidth]{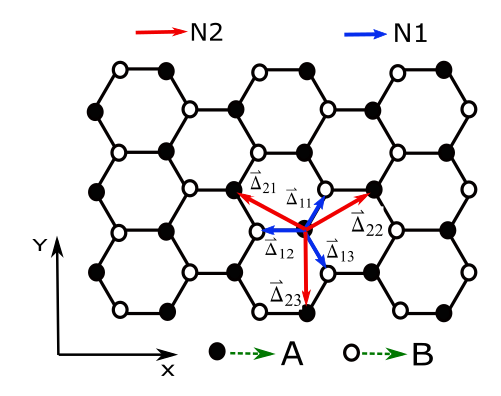}
\label{fig_1a_ap}} \quad\quad
\subfigure[]{
\includegraphics[width=0.8\columnwidth]{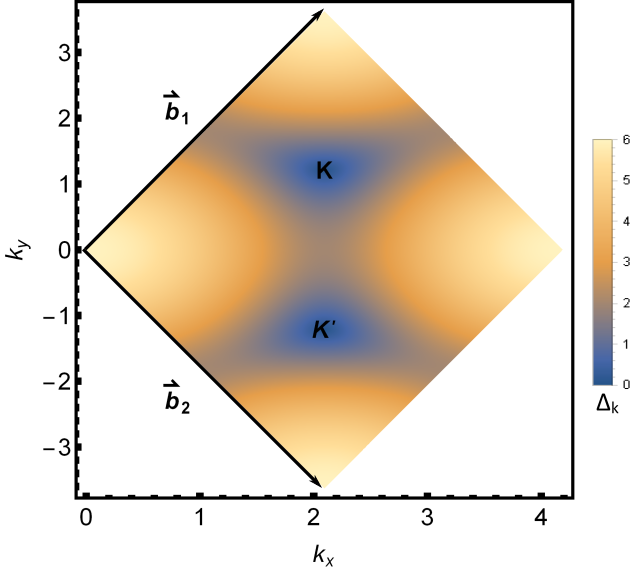}
\label{fig_1b_ap}}
\caption{(a) Hexagonal graphene lattice showing the nearest-neighbor (N1) and next-nearest-neighbor (N2) hopping vectors $\vec{\Delta}_{1i}$ and $\vec{\Delta}_{2i}$, respectively, where the lattice constant is set to be $a=1$. The hollow and the filled atoms represent the B and A sublattices respectively. (b) The Brillouin zone of graphene spanned by the reciprocal lattice vectors $\vec{b}_1$ and $\vec{b}_2$ containing two inequivalent Dirac points $K$ and $K^{\prime}$ (the cartesian directions has been labeled by $k_x$ and $k_y$ respectively). The color density shows the absolute value of the band gap 
$\Delta_k$ of the reciprocal space graphene Hamiltonian showing vanishing gaps at the Dirac points for a $600\times 600$ lattice size having the N1 hopping strength $t_1=1.0$ and the N2 hopping $t_2=0$.}
\end{figure*}

\begin{widetext}
	\begin{equation}
	H_{\alpha,\beta,n,m}=-t_1\sum\limits_{\left<m\alpha,n\beta\right>}a_{m,\alpha}^{\dagger}a^{}_{n,\beta}+M\sum\limits_{n} a_{n,A}^{\dagger}a_{n,A}^{}
	-M\sum\limits_{n} a_{n,B}^{\dagger}a_{n,B}^{}- t_2 {\sum\limits_{\left<\left<m\alpha,n\alpha\right>\right>} [e^{i\phi_{m,n}}a_{m,\alpha}^{\dagger}
a_{n,\alpha}^{}}+{\rm H.c.}], \end{equation}
\end{widetext}
where the real nearest-neighbor (N1) hopping $t_1$ (with $t_2=0,M=0$) comprises the bare graphene Hamiltonian; the indices $n$ and $\alpha$ represent site and sublattice respectively. The diagonal staggered mass (Semenoff mass) $M$ explicitly breaks the sublattice symmetry of the model. {Further, the complex phase $\phi_{m,n}$ is chosen such that within each hexagonal plaquette, $\phi_{m,n}=\phi$ if the hopping is in the clockwise sense while $\phi_{m,n}=-\phi$ if the corresponding hopping is in the anti-clockwise sense.} The complex phase thus breaks the time-reversal symmetry in the next nearest-neighbor hopping. However, the net flux through each hexagonal plaquette remains zero.
{The Haldane model does not respect any of the $\mathcal{S}$, $\mathcal{P}$ or $\mathcal{T}$ symmetries. Depending on the parameters $M$, $t_1$, $t_2$ and $\phi$, it is known to exhibit non-trivial Chern topology in its ground state when the lower energy band is completely filled}. This is somewhat different {from the topology manifested in the SSH and Kitaev models considered earlier, where the topology of those models were protected by discrete symmetries}. However, the Haldane Chern insulator explicitly breaks all symmetries yet being a two-dimensional system, can still host a non-trivial topology.

\begin{figure} 
\ig[width=\columnwidth]{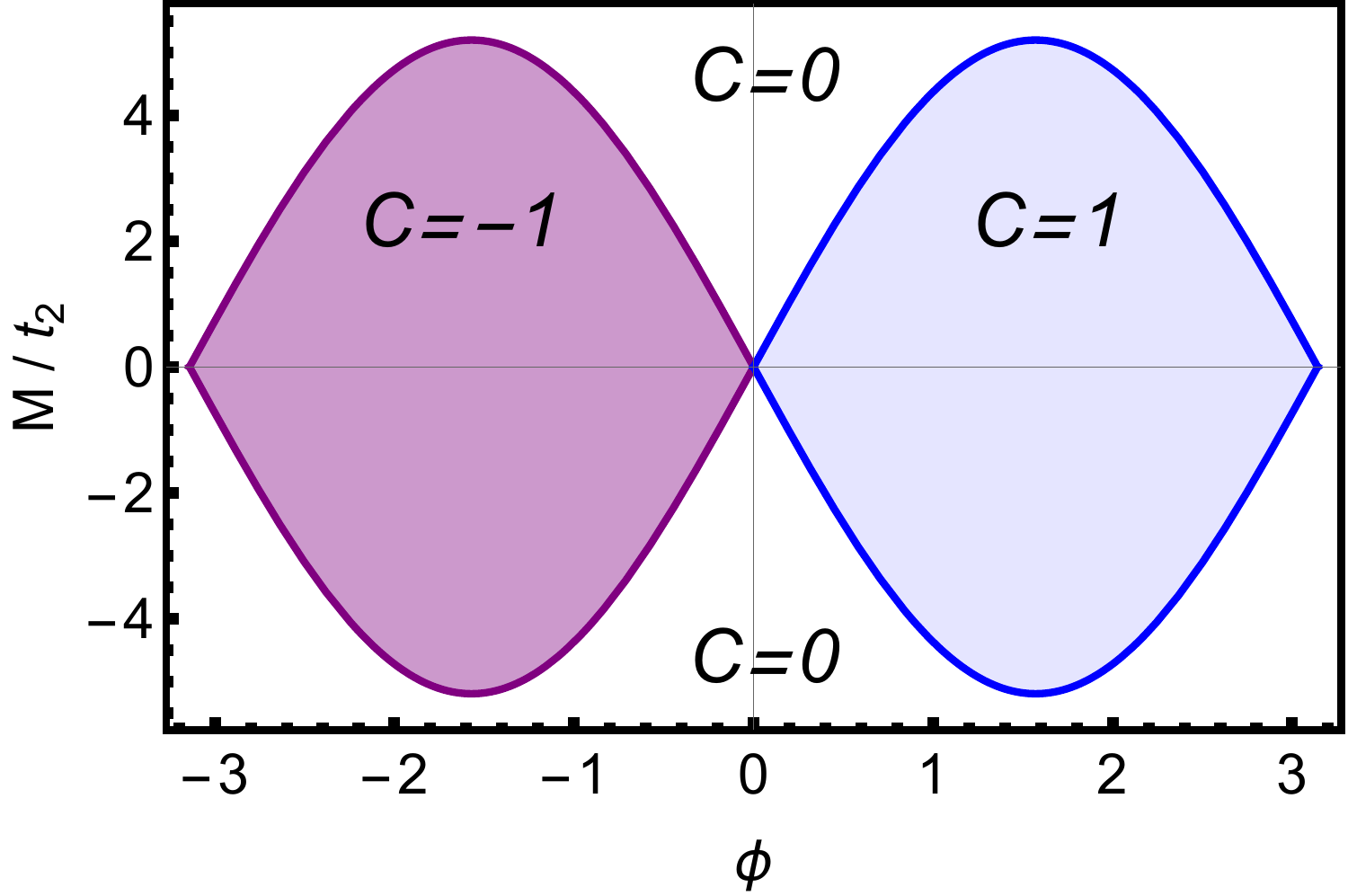}
\caption[]{Topological phase diagram of the Haldane model of graphene where the inequivalent phases (with different Chern numbers) are separated by gapless critical lines (solid lines in the figure).} \label{figchern} 
\end{figure}

For a half-filled Haldane system, the topology of the Hamiltonian with periodic boundary conditions is precisely characterized by a gauge invariant Chern topological invariant (see Eq.~\eqref{chern_number} of Sec.~\ref{sec:topology}),
\begin{equation}\label{haldane_chern}
C=\frac{1}{2\pi}\oint_{BZ}dk_1dk_2\mathcal{F}_{12}(\ket{\psi_k}),
\end{equation}
where $\mathcal{F}_{12}(\ket{\psi_k})$ is the $U(1)$ curvature defined over the ground state $\ket{\psi_k}$ of the Hamiltonian $H^k$, i.e.,
\begin{equation}
\begin{split}
\mathcal{F}_{12}(\ket{\psi_k})= i\partial_{k_2}\langle{\psi_k|\partial_{k_1}|\psi_k}\rangle-i\partial_{k_1}\langle{\psi_k|\partial_{k_2}|\psi_k\rangle}.
\end{split}
\end{equation}
The Chern invariant is integer quantized as long as {the system} does not approach a QCP where the Chern number becomes ill-defined. Different integer values of the Chern number characterize distinct topological phases separated by QCPs (see Fig.~\ref{figchern}). 

Each point on the {underlying Bravais lattice of the graphene honeycomb lattice can be referenced in terms of the vector,}
{\begin{equation}
	\vec{r}=n_1\vec{a}_1+n_2\vec{a}_2,
	\end{equation}}
{where $\vec{a}_1$ and $\vec{a}_2$ are the primitive vectors and $n_1,n_2$ are integers.
	We choose the vectors $\vec{a}_1$ and $\vec{a}_2$ to be the next-nearest-neighbor (N2) hopping vectors such that
	\begin{eqnarray}\label{eq:lattice:ap}
	\vec{a}_1&=&\vec{\Delta}_{22}, \non \\
	\vec{a}_2&=&-\vec{\Delta}_{21},
	\end{eqnarray}
	where $\vec{\Delta}_{2i}$ are the $N2$ vectors as shown in Fig.~\ref{fig_1a_ap}. Their values are found to be,
	\begin{equation}
	\begin{split}
	\vec{\Delta}_{11}=\frac{a}{2}\{1,\sqrt{3}\},~~\vec{\Delta}_{12}=\{-a,0\},~~\vec{\Delta}_{13}=\frac{a}{2}\{1,-\sqrt{3}\}, \\
	\vec{\Delta}_{21}=\frac{a}{2}\{-3,\sqrt{3}\},~~\vec{\Delta}_{22}=\frac{a}{2}\{3,\sqrt{3}\},~~\vec{\Delta}_{23}=\{0,-a\sqrt{3}\},
	\end{split}
	\end{equation}
	where $a$ is the distance between neighboring sites.}

Invoking the discrete translational invariance of the Hamiltonian, one can employ a discrete Fourier transform to decouple the Hamiltonian $H$ in momentum space. The reciprocal space is spanned by the vector $\vec{b}=k_1\vec{b}_1+k_2\vec{b}_2$, where the primitive vectors are given by
\begin{equation}
\vec{b}_1=\frac{2\pi}{3a}\{1,\sqrt{3}\}~~~\text{and}~~~\vec{b}_2=\frac{2\pi}{3a}\{1,-\sqrt{3}\},
\end{equation}
{where $k_1$ and $k_2$ assume integer values.} We choose a rhomboidal BZ spanned by reciprocal lattice vectors $\vec{b}_1$ and $\vec{b}_2$ (see Fig.~\ref{fig_1b_ap}). It contains two independent Dirac points $K$ and $K^{\prime}$ see (Fig.~\ref{fig_1b_ap}),
\begin{equation}
K=\frac{2\pi}{3}\left(1,\frac{1}{\sqrt{3}}\right)~~\text{and}~~K^{\prime}=\frac{2\pi}{3}\left(1,-\frac{1}{\sqrt{3}}\right),
\end{equation}
where we have set $a=1$.

The bare Haldane Hamiltonian gets decoupled in the momentum space where $H(k)$ can be written in the basis $\ket{k,A}$ and $\ket{k,B}$ as,
\begin{equation}
H(k)=\vec{h}(k)\cdot\vec{\sigma} =h_x(k) \sigma^x+h_y(k) \sigma^y+h_z(k) \sigma^z,
\label{eq_hamil_k_ap}
\end{equation}
such that
\begin{equation}\label{eq:bloch_ham_ap}
\begin{split}
h_x(k)=-t_1\sum\limits_{i=1}^{3}\cos{\left(\vec{k}\cdot\vec{\Delta}_{1i}\right)},\\
h_y(k)=-t_1\sum\limits_{i=1}^{3}\sin{\left(\vec{k}\cdot\vec{\Delta}_{1i}\right)},\\
{h_z(k)=M-2t_2\sin{\phi}\sum\limits_{i=1}^{3}\sin{\left(\vec{k}\cdot\vec{\Delta}_{2i}\right)},}
\end{split}
\end{equation}
\\

Similar to the one-dimensional models, the Chern insulator also {exhibits a topological bulk-boundary correspondence -- the topological non-triviality of the bulk system with periodic boundary conditions is reflected as {dissipation-less conducting states} on the boundary surface under open boundary conditions}. This is seen from the Kubo formula which explicitly connects the bulk Chern number with the transverse Hall conductivity in clean systems. Using linear response theory, it can be rigorously established that at sufficiently low temperatures, the transverse Hall conductivity is given by\ct{hajdu87,vasilopoulos11},
	\begin{equation}
	\sigma_{xy}=-\frac{e^2}{2\pi\hbar}C,
	\end{equation}
	where $e$ is the electronic charge, $\hbar$ is the Planck's constant and $C$ is the integer-quantized topological Chern number defined in Eq.~\eqref{haldane_chern}. This connection lies at the root of the highly stable integer quantized anomalous quantum Hall effect manifested in such systems without the presence of any external magnetic fields. We have thus found a robust physical manifestation of the complete geometric 
and topological structure of quantum states which can be experimentally probed. Indeed, the anomalous quantum Hall effect and the quantum Hall effect are some
of the benchmark experiments in physics till date which show near perfect quantization of the transverse Hall conductivity thus firmly establishing the role of topology in quantum many-body systems. Interestingly, the mere semblance between the electromagnetic theory and the geometry of quantum states that we discussed extensively earlier has given rise to a completely new field called {\it Berry electrodynamics} where the action of the gauge field $F_{\mu\nu}$ on moving charges is studied. In simple terms, it concerns the motion of charges on the curved parameter-manifold and hence, one must account for its curvature which itself appears as the gauge field, similar to the "magnetic field" in electrodynamics.

{We have reached the end of the} first part of our review whose primary goal was to introduce the rich concept of topology in quantum many-body systems. {We next move on to} the second part in which we will be focusing on the topological physics of dynamical quantum many-body systems and their experimental realization.\\


\section{Periodically driven topological systems}\label{sec_periodic}

It has been seen that irradiated quantum many-body systems develop novel out-of equilibrium properties which are absent in the equilibrium system. This has initiated a plethora of studies in periodically driven quantum many-body systems to explore exciting new phases of matter \cite{oka08, kitagawa11, lindner11, moessner13}. For, example, it has been seen that bare graphene which does not have a well-defined topological classification in equilibrium, may exhibit non-trivial topological properties when irradiated with light\cite{kitagawa11}. In this section, we will discuss two such interesting systems, i.e., periodically driven Kitaev models both as a Kitaev chain\ct{thakurathi13,smitha13} and the Kitaev honeycomb lattice\ct{thakurathi14}.

\subsection{Periodically driven Kitaev chain}\label{subsec_kitaev_chain_dyn}

Let us first consider the case of a periodically driven Kitaev chain \cite{thakurathi13}. In particular, we explore the simple situation where the chemical potential $\mu$ in the Kitaev chain Hamiltonian, given in Eq.~\eqref{ham1}, is periodically kicked in time, 
\beq \mu (t) ~=~ c_0 ~+~ c_1 \sum_{n=-\infty}^\infty \de (t - nT). 
\label{ht1} \eeq
Here, $T=2\pi/\om$ is the time interval between subsequent kicks. In the Majorana representation, the Hamiltonian can be represented in the general form
\beq H(t) ~=~ i \sum_{m,n=1}^{2N} ~a_m M_{mn}(t) a_n, \label{ham4} \eeq
where $M(t)$ is a real antisymmetric matrix satisfying $M(t+T)=M(t)$. Noting that the Heisenberg operators $a_n (t)$ satisfy,
\beq \frac{da_m(t)}{dt} ~=~ 4 \sum_{n=1}^{2N} ~ M_{mn} (t) ~a_n (t), \eeq
the time-evolution operator is found to be,
\bea U(t_2,t_1) &=& {\cal T} e^{4 \int_{t_1}^{t_2} dt M(t)}, \eea
where $\cal T$ denotes the time-ordering symbol. 

We will restrict ourselves to observing the system only at stroboscopic instants, separated by time interval $T$. Floquet theory dictates that the evolution of a quantum system at such instants is generated by the Floquet operator $U(T,0)$ \cite{floquet_sens_47, shirley65, bukov_ap_139, eckardt_rmp_011004}. It is important to note that the eigenvalues of $U(T,0)$ are given by phases, $e^{i\ta_j}$, and they come in complex conjugate pairs if $e^{i\ta_j} \ne 1$. This is because $U(T,0) \psi_j = e^{i\ta_j} \psi_j$ implies that $U(T,0) \psi_j^* = e^{-i\ta_j} \psi_j^*$. 
For eigenvalues $e^{i\ta_j} = \pm 1$ (these eigenvalues may, in principle, 
appear with no degeneracy), the eigenvectors can be chosen to be real; this can
be seen by noting that $U(T,0)\psi_j = \pm\psi_j$ also implies $U(T,0)\psi_j^* = \pm\psi_j^*$ (as $U(T,0)$ is a real orthogonal matrix) and one can choose the eigenvectors to be the real combinations $\psi_j + \psi^*_j$ and $i(\psi_j - \psi^*_j)$.

For the periodic $\de$-function kick in Eq.~\eqref{ht1}, the Floquet operator assumes the form
\beq U(T,0) ~=~ e^{4 M_1} ~e^{4 M_0 T}, \label{flo1x} \eeq 
where $M_{0/1}$ are $(2N)$-dimensional antisymmetric matrices whose non-zero 
matrix elements can be found using Eqs.~\eqref{ham2}, \eqref{ht1} and \eqref{ham4}:
\bea (M_0)_{2n+1,2n} &=& - ~(M_0)_{2n,2n+1} ~=~ - \frac{1}{4} (\ga - \De), 
\non \\
(M_0)_{2n-1,2n+2} &=& - ~(M_0)_{2n+2,2n-1} ~=~ - \frac{1}{4} (\ga + \De), 
\non \\
(M_0)_{2n-1,2n} &=& - ~(M_0)_{2n,2n-1} ~=~ \frac{c_0}{2}, \non \\
(M_1)_{2n-1,2n} &=& - ~(M_1)_{2n,2n-1} ~=~ \frac{c_1}{2}, \label{mt1} \eea
for an appropriate range of values of $n$. A more convenient form is to use the symmetrized expression 
\beq U(T,0) ~=~ e^{2 M_1} ~e^{4 M_0 T} ~e^{2 M_1}. \label{flo2x} \eeq
It is easy to show that the Floquet operators in Eqs.~\eqref{flo1x} and 
\eqref{flo2x} have the same eigenvalues, while their eigenvectors are related
by a unitary transformation. We will see below that the symmetrized form in 
Eq.~\eqref{flo2x} leads to some simplifications when we derive an effective 
Hamiltonian and a topological invariant.

The inverse participation ratio (IPR) serves as a useful probe to find the eigenvectors of $U(T,0)$ which are localized at the ends. Assuming that the eigenvectors are normalized so that $\sum_{m=1}^{2N} |\psi_j (m)|^2 = 1$ for each value of $j$, where $m=1,2,\cdots,2N$ labels the components of the eigenvector, the IPR of an eigenvector is defined as $I_j = \sum_{m=1}^{2N} |\psi_j (m)|^4$. If $\psi_j$ is extended in real space, so that 
$|\psi_j (m)|^2 \approx 1/(2N)$ for each $m$, then $I_j = 1/(2N)$;
which vanishes as $N \to \infty$. On the contrary, if $\psi_j$ is localized over
a distance $\xi$ (which is of the order of the decay length of the eigenvector
and remains constant as $N \to \infty$), then $|\psi_j (m)|^2 
\sim 1/\xi$ in a region of length $\xi$ and $\sim 0$ elsewhere; the IPR thus remains finite as $N \to \infty$. For sufficiently large $N$, one can hence identify localized states, if any. A subsequent analysis of the plot of the probabilities $|\psi_j (m)|^2$ versus $m$ reveals whether the localized state is indeed an end state.

\begin{figure*}
\subfigure[]{
\ig[width=\columnwidth]{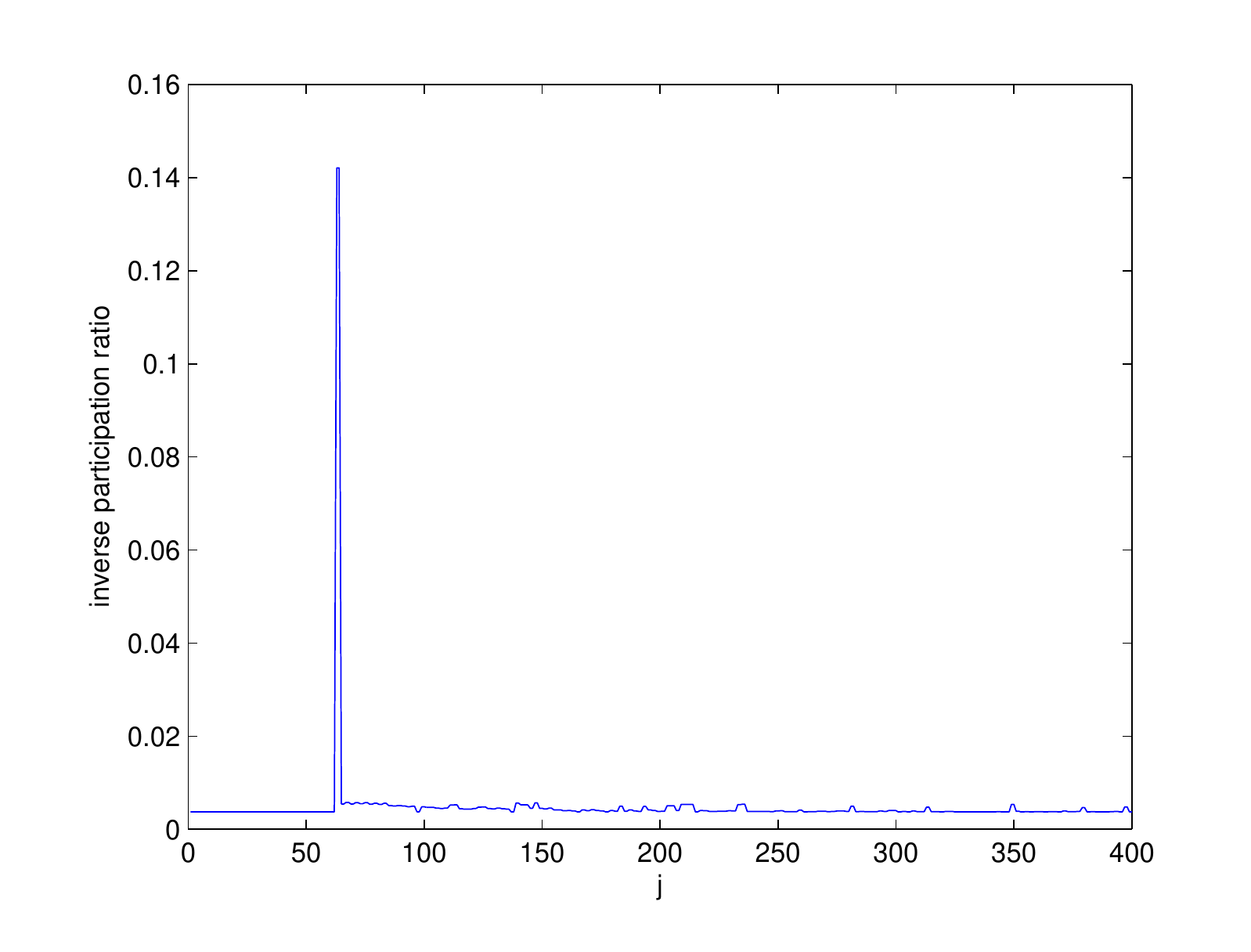}
\label{flo1}} 
\subfigure[]{
\ig[width=\columnwidth]{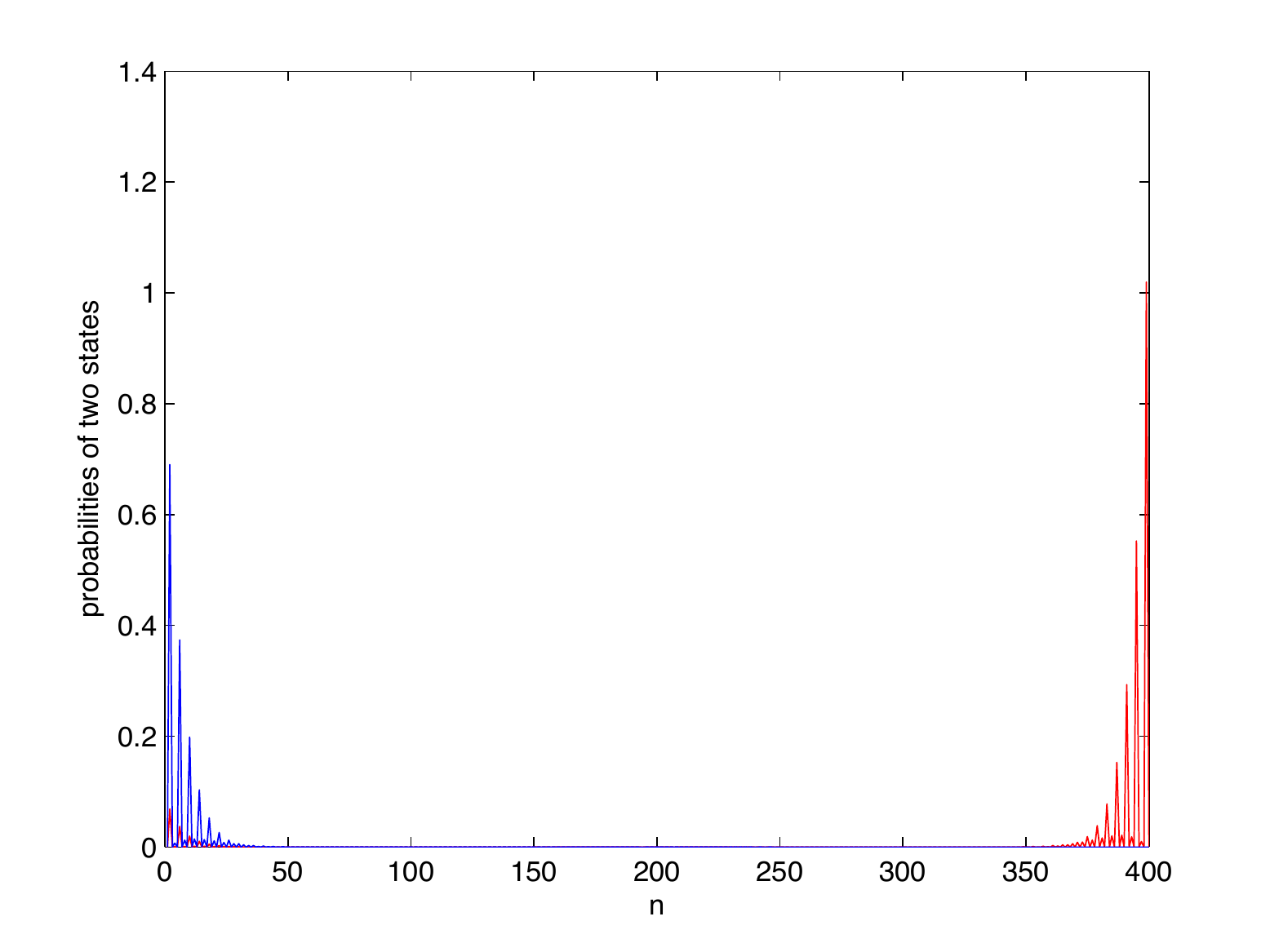}
\label{flo2}} 
\subfigure[]{
\ig[width=\columnwidth]{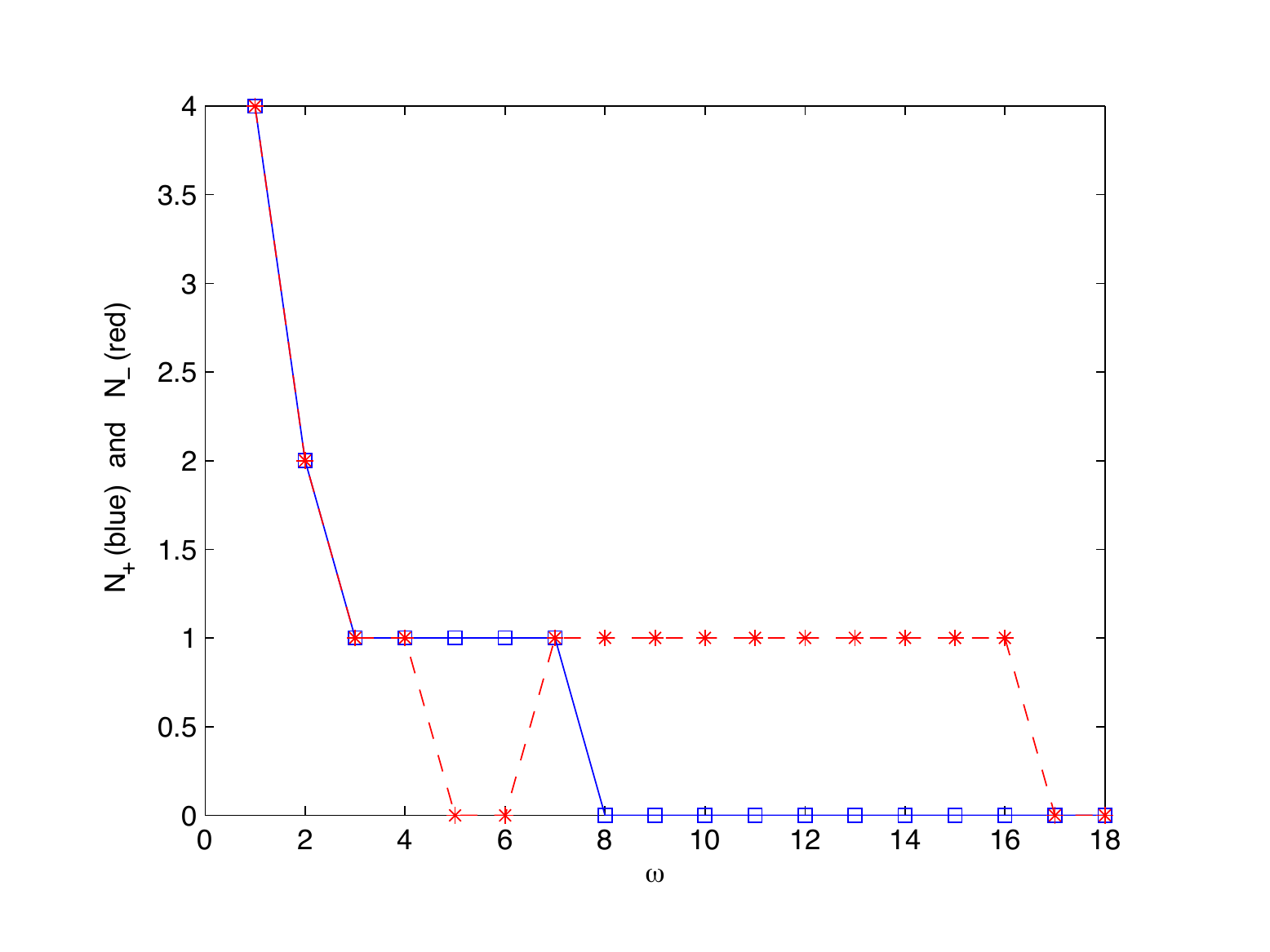}
\label{flo3}} 
\caption{(a) IPRs of different eigenvectors of the Floquet operator for a 200-site system with a periodic $\de$-function kick with $\ga =1$, $\De = - 1$, $c_0=2.5$, $c_1=0.2$ and $\om=12$. The two eigenvectors with the largest IPRs both have an IPR equal to $0.142$ and Floquet eigenvalue equal to $-1$. (b) The corresponding Majorana end states of the two eigenvectors with the largest IPRs. (c) Plot of the number of end states versus $\om$ for a 200-site system, with Floquet eigenvalues $+1$ ($N_+$, blue squares) and $-1$ ($N_-$, red stars). Values of other relevant parameters are same as that in (a) (See Ref.~\onlinecite{thakurathi13}).}
\end{figure*}

Fig.~\ref{flo1} shows the IPRs of the different eigenvectors for a finite system with $N = 200$ with parameters chosen as $\ga =1$, $\De = -1$, $c_0=2.5$, $c_1=0.2$ and $\om = 12$. Two of the IPRs clearly stand out with a value of $0.142$ each. The corresponding eigenvectors are found to have Floquet eigenvalues $e^{i\ta} = -1$; importantly, the value of $\ta=\pi$ is separated by a gap of $0.148$ from the values of $\ta$ for all the other eigenvalues. These eigenvectors are localized at the two ends of the system and are real; the corresponding probabilities are shown in Fig.~\ref{flo2}. The 
state at the left end has non-zero $a_m$ only if $m$ is even, while the state 
at the right end has non-zero $a_m$ only for $m$ odd. Thus, the periodic driving produces Majorana end modes even though for the parameter values given above, the ground state of the instantaneous Hamiltonian is at all times in topologically trivial phase. Remarkably, the number of Majorana end modes is crucially dependent on the driving frequency $\om$. This is illustrated in Fig.~\ref{flo3}, where the number of eigenvalues lying near $+1$ and $-1$, denoted by $N_+$ and $N_-$, respectively, are plotted as a function of $\om$ for a finite system of size $N=200$ with $\ga =1$, $\De = -1$, $c_0=2.5$ and $c_1=0.2$. Although the number of end modes is not a monotonic function of $\om$, the number generally increases as $\om$ decreases. Recently, it has also been shown that the number and locality of these Majorana modes can be controlled if the system is driven by multiple periodic drives having commensurate frequencies\cite{pablo20}.

\subsection{Topological invariants for periodic driving}

As in the equilibrium situation, to define a topological invariant, let us consider a system with periodic boundary conditions. Once again, Fourier transforming to the quasimomentum basis decouples each momentum mode; for each value of $k$, one can define a Floquet operator $U_k (T,0)$ which is a $2 \times 2$ unitary matrix. Using Eqs.~\eqref{hk}, \eqref{ht1} and \eqref{flo2x}, it follows that 
\beq U_k (T,0) = e^{ic_1 \tau^z} e^{-i2T[(\ga \cos k -c_0) \tau^z + \De \sin k
	\tau^y]} e^{ic_1 \tau^z}, \label{uk1} \eeq
where $k$ is taken to lie in the full range $- \pi \le k \le \pi$. A couple of remarks about the Floquet operator $U_k(T,0)$ are in order. First, assuming that $\De \ne 0$, it is easy to show that $U_k (T,0) \ne \pm \mathbb{I}$ for any value of $k \ne 0$ or $\pi$, where $\mathbb{I}$ is the identity operator. Second, for $k = 0$ or $\pi$, $U_k (T,0) \ne \pm \mathbb{I}$ unless
$2T (c_0 \pm \ga) + 2c_1 = n \pi$, or equivalently unless
\beq \om ~=~ \frac{4\pi (c_0 \pm \ga)}{n \pi - 2c_1} \label{omeq} \eeq
for some integer value of $n$. The $\pm$ sign in Eq.~\eqref{omeq} corresponds
to $k=\pi$ and $0$ respectively. In other words, if the condition set by Eq.~\eqref{omeq} is not satisfied, the eigenvalues of the Floquet operator are non-degenerate for all values of $k$ and thus the `Floquet eigenspectrum' is gapped. When Eq.~\eqref{omeq} is satisfied, the spectrum is gapless at modes $k=0$ or $\pi$, which can be associated with a non-equilibrium phase transition.

Utilizing the properties of the Floquet operator discussed above, one can now construct a topological invariant as follows. First, let us define an effective Hamiltonian $h_{\mathrm{eff},k}$ as 
\beq U_k (T,0) ~=~ e^{-ih_{\mathrm{eff},k}}. \label{heff1} \eeq
The symmetric structure of Eq.~\eqref{uk1} (as chosen in Eq.~\eqref{flo2x}) ensures that $h_{\mathrm{eff},k}$ assumes the form
\beq h_{\mathrm{eff},k} ~=~ h^y(k) ~\tau^y ~+~ h^z(k) ~\tau^z \label{heff2} \eeq
as in Eq.~\eqref{hk}. Further, to define $h_{\mathrm{eff},k}$ uniquely, we impose 
the condition that the coefficients in Eq.~\eqref{heff2} satisfy
$0 < \sqrt{h^y(k)^2 + h^z(k)^2} < \pi$, for $U_k (T,0) \ne \pm \mathbb{I}$. A winding number $W$, analogously to the equilibrium situation, can thus be computed as described in Eq.~\eqref{wind}.

An useful interpretation of the winding number defined above can be made by noting that, the condition $0 < h^y(k)^2 + h^z(k)^2 < \pi$ implies that $h_{\mathrm{eff},k}$ can be mapped to a point on the surface of a sphere whose polar angles $(\al,\beta)$ are given by $\al = \sqrt{h^y(k)^2 + h^z(k)^2}$ and $\beta = \tan^{-1} (h^z(k)/h^y(k))$. As $k$ goes from 0 to 
$2\pi$, we obtain a closed curve which does not pass through the north and 
south poles. The integer $W$ can then be related to the winding number of 
this curve around either the north pole or the south pole. Note that the
winding numbers around the north and south pole are given by the same integer.

\begin{figure}[htb]
\ig[width=0.95\columnwidth]{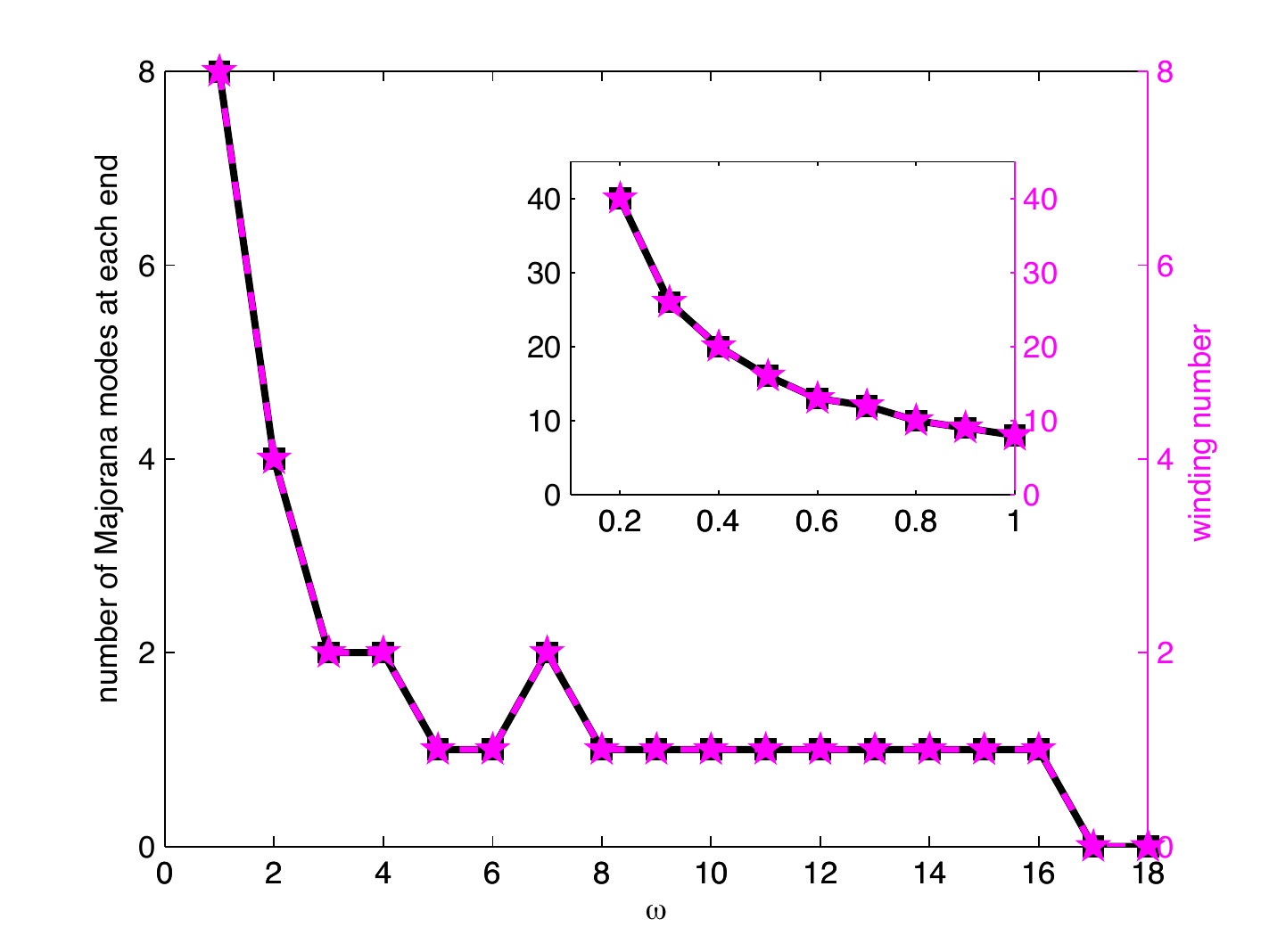}
\caption[]{Comparison of the number of Majorana modes at each end of a 200-site system (black solid, $y$-axis on left) and the winding number (magenta dashed, $y$-axis on right) as a function of $\om$ from 1 to 18, for $\ga =1$, $\De = -1$, $c_0=2.5$ and a periodic $\de$-function kick with $c_1=0.2$. The inset shows a range of $\om$ from $0.2$ to 1 where there is a large number of Majorana modes (See Ref.~\onlinecite{thakurathi13}).} \label{flo5} 
\end{figure}

Remarkably, the winding number exhibits a bulk-boundary correspondence, akin to its equilibrium counterpart. This correspondence can be seen in Fig.~\ref{flo5} which compares the number of Majorana modes at each end of a 
chain and the winding number as a function of $\om$. Note that in the limit $\om \to \infty$, i.e., $T \to 0$, Eq.~\eqref{uk1} becomes independent of $k$, and we therefore find a single point in the $(h^y(k), h^z(k))$ plane. This corresponds to a curve with zero winding number; thus there is a maximum value of $\om$ beyond which no Majorana end modes exist. On the other hand, for very small $\om<0.2$, the number of Majorana end modes appear to increase drastically and diverge in the limit $\om\to0$.

It turns out that it is possible to define an alternate topological invariant that not only exhibits a direct correspondence with the Majorana end modes but can also distinguish between end modes with eigenvalues $+1$ and $-1$. To elucidate, let us first define a set of parameters, 
\bea b_0 &=& \frac{4(c_0 - \ga)}{\om} ~+~ \frac{2c_1}{\pi}, \non \\
b_\pi &=& \frac{4(c_0 + \ga)}{\om} ~+~ \frac{2c_1}{\pi}, \label{b0p} \eea
so that the Floquet operators at $k=0,\pi$ can be recast as $U_0 (T,0) = e^{i\pi b_0 \tau^z}$ and $U_\pi (T,0) = e^{i\pi b_\pi \tau^z}$. Next, we define a finite line segment, $L_\om$, which goes from $b_0$ to $b_\pi$ in one dimension; this line segment is further assumed to coincide with the $z$-axis. 

For $\om \to \infty$, the line $L_\om$ collapses to a single point given by 
$z=2c_1/\pi$. As $\om$ is decreased, $L_\om$ moves and is also increased in length. With the system parameters chosen as $\ga = 1$, $c_0 = 2.5$ and $c_1 = 0.2$, the {\it right} end of $L_\om$, given by $b_\pi$ in Eq.~\eqref{b0p}, crosses the point $z=n$ with $n=1$ when $\om\simeq 16.04$. at this point, the Floquet eigenvalue at $k=\pi$ thus assumes the value $e^{in\pi} = -1$ and is also degenerate. On further decreasing $\om$, the point $z=1$ enters the segment $L_\om$. Importantly, as can be seen from Fig.~\ref{flo5}, a Majorana end 
mode first appears within the range $16 \le \om \le 17$ and also has a Floquet 
eigenvalue equal to $-1$. As $\om$ is decreased further, the right end of 
$L_\om$ given by $b_\pi$ crosses the point $z=n$ with $n=2$ when $\om \simeq 7.48$; Eq.~\eqref{uk1} then shows that the Floquet eigenvalue at 
$k=\pi$ is equal to $e^{in\pi} = 1$. Once again, it can be seen from Fig.~\ref{flo5} that a Majorana end mode appears in the range $7 \le \om \le 8$ with a Floquet eigenvalue equal to $1$. Similarly, as $\om$ is decreased even further, the {\it left} end of $L_\om$, given by $b_0$ in Eq.~\eqref{b0p},
crosses the point $z=n$ with $n=1$ at when $\om = 6.88$,
where the Floquet eigenvalue at $k=0$ equals $e^{in\pi} = -1$. Note that as $\om$ is decreased a little more, $L_\om$ no longer includes the point $z=1$. This is reflected in Fig.~\ref{flo5} in the form of a disappearing Majorana end mode with Floquet eigenvalue equal to $-1$, within the range $6 \le \om \le 7$.

In simpler terms, for any value of $\om$,
the number of points $z=n$ (where $n$ is an integer) which lie inside the line 
segment $L_\om$ is equal to the number of Majorana modes at each end of a 
chain. Further, the numbers of points with $n$ odd and even will give the 
numbers of end modes with Floquet eigenvalue equal to $-1$ and $1$ 
respectively. It is therefore clear that the numbers of odd and even integers lying inside $L_\om$ form a set of topological invariants which can change only at values of $\om$ where either $b_0$ or $b_\pi$ acquires an integer value. When that happens, the Floquet spectrum at either $k=0$ or $\pi$ becomes gapless, thus bearing the hallmark of a topological phase transition.

The above results can be summarized as follows. Assuming that $b_{0/\pi}$ are not integers, we consider all the integers lying between $b_0$ and $b_\pi$. Of these, let $n_e^>$ ($n_o^>$) and $n_e^<$ ($n_o^<$) respectively denote the numbers of even (odd) integers which are greater than and less than $2c_1 /\pi$. Then the numbers $N_+$ and $N_-$ of
modes at each end of a chain with Floquet eigenvalues $+1$ and $-1$ are given 
by $N_+ ~=~ |n_e^> - n_e^< |$ and $N_- ~=~ |n_o^> - n_o^< |$. Also, note that the winding number $W$, given by $|W| = |n_e^> - n_e^< + n_o^> - n_o^<|$, is generally {\it not} equal to the total number
of modes, $N_+ + N_-$, at each end of a chain (although
$|W| - (N_+ + N_-)$ is always an even integer). In Table I, we list the
values of $N_+$, $N_-$ and $|W|$ versus $\om$ for a 200-site system
with $\ga = 1$, $\De = -1$, $c_0 = 0.5$ and $c_1 = 0.2$.

\begin{table}[htb]
	\begin{center} \begin{tabular}{|c|c|c|c|c|c|c|c|c|c|c|c|c|c|c|c|c|c|c|}
			\hline
			$\om$ & 1 & 2 & 3 & 4 & 5 & 6 & 7 & 8 & 9 & 10 & 11 & 12 & 13 & 14 & 15 &
			16 & 17 & 18 \\
			\hline
			$N_+$ & 2 & 0 & 0 & 1 & 1 & 1 & 1 & 1 & 1 & 1 & 1 & 1 & 1 & 1 & 1 &
			0 & 0 & 0 \\
			\hline
			$N_-$ & 2 & 2 & 1 & 1 & 1 & 1 & 0 & 0 & 0 & 0 & 0 & 0 & 0 & 0 & 0 &
			0 & 0 & 0 \\
			\hline
			$|W|$ & 4 & 2 & 1 & 0 & 0 & 0 & 1 & 1 & 1 & 1 & 1 & 1 & 1 & 1 & 1 &
			0 & 0 & 0 \\
			\hline
	\end{tabular} \end{center}
	\caption{Values of $N_+$, $N_-$ and $|W|$ versus $\om$ for $\ga = 1$, $\De 
		= -1$, $c_0 = 0.5$ and $c_1 = 0.2$. $|W| \ne N_+ + N_-$ for $4 \le
		\om \le 6$.} \label{tab2} \end{table}

Finally, in the limit $\om \to 0$, it is clear from Eq.~\eqref{b0p} that the number of Majorana end modes diverges asymptotically as $8|\ga|/\om$ if $c_0 \pm 
\ga$ have the same sign and as $8|c_0|/\om$ if $c_0 \pm \ga$ have opposite 
signs. 

Before concluding this subsection, we would also like to note that several other variants of topological invariants have been proposed over the years which capture the topological features of periodically driven systems \cite{kundu13, asboth14, yao17, vega18}. Finally, we would like to mention that the topological properties of a (integrable) Kitaev chain with long range interactions has been extensively studied (see Ref.~\onlinecite{maity_review} and references therein). Similarly, the emergent topological features of the long range chain with periodic drives has also been investigated \cite{maity_periodic}.

\subsection{Periodically driven Kitaev honeycomb model}

Finally, we consider a periodically driven Kitaev honeycomb model \cite{thakurathi14}, the equilibrium situation of which was discussed in Sec.~\ref{subsec_kitaev_honey}. As with the periodically driven Kitaev chain, we analyze the situation in which one of the coupling parameters is repeatedly kicked after equal time intervals. For simplicity, consider the case in which the driven parameter is $J_3$, so that, 
\beq J_3 (t) ~=~ J_0 ~+~ J_p ~\sum_{n=-\infty}^\infty ~\de (t - nT),
\label{ht11} \eeq
where the $T=2 \pi/\om$ is the time-period.

As in the case of the Kitaev chain, the Floquet operator $U(T,0)$ governs the 
evolution of the system at stroboscopic times, and this can be evaluated by recasting Eq.~\eqref{eom1} as follows. Given a system with $N=N_x N_y$ sites, we introduce a column vector $\hat c = (\hat a_1,\hat a_2,\cdots,\hat a_N, \hat b_1,\hat b_2,\cdots, \hat b_N)^T$. Eqs.~\eqref{eom1} can then be written as $d\hat c(t)/dt = 4 M(t) \hat c(t)$, where
\beq M ~=~ \left( \begin{array}{cc}
	0 & - L^T \\
	L & 0 \end{array} \right). \eeq

The periodicity of $M(t)$ in time implies that the evolution of $\hat{c}$ over an interval between subsequent kicks 
is given by \bea \hat c(T) &=& U(T,0) ~\hat c(0), \non \\
{\rm where}~~~ U(T,0) &=& {\cal T} e^{4 \int_0^T dt M(t)}, \eea
and $\cal T$ denotes the time-ordering symbol. For the case of a periodically kicked system as in Eq.~\eqref{ht11}, we obtain,
$U(T,0) = e^{4M_1} e^{4M_0 T}$, 
where $e^{4M_0 T}$ is the operator which time evolves from $t=0$ to $t=T$,
and $e^{4M_1}$ then evolves across the $\de$-function at $t=T$.

\begin{figure}[htb]
\ig[width=0.9\columnwidth]{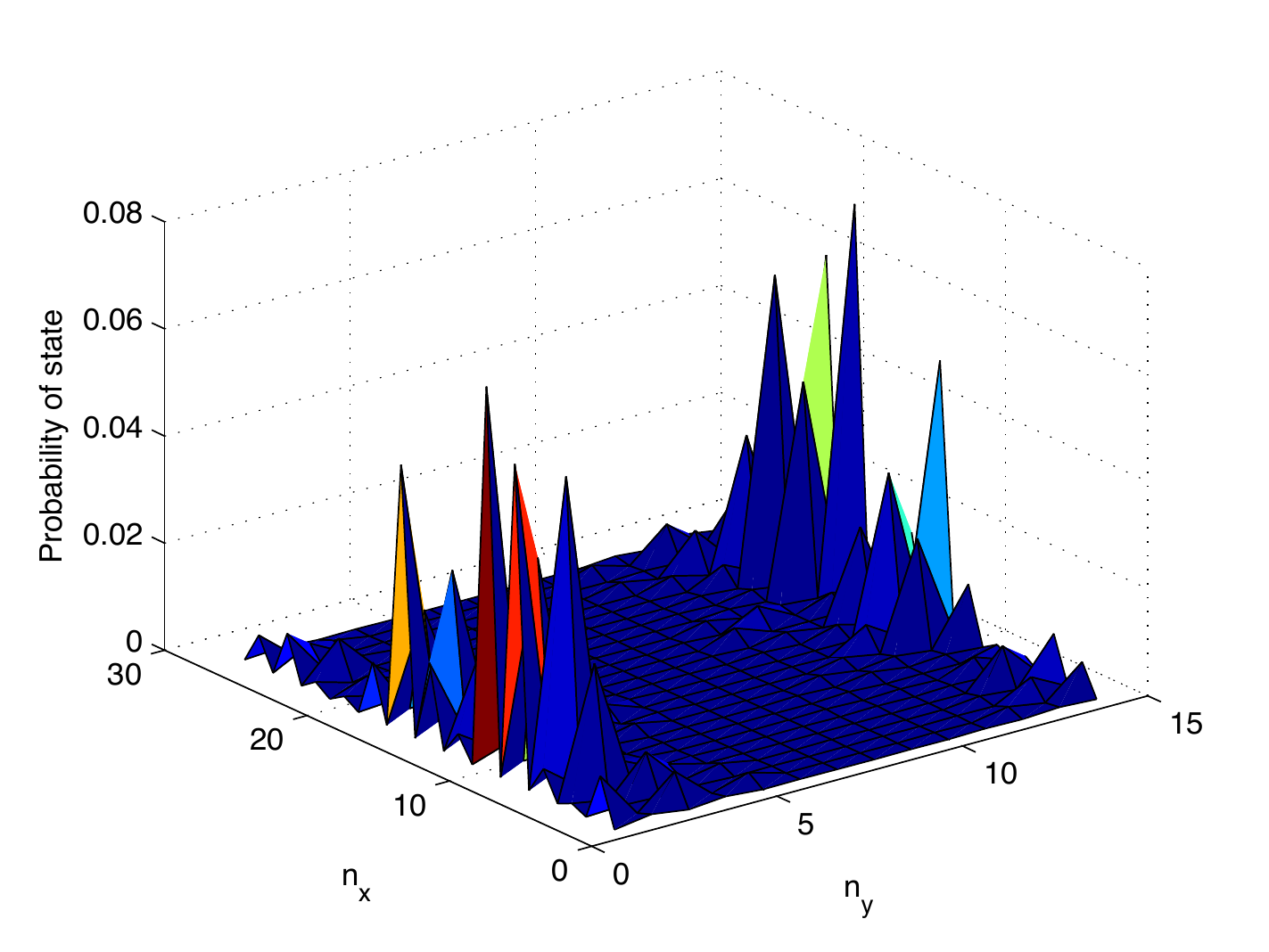}
\caption[]{Zigzag edge states for the Kitaev honeycomb model with $N_x \times N_y = 27 \times 14$, $J_1=0.7, ~J_2=0.15, ~J_0=0.15, ~J_p=0.2$ and $\om=3$ (See Ref.~\onlinecite{thakurathi14}).} \label{fig08} \end{figure}

Proceeding as in the case of the Kitaev chain, one can now look for the presence of zero energy Majorana edge modes in the Floquet operator. Remarkably,
edge modes are found to appear on both zigzag and armchair edges. (An example of a Floquet zigzag edge mode is shown in Fig.~\ref{fig08}). This is in contrast to the time-independent version of the model in which all such modes appear for the armchair edges.

To explore the emergence of the Floquet Majorana modes in the context of a bulk-boundary correspondence, let is consider an infinite system, so that the momentum modes are decoupled. The Floquet operator for each mode then assumes the form
\bea U_{\vec k} (T,0) &=& e^{-i2J_p \tau^y} ~e^{-iT (X_{\vec k} \tau^z ~+~
	Y_{\vec k} \tau^y)}, \non \\
X_{\vec k} &=& 2 ~[J_1 \sin({\vec k} \cdot {\vec M}_1) ~-~ J_2 \sin({\vec k}
\cdot {\vec M}_2 )], \non \\
Y_{\vec k} &=& 2~[ J_0 ~+~ J_1 \cos({\vec k} \cdot {\vec M}_1 ~+~ J_2
\cos({\vec k} \cdot {\vec M}_2)]. \non \\
&& \label{bulkeq1} \eea

As Majorana modes are associated with the existence of Floquet eigenvalues equal to $\pm1$, one can analytically derive the critical drive frequencies at which Majorana edge modes appear or disappear,
\beq \om_{\vec k} ~=~ \frac{4 \pi ~[~J_0 ~\pm~ \sqrt{J_1^2 + J_2^2 + 2 J_1
		J_2 \cos (k_x)}~]}{n\pi ~-~ 2 J_p}. \label{cond3} \eeq

A deeper insight into the appearance of Majorana modes can be obtained by mapping the two-dimensional Kitaev model to a chain as follows.
Consider a system which has a finite width in the $y$-direction (with zigzag
edges along the top and bottom as indicated in Fig.~\ref{fig02}) and is
infinitely long in the $x$-direction. The momentum $k$ along the $x$-axis is
a good quantum number. We now use the Heisenberg equations of motion
\bea \frac{d\hat a_m}{dt} &=& (J_1 e^{-ik/2} ~+~ J_2 e^{ik/2}) \hat b_m ~+~ 
J_3 \hat b_{m+1}, \non \\
\frac{d\hat b_m}{dt} &=& -(J_1 e^{ik/2} + J_2 e^{-ik/2}) \hat a_m - J_3 
\hat a_{m-1}, \label{eom2} \eea
for all $m \ge 1$, with the understanding that $\hat a_0 = 0$. Next we rewrite Eq.~\eqref{ham2} as,
\bea H &=& i \sum_{n=1}^\infty ~[ - J_x \hat b_{n+1} \hat a_n - J_y \hat b_n 
\hat a_{n+1} + \mu \hat b_n \hat a_n ], \label{ham2a} \eea
which becomes identical to Eq.~\eqref{ham2} on substituting $a_{n}\to a_{2n}, b_{n}\to a_{2n-1}$. For $k=0$, it is straightforward to see that the Heisenberg equations of motion of the operators $\hat a_n$ and $\hat b_n$ in Eq.~\eqref{ham2a} agree with Eqs.~\eqref{eom2} when,
\beq J_x ~=~ J_3/2, ~~J_y ~=~ 0, ~~{\rm and}~~ \mu ~=~ -(J_1 + J_2)/2.
\label{jxy} \eeq

We now see the situation of $J_3$ being periodically kicked corresponds, for $k=0$, to the situation in which the parameter $J_x$ of the chain is kicked periodically. As discussed in Sec.~\ref{subsec_kitaev_chain_dyn}, the $\de$-function kicks can produce Floquet Majorana modes at the ends of the system, which correspond to the zigzag edges of the two-dimensional
model. This idea can also be extended for $k \ne 0$, where two of the parameters appearing in Eqs.~\eqref{eom2} can be written as,
\bea & & J_1 e^{\pm ik/2} ~+~ J_2 e^{\mp ik/2} ~=~ J_k e^{\pm i\phi_k}, \non \\
{\rm where~} && J_k ~=~ \sqrt{J_1^2 ~+~ J_2^2 ~+~ 2 J_1 J_2 \cos k}.
\label{jjj} \eea
The phase $\phi_k$ can be removed from Eqs.~\eqref{eom2} by a unitary transformation; this unitary transformation is independent of
$J_3$ and is therefore not affected by the periodic kicks in $J_3$. The situation thus becomes similar to that for $k=0$, except that the
parameter $\mu$ in Eq.~\eqref{jxy} is now given by $\mu_k = - J_k/2$. We thus have a family of one-dimensional problems which are labeled by the parameter $k$ and the appearance of Majorana modes can be similarly explained for each of them separately.

\begin{figure} 
\ig[width=3.2in]{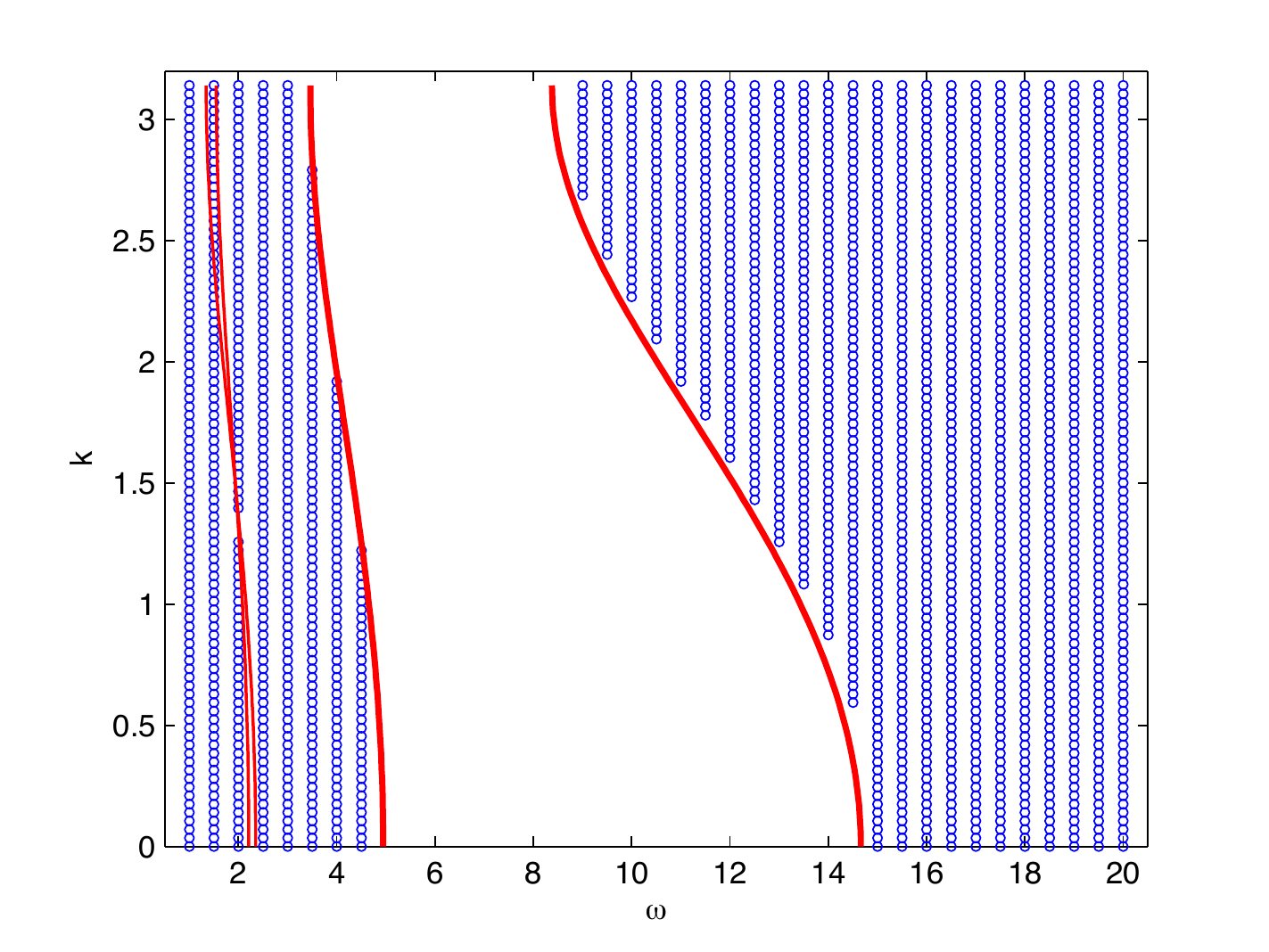}
\caption[]{Blue regions in the $(\om,k)$ space where Majorana states appear on the zigzag edges when the parameter $J_3$ is given periodic $\de$-function kicks. The system being considered has a width of $100$ sites, and $J_1=0.7, ~J_2=0.15, ~J_0=0.15$ and $J_p = 0.3$. The empty region in the middle is bounded on the right and left by two solid red lines which show the analytical results given in Eqs.~\eqref{cond6} and \eqref{cond7} respectively. Two more red solid lines corresponding to $n=-1$ and 2 are shown. They almost coincide with each other and appear within the blue regions on the left; they cross near $\om=2$ and $k \simeq 1.3$ (See Ref.~\onlinecite{thakurathi14}).}\label{fig10} \end{figure}	

The mapping to a chain discussed above allows us to precisely identify the conditions for the appearance or disappearance of Majorana modes. According to Eqs.~\eqref{cond3} and \eqref{jjj}, Floquet edge modes with a given momentum $k$ and FE equal to $(-1)^n$ should appear or disappear when
\beq \om_k ~=~ \frac{4\pi ~[J_0 ~\pm~ J_k]}{n\pi ~-~ 2J_p}, \label{cond5} \eeq
where $n$ is an integer. As the kicking frequency $\om$ is decreased, Eq.~\eqref{cond5} gives the red solid line on the right side of the empty region in Fig.~\ref{fig10} where a Floquet edge mode disappears with $n=0$, namely,
\beq \om_k ~=~ \frac{4 \pi ~[J_k ~-~ J_0]}{2 J_p}, \label{cond6} \eeq
and the red solid line on the left side of the empty region in Fig.~\ref{fig10} where a Floquet edge mode appears with $n=1$, namely,
\beq \om_k ~=~ \frac{4 \pi ~[J_0 ~+~ J_k]}{\pi ~-~ 2 J_p}. \label{cond7} \eeq
In general, for $n \le 0$, we have a line given by
\beq \om_k ~=~ \frac{4 \pi ~[J_k ~-~ J_0]}{2 J_p - n \pi}, \label{cond8} \eeq
while for $n \ge 1$, we have a line given by
\beq \om_k ~=~ \frac{4\pi ~[J_0 ~+~ J_k]}{n\pi ~-~ 2J_p}, \label{cond9} \eeq
where Majorana modes appear or disappear. Fig.~\ref{fig10} also show the red solid lines for $n=-1$ and 2. These appear within the blue regions; they cross near $\om=2$ and $k \simeq 1.3$ where we see a small gap indicating that there are no Majorana modes in that region. When $\om$ is decreased below these two lines, the Majorana mode with FE equal to $-1$ disappears and a mode with FE equal to $+1$ appears. When $\om$ is decreased even further, more modes start appearing which correspond to $n > 2$ and $n < -1$.

Before concluding this section, we remark here that the existence of localized eigenstates of the Floquet operator does not guarantee that the time-evolved state of the system would also exhibit localized Majorana edge modes. This is in contrast with equilibrium systems where the topological features are manifested in the ground state of the system. In the periodically driven scenarios discussed above, the time-evolved state, in general, has finite overlap with a large number of Floquet eigenstates. Hence, the localized Majorana modes can only manifest if one can ensure that the time-evolved state maintains a high fidelity with the ground state of the effective Floquet Hamiltonian throughout the evolution. This difficulty is a part of the broader challenge of dynamically preparing topological states which we will discuss in the next section. However, the signature of these non-equilibrium Majorana modes is expected to manifest in stroboscopic heat transport through the system and the spin density in the $XY$ spin-$1/2$ chain (see Ref.~\onlinecite{paolo17}).
	
\section{Generically driven one-dimensional topological systems}

In the previous section, we saw that in periodically driven systems, the Floquet Hamiltonian generating the stroboscopic dynamics may be topologically non-trivial. In fact, there exist countless scenarios when in a driven system the effective Hamiltonian or simply a quenched Hamiltonian generating the dynamics becomes topologically non-trivial despite the triviality of the equilibrium system. However, to observe such dynamical topological phases, one must figure out a way to dynamically prepare the non-equilibrium thermodynamically large system in the ground state of the non-trivial effective Hamiltonian. As discussed in Sec.~\ref{sec:models}, the thermodynamic limit is required for the perfect exponential localization of the edge states. Such an endeavor is not simple, as in a generic situation the out-of-equilibrium dynamics is bound to generate excitations except when the dynamics can be considered adiabatic. At the same time, it turns out that such adiabatic transformations connecting inequivalent topological phases is not possible in thermodynamically large systems as the adiabaticity necessarily breaks down at the critical point separating the two phases (see Sec.~\ref{sec:adiabatic} for details). Clearly, the thermodynamic limit and the adiabaticity seems too much to ask for while attempting to dynamically prepare topologically non-trivial quantum {\it states} from trivial ones. In this section, we will focus on the progress made in the attempt to address this particular bottleneck. We will also see that it is possible to topologically classify out-of-equilibrium quantum states without referring to the Hamiltonian generating the dynamics.

\subsection{Explicit symmetry breaking: unitary preparation of topological states}
\label{unitary_prep}

In the previous sections we have seen that the inequivalent SPT phases cannot be connected adiabatically in thermodynamically large systems as they are separated by a gapless quantum critical point. However, it is the protecting symmetry (such as $S$ in the case of the SSH model) which allows for the existence of distinct inequivalent topological phases in such models. This suggests that explicitly breaking the protecting symmetry might let us connect inequivalent topological phases adiabatically even in thermodynamically large systems. To visualize this protocol, we first look into the dynamical change of the winding number when the system is subjected to an external time-dependent driving.\\

We consider the SSH model and study the temporal evolution\ct{utso19} of the equilibrium topological invariant, i.e., the winding number under a generic unitary drive. We begin with an initial state $|\psi_{k}(0)\rangle$, the system is allowed to evolve under the driven Hamiltonian $H_k(t)$. The state $|\psi_{k}(0)\rangle$ therefore evolves with time as
\begin{eqnarray}\label{eq:evolve}
|\psi_{k}(t)\rangle &=& \mathbb{T} e^{-i\int_{0}^{t}H_{k} (t^{'})dt^{'}}
|\psi_{k}(0)\rangle \non \\ 
&\equiv& e^{-iH_{k}^{\rm eff}(t) t}|\psi_{k}(0)\rangle \non \\
&=&U_{k}(t)|\psi_{k}(0)\rangle, \end{eqnarray}
where $H^{\rm eff}_k(t)$ is the time-dependent effective Hamiltonian acting as a generator of the unitary evolution acting on the driven system and $\mathbb{T}$ denotes the time ordering operator. We now investigate the fate of the winding number under such a time-dependent dynamics. To analyze this, let us recall the time-dependent or dynamical connection as
\begin{equation}\label{eq:db}
A_{k}(t)\equiv\left[\langle\psi_{k}(0)|U_k^{\dagger}\right]\partial_{k}\left[ U_{k}|\psi_{k}(0)\rangle \right],
\end{equation}
which evolves in time as,
\begin{eqnarray}\label{eq:conn}
A_{k}(t)&=&\langle\psi_{k}(0)|\partial_{k}|\psi_{k}(0)\rangle + \langle\psi_{k}(0)|U_k^{\dagger}(\partial_{k}U_{k})|\psi_{k}(0)\rangle \non \\
&=& A_{k}(0)+ \langle\psi_{k}(0)|U_k^{\dagger}(\partial_{k}U_{k})|\psi_{k}(0)\rangle.
\end{eqnarray}
Hence, the change in the connection at a later time is given by
\begin{equation}
\Delta A_{k}=A_{k}(t)-A_{k}(0)=\langle\psi_{k}(0)|U_k^{\dagger}\partial_{k}U_{k}|\psi_{k}(0)\rangle.
\label{eq_connection} \end{equation}
Recasting the effective Hamiltonian to the following form, $H_{k}^{\rm eff}(t)=|m(k,t)|\left(\hat{m}(k,t).\vec{\sigma}\right)$ and
also denoting $|m(k,t)|$ simply as $m$, we obtain
\begin{eqnarray}\label{eqn_udu}
&& U_k^{\dagger}\partial_{k}U_{k} \non \\
&& =\partial_{k}m ~\sin^2 (mt) ~\{-it \hat{m}\cdot {\vec \sigma} 
+i \hat{m}\times\partial_{k}\hat{m}\cdot {\vec \sigma} \} \non \\
&& ~~~-i ~\sin (mt)\cos (mt) ~\partial_{k}\hat{m}\cdot {\vec \sigma}.
\end{eqnarray}

The initial state $|\psi_{k}(0)\rangle$ that we consider happens to be the ground state of the SSH Hamiltonian (belonging BDI class) which can be chosen to be of the form of Eq.~\eqref{eq:eg} where $\phi(k)$ is an odd function of $k$. Interestingly, the terms on the right hand side of the Eq.~\eqref{eqn_udu} can be shown to vanish individually when integrated over the entire BZ, pertaining to certain conditions imposed upon the effective Hamiltonian $H^{\rm eff}_{k}$ as discussed below.\\

Let us now analyze the implications of Eqs.~ \eqref{eq_connection} and \eqref{eqn_udu}. Taking the expectation value of the first term of the above equation with respect to the state $|\psi_{k}(0)\rangle$, one observes 
that the integral of this quantity over the full BZ vanishes identically if $m_{x}(k)$ is an even function of $k$ and $m_{y}(k)$ is an odd function of $k$. Similarly, analyzing the integral of the next two terms over the full BZ, we see that both of them vanish identically if $m_{z}(k)$ is an odd function of $k$ or zero in addition to the above constraints imposed on $m_{x}(k)$ and $m_{y}(k)$. If the above conditions are satisfied by the effective Hamiltonian then the winding number must remain invariant in time.\\

It is evident from Eq.~\eqref{eq:symm} that the above constraints on the single particle Hamiltonian in $k$ space, demand the presence of certain symmetries of the effective Hamiltonian.
Namely, one concludes that the equilibrium winding number remains invariant under temporal evolution if the effective dynamical Hamiltonian $(H^{\rm eff}_{k})$ respects either of the symmetry combinations, $\mathcal{T}$ and $\mathcal{P}$ ($m_x(k)\rightarrow$ even, $m_y(k)\rightarrow$ odd, $m_z(k)\rightarrow 0$) simultaneously or just $\mathcal{P}$ ($m_x(k)\rightarrow$ even, $m_y(k)\rightarrow$ odd, $m_z(k)\rightarrow$ odd). The SSH Hamiltonian described in Eq.~\eqref{eq:H} with real hopping amplitudes ($v,w\in\mathbb{R}$) lies in the BDI class of topological Hamiltonians. This suggests that the topological invariant is temporally invariant under all unitary dynamics unless the BDI symmetries (here particularly the particle-hole symmetry) are explicitly broken by the effective Hamiltonian generating the dynamics.\\

Nonetheless, there appears to be a more serious issue with the above procedure. As discussed so far, it is clear that in SPT phases, it is the protecting symmetry which is behind the quantization of the topological invariant and thus the quantized numbers of experimentally verifiable edge states. In fact, if the symmetry is explicitly disrespected in the dynamics, even if one is able to dynamically tune the value of the topological invariant, it will no longer remain quantized and the topological classification is rendered useless. In this regard one needs to connect the changing winding number with an observable physical phenomena in the non-equilibrium system. This calls for a physical significance of the topological winding number in one-dimensional systems which turns out to be intricately connected with the macroscopic dipole electric polarization of the chain. The macroscopic electric polarization of the system is defined as,
\begin{equation}\label{polarization}
P=e\braket{\hat{X}},
\end{equation}
$e$ being the electronic charge which we set equal to unity and $\hat{X}$ being the position operator summed over all sites of the chain, i.e., $\hat{X}=\sum\limits_i\hat{x}_i$. However, since the momentum space and real space operators are related by a Fourier transform, we can replace the position operator\ct{resta93,bardyn18} as
\begin{equation}\label{XV}
\hat{X}\rightarrow\frac{i}{2\pi}\int dk \partial_k,
\end{equation}
which suggests that the winding number is nothing but the macroscopic electric polarization of the system. The equivalence in Eq.~\eqref{XV} may not be immediately apparent, as with periodic boundary conditions the operator $\hat{X}$ is not manifestly periodic and one needs to compactify the defining Eq.~\eqref{polarization} before proceeding. We will discuss the derivation in detail in a later section. Now, furthering this correspondence, it is straightforward to show that in an arbitrary time-dependent situation the bulk polarization current density $j(t)$ of the SSH chain is directly proportional to the rate of change of the topological winding number $(\nu)$ (see Refs.~\onlinecite{utso19} and \onlinecite{cooper18}):
\begin{equation}\label{w_j}
j(t)=\frac{dP(t)}{dt}=\frac{d\nu}{dt}=\frac{1}{2\pi}\int_{BZ}dk\bra{\psi_k(t)}\partial_{k}H_k(t)\ket{\psi_k(t)},
\end{equation}
where $\ket{\psi(t)}$ is the time-evolved state for each quasimomenta mode $k$ and $H_k(t)$ is the instantaneous time-dependent Hamiltonian. Particularly, in the case of a time periodic drive with a period $T$, the stroboscopic (measured after a complete period), the variation of the winding number denoted as $\Delta\nu_{m}$ for the $m-$th stroboscopic interval, is related to the average change in the bulk polarization density of the chain within the $(m-1)-$th and the $m-$th period of evolution i.e.,
\begin{equation}\label{avgJ}
\Delta\nu_{m}=\frac{\nu(mT)-\nu((m-1)T)}{T}=\frac{1}{T}\int_{(m-1)T}^{mT} dt 
j(t), \end{equation}
which is nothing but the net charge transferred through the chain in one cycle of the drive. Moreover, since it is the Floquet Hamiltonian $H_F$ which generates the stroboscopic time evolution, the above discovered symmetry constraints must be imposed on $H_F$ to keep the topological index invariant in stroboscopic time. However, it might be noted that the Floquet Hamiltonian is responsible only for the stroboscopic evolution. The micromotion within consecutive stroboscopic intervals is governed by the most general time-dependent Hamiltonian itself. It turns out, that there might arise scenarios when the instantaneous Hamiltonian explicitly breaks the protecting symmetries while the Floquet Hamiltonian still respects it. In those situations although the stroboscopic winding number remains invariant, it might vary in micromotion.\\

\subsubsection{Preparation of Floquet topological state}

To demonstrate the applicability of explicit symmetry breaking in the dynamics, we will demonstrate a protocol which adiabatically prepares a Floquet topological state from a trivial one.
Starting from the trivial ground state of a BDI symmetric SSH Hamiltonian, we periodically drive
the system. Choosing a sufficiently high driving frequency $(\omega=2\pi/T)$, the periodic perturbation is switched on adiabatically, allowing the instantaneous state of the system to follow the instantaneous ground state of the Floquet Hamiltonian.
We choose the protocol\ct{dutta19}
\begin{eqnarray}
H_k(t) &=& H_0 + \lambda_k (t) \non \\
&=& H_0(k)+A_1(t)V_0 \sigma^x + A_2(t) V_k(t), \end{eqnarray}
with $V_k(t+T)=V_k(t)$ being periodic in time having a time period of $T$. The amplitudes $A_1(t)$ and $A_2(t)$ are adiabatic ramps which slowly switch on and off the periodic perturbation,
\begin{eqnarray}\label{protocol}
A_1(t) &=& A_2(t)= -t/\tau\text{~~~for~~~} 0\le t \le \tau, \non \\
A_1(t) &=& 1 \text{~~and~~} A_2(t) = \frac{t}{\tau} - 2 
\text{~~for~~} \tau\le t \le 2\tau, \non \\
A_1(t) &=& 1 \text{~~and~~} A_2(t) = 0 \text{~~~for~~~} t\ge 2\tau,
\end{eqnarray}
allowing the Floquet Hamiltonian to get adiabatically modified. The periodic perturbation $V_k(t)$ is chosen to be sinusoidal,
\begin{equation}\label{Vkt}
V_k(t)=V_1\sin(\omega t)\sigma^x+V_1\cos(\omega t)\sigma^y.
\end{equation}

We note that in the interval $0 \le t \le \tau$, the periodic perturbation $V_k(t)$ is switched on adiabatically and in the subsequent interval $\tau \le t \le 2\tau$ a part of the periodic drive is switched off, again adiabatically, to restore all the symmetries.\\
Evidently, due to the presence of the linear ramping, the protocol is not perfectly periodic in the time interval $0 \le t \le 2\tau$. However, as the ramping is much slower than the time period of the periodic perturbation, the complete evolution may be partitioned into the `fast' and the `slow' time variables \ct{kitagawa11}: the former is the time scale of the high frequency periodic drive and the later represents the time scale of the linearly ramped amplitude. Therefore, for a fixed value of the amplitude, the drive is completely periodic and a Floquet picture holds. \\ 
If the frequency of the periodic drive is much higher than the characteristic bandwidth of the system and the amplitude of the drive is small enough $\left(\mathcal{A}^2/\omega\ll 1, \mathcal{A}~\text{being the {\it effective} amplitude of the drive}\right)$, the drive is off resonant. The effective Floquet Hamiltonian may be approximated just to incorporate the single-photon virtual processes or nearest-neighbor hopping in the Floquet Bloch lattice i.e.,
\begin{equation}
H_F(k)\simeq \frac{1}{T}\int_0^T H_k(t) dt+\frac{1}{\omega}\left[H_{-1}(k),H_{+1}(k)\right]+\mathcal{O}\left( \frac{\mathcal{A}^4}{\omega^2}\right),
\label{eq_floquet_ham}
\end{equation}
where
$H_{\pm 1}(k)=\frac{1}{T}\int_0^T H(t)e^{\pm i\omega t}dt.$
We now further assume that $A(t)$ changes very slowly $(\tau \gg T)$ such that its change over a time period $T$ may be neglected at very high frequencies. Using Eq.~\eqref{eq_floquet_ham}, the effective Floquet Hamiltonian takes the following form,
\begin{equation}
\label{eq:floquet}
H_F^t(k)\simeq H_0(k)-A_1(t)V_0\sigma^x+\frac{\left[A_2(t)V_1\right]^2}{2\omega}\sigma^z ,
\end{equation}

The slowly varying amplitude can now be interpreted as an adiabatic deformation of the Floquet Hamiltonian $H_F^t(k)$. The magnitude of $V_0$ is so chosen that the term proportional to $\sigma^x$ in the effective Floquet Hamiltonian Eq.~\eqref{eq:floquet} drives the Floquet Hamiltonian across the QCP to render the final Floquet Hamiltonian topologically non-trivial at time $t=2\tau$.

The diagonal term on the other hand serves two purposes. It generates a staggered mass term which opens up a gap between the Floquet eigenstates and keeps the floquet spectrum gapped at all times. This gap in the Floquet spectrum protects the instantaneous ground states against the generation of excitations particularly near the QCP where the gap otherwise vanishes in the absence of the staggered mass. The mass term also explicitly breaks the $\mathcal{P}$ and $\mathcal{S}$ symmetry at the same time in the Floquet Hamiltonian $H_F(k)$ causing the bulk topological invariant to change stroboscopically. At the end of the switching off (i.e., at $t=2\tau$), the particle hole symmetry is completely restored in the final Floquet Hamiltonian $H_F^t(k)$ which now belongs to a completely symmetric BDI SSH model in the topologically non-trivial sector. 

\begin{figure}
\begin{center}
\includegraphics[width=0.47\textwidth,height=0.64\columnwidth]{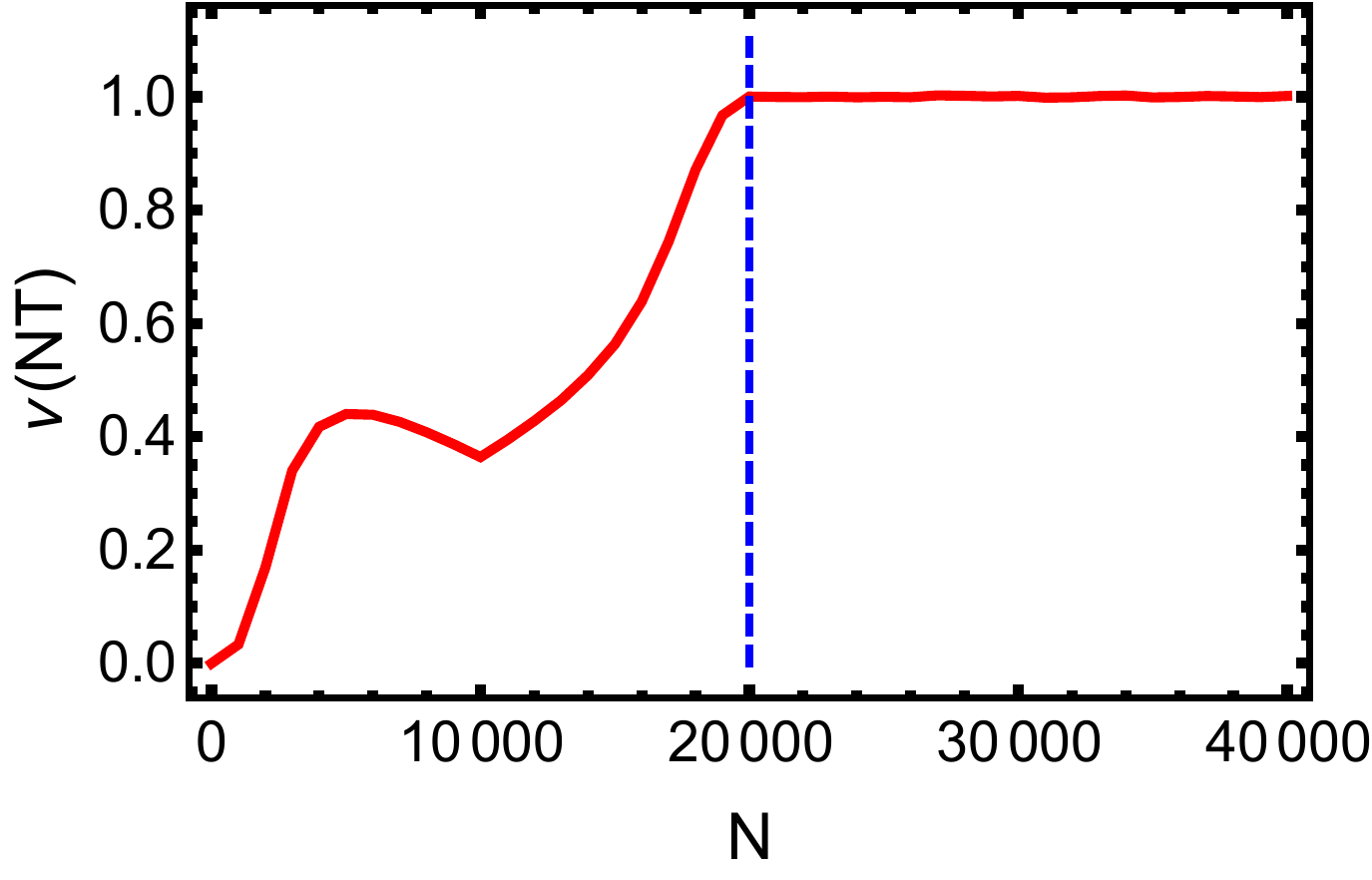}
\caption{Temporal variation of the winding number as a function of stroboscopic time period for a periodic boundary condition. The winding number assumes a constant value (unity) after the $\mathcal{P}$ symmetry have been completely restored in the Floquet Hamiltonian at $t=2\tau$, with $\tau=10^4T$. (See Ref.~\onlinecite{dutta19}).
\label{1a}}
\end{center}
	
\end{figure}
\begin{figure}[ht]
\begin{center}
\includegraphics[width=0.47\textwidth,height=0.64\columnwidth]{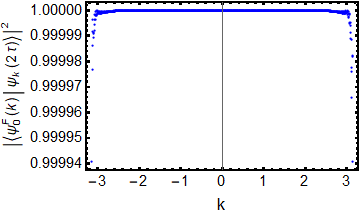}
\caption{Overlap of the evolved state $\ket{\psi_k(2\tau)}$ at $t=2\tau$ with the Floquet eigenstate at the same time $\ket{\psi_0^F(k)}$ for all $k\in \left[-\pi,\pi\right]$. For a sufficiently large $\tau$ ($\tau=10^4 T$), the instantaneous state of the system has a significant overlap with the Floquet topological ground state in the high frequency limit. The initial parameters are $v=1.55$, $w=1.50$ and $L=1000$; $\omega=100.0$ and amplitudes $V_0=0.1$, $V_1=7.0$. (See Ref.~\onlinecite{dutta19}).
\label{fidelity}
}
\end{center}
\end{figure}

The winding number calculated from the non-equilibrium state of the driven chain is also observed to become topologically non-trivial (see Fig.~\ref{1a}). In response to the explicit breaking of the $\mathcal{P}$ symmetry and the consequent variation of the winding number, we observe a stroboscopic generation of polarization current in the bulk which vanishes just after the BDI symmetries are restored in the Floquet Hamiltonian at $t=2\tau$. Since the $\mathcal{S}$ symmetry has been restored finally, the winding number also remains invariant further and the final Floquet Hamiltonian indeed becomes topologically inequivalent to the trivial Hamiltonian. The mass term necessarily avoids the crossing of a gapless QCP and therefore under an adiabatic protocol, the instantaneous state of the system $\ket{\psi_k(2\tau)}$ is seen to dynamically follow the ground state $\ket{\psi_0^F(k)}$ of the high frequency Floquet Hamiltonian.\\

{The adiabatic ramping functions $A(t)$ in Eq.~\eqref{protocol} have been chosen so that its rate of change is much slower than the time period of the periodic perturbation (i.e., $\tau \gg T$). This ensures that the dynamical state of the system at stroboscopic intervals of time have a significant fidelity to the topological eigenstate of the Floquet Hamiltonian. In Fig.~\ref{fidelity}, we see that at $t=2\tau$, the stroboscopic state $\ket{\psi_k(2\tau)}$ has a significant overlap with the eigenstate $\ket{\psi_0^F(k)}$ of the Floquet}
{Hamiltonian for all $k\in\left[-\pi,\pi\right]$.}

\subsubsection{Stroboscopic emergence of edge states}

In this section we study the time periodic driving protocol given in Eq.~\eqref{protocol} in a SSH chain with open boundary conditions. The static SSH Hamiltonian $H(v_0,w)$ as in Eq.~\eqref{eq:H}, with $\mathcal{N}$ unit cells under open boundary conditions (i.e., $c^{\dagger}_{2\mathcal{N}+1}=0$), is a $2\mathcal{N}\times2\mathcal{N}$ Hermitian matrix in the single-particle basis $\mathcal{B}\equiv\{\ket{i}\}_{i=1}^{2\mathcal{N}}$. Here $\ket{i}$ denotes the $i^{th}$ site in the chain.
We start with an eigenstate $\ket{\psi(0)}$ of the initial topologically trivial Hamiltonian $H(v_0,w)$ having $\mathcal{N}=50$ unit cells and adiabatically switch on an unitary time dependent periodic perturbation,
\begin{equation}
H(t)=H(v_0,w)+v(t)\left(\sum_{i=1}^{2\mathcal{N}}c^{\dagger}_{n,1}c_{n,2}+{\rm 
H.c.}\right),
\end{equation}
where
\begin{eqnarray}\label{drive}
v(t) &=& - \frac{t}{\tau}\left[V_0+V_1 e^{-i\omega t} \right] \text{~~for~~} 0\le t \le \tau , \non \\
v(t) &=& -V_0 +\left(\frac{t}{\tau} -2 \right) V_1 e^{-i\omega t} \text{~~for~~} \tau\le t \le 2\tau,\non \\
v(t) &=& -V_0 ~~~~\text{~~for~} t\ge 2\tau, \\
\end{eqnarray}
and $\omega=2\pi/T$ is the frequency of the drive while $T$ being it's time period.
The matrix elements of the dynamical unitary propagator $U(t)$ in the time-independent basis $\mathcal{B}$ under the protocol Eq.~\eqref{drive}, satisfies a linear system of 
$2\mathcal{N}\times 2\mathcal{N}$ differential equations,
\begin{equation}\label{schrodinger}
i\frac{dU(t)}{dt}=H(t)U(t).
\end{equation}
To find the stroboscopic propagator $U(nT,(n-1)T)$, which dynamically evolves the state of the system from the time $t=(n-1)T$ to $t=nT$, we iteratively solve Eq.~\eqref{schrodinger} with appropriate boundary conditions in each interval $\left[(n-1)T,nT\right]$.
By a repetitive application of the unitary propagator $U(nT,(n-1)T)$ for each stroboscopic interval, we obtain the time-evolved state of the system at $t=NT=2\tau$,
\begin{equation}
\ket{\psi(2\tau)}=\prod_{n=1}^{N=\frac{2\tau}{T}} U(nT,(n-1)T)\ket{\psi(0)}.
\end{equation}

\begin{figure}
\includegraphics[width=0.47\textwidth,height=0.65\columnwidth]{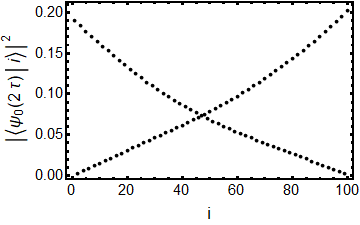}
\caption{Under open boundary conditions, at $t=2 \tau$, the stroboscopic time-evolved state shows localization at the ends of the chain after the BDI symmetries have been restored in the topological Floquet Hamiltonian. The energy of the localized state $E=0.003$ for a finite chain with 100 lattice sites. The initial state (same as in (a)) is subjected to a unitary drive as in Eq.~\ref{protocol} with $V_0=0.1$, $V_1=2.0$, $\tau=500T$ and $\omega=100.0$, where $i$ denotes the lattice site. (See Ref.~\onlinecite{dutta19}).} \label{2b} \end{figure}

To maintain adiabaticity of the deformation of the Floquet Hamiltonian {which is essential for the adiabatic preparation}, we choose $\tau \gg T$ ($\tau=500T$) and a sufficiently high frequency $\omega=100$ for the periodic perturbation. The initial parameters of the topologically trivial SSH chain $H(v_0,w)$ were chosen to be $v_0=1.55$ and $w=1.50$. With the choice of the amplitudes $V_0=0.1$ and $V_1=2.0$ as specified in the (see Fig.~\ref{2b}), we study the support of the stroboscopic state $\ket{\psi(2\tau)}$ of the system at the $i^{th}$ site of the chain through the quantity $|\langle i\ket{\psi(2\tau)}|^2$. We observe localization of the state $\ket{\psi(2\tau)}$ to the two ends of the finite chain corresponding to the topologically non-trivial winding number of the bulk Floquet Hamiltonian depicted in Fig.~\ref{1a}.
We note that although the generated stroboscopic end modes de-localize considerably into the bulk of the chain, in the thermodynamic limit the hybridization between the end modes will eventually vanish and they will get strictly localized at the edge.
 
The stroboscopic state thus prepared at $t \geq 2\tau$, is observed to host topologically protected zero energy modes localized at the ends of a finite chain with open boundary conditions (see Fig.~\ref{2b}). As the parameters of the final Floquet Hamiltonian are close to the critical values, the localization length of the zero energy edge states are large. The end states therefore hybridize near the midpoint of the chain which consequently lifts the energy of the localized states from exactly zero; this is also true for an undriven SSH chain for the same set of parameters. However, for a sufficiently long chain, the edge states formed would be highly localized and will be perfectly localized in the thermodynamic limit. It can also be shown that higher winding number Floquet states can be prepared adiabatically using similar dynamical protocols starting from trivial states. \\

\subsection{Dynamical symmetry breaking: A symmetry classification of non-equilibrium systems}

In the previous section we discussed the applications of explicitly breaking the protective symmetries of a SPT insulator in dynamically preparing non-trivial states. There is however a different kind of symmetry breaking in such systems which arise purely out of the non-equilibrium dynamics while the Hamiltonian generating the dynamics still respects all the symmetries of the initial system\ct{cooper18}. To be clear, this will be addressed as {\it dynamical symmetry breaking} in contrast to the case when the symmetries are explicitly broken in the Hamiltonian itself. To precisely discuss these scenarios one must therefore find a way to develop a symmetry classification of the non-equilibrium states themselves irrespective of the symmetries of the Hamiltonian that generates the dynamics. To achieve this, we must first look into the symmetry classification of equilibrium states and how they are inherited from the symmetries of the Hamiltonian itself. As demonstrated in Sec.~\ref{sec:models}, all non-interacting quantum systems can be classified with respect to two anti-unitary symmetries, i.e., charge conjugation/particle-hole $\mathcal{P}$, time-reversal $\mathcal{T}$ and an unitary symmetry, i.e., chiral/sublattice $\mathcal{S}$. In an equilibrium system, if the ground state is non-degenerate, it is naturally expected that the ground state will directly inherit all the symmetries respected by the Hamiltonian,
\begin{equation}
\mathcal{H}=\sum\limits_{ij}\psi^{\dagger}_iH_{ij}\psi_j,
\end{equation} 
where $\psi_j$ and $\psi_j^{\dagger}$ are fermionic annihilation and creation operators respectively and $H$ is the single-particle Hamiltonian. As discussed previously, the symmetry operators can be re-written as $\mathcal{P}=P\mathcal{K}$, $\mathcal{T}=T\mathcal{K}$ and $\mathcal{S}$ where $P$, $T$ and $S$ are unitary. The symmetry transformations corresponding to is said to respect the time-reversal, charge conjugation and chiral symmetries can then be represented to act on the creation and annihilation operators such that the fermionic anti-commutators are preserved,
\begin{eqnarray}
\mathcal{T}\psi_i\mathcal{T}^{-1} &=& T_{ij}\psi_j ~~~{\rm and}~~~\mathcal{T}i\mathcal{T}^{-1}=-i, \non \\
\mathcal{P}\psi_i\mathcal{P}^{-1} &=& P_{ij}\psi_j ~~~{\rm and} ~~~\mathcal{P}
i\mathcal{P}^{-1}=-i,\non \\
\mathcal{S} &=& \mathcal{T}\mathcal{P} ~~~{\rm and} ~~~\mathcal{S}i
\mathcal{S}^{-1}=i. \end{eqnarray} 
Since the charge conjugation operator transforms a particle into its anti-particle, it acts on the annihilation operators to produce linear combinations of creation operators. 
If the system respect theses discrete symmetries then one obtains,
\begin{equation}
\begin{split}
\mathcal{P}\mathcal{H}\mathcal{P}^{-1}=H,
\end{split}
\end{equation}
and equivalently for the other two. Similarly, the single-particle density matrix $\rho$ can be constructed out of the single-particle fermionic correlation functions in a many-body state $\ket{\Phi}$ as,
\begin{equation}
\rho_{ij}=\braket{\Phi|\psi_i^{\dagger}\psi_j|\Phi}.
\end{equation}
 It is then straight forward to show that if the equilibrium state $\rho$ respects the three discrete symmetries, one must obtain,
\begin{equation}\label{state_symm}
\begin{split}
T\rho^*T^{\dagger}=\rho,\\
P\rho^*P^{\dagger}=\mathbb{I}-\rho,\\
S\rho S^{\dagger}=\mathbb{I}-\rho.
\end{split}
\end{equation}

The density matrix thus evolves in time as,
\begin{equation}\label{heis_evo}
\rho(t)=U(t)\rho(0)U^{\dagger}(t),
\end{equation}
where $U(t)$ is the temporal propagator and is generated by a time-dependent effective Hamiltonian
$H^{eff}(t)$ (see Eq.~\eqref{eq:evolve}). Unlike our discussion in the previous section, we further assume that the symmetries of the initial Hamiltonian are not explicitly broken in the effective Hamiltonian generating the dynamics and therefore satisfies the constraints in Eq.~\eqref{eq:symm}. It can then be directly inferred from Eq.~\eqref{state_symm}-\eqref{heis_evo} that the dynamical state $\rho(t)$ must satisfy the following constraints under the same symmetry operations,
\begin{equation}\label{dymm_symm}
\begin{split}
T\rho^*(t)T^{\dagger}=\rho(-t),\\
P\rho^*(t)P^{\dagger}=\mathbb{I}-\rho(t),\\
S\rho^*(t)S^{\dagger}=\mathbb{I}-\rho(-t).
\end{split}
\end{equation}
However, in generic out-of-equilibrium systems the equivalence of $\rho(t)$ and $\rho(-t)$ does not hold, i.e., in most situations $\rho(t)\neq\rho(-t)$. A comparison with Eq.~\eqref{state_symm} reveals that the dynamical state no longer respects the time-reversal and the chiral symmetry even while the effective time-dependent Hamiltonian stays symmetric. This shows that although the charge conjugation symmetry is preserved, the chiral and time-reversal and symmetries are generically broken in the dynamical state without explicitly breaking the symmetries of the effective Hamiltonian. This phenomena will be termed as a {\it dynamical symmetry breaking} in all the subsequent discussions. Consequently, if the topological phases are protected by the particle-hole symmetry, the time-dependent topological invariant should stay invariant as the dynamics preserves $\mathcal{P}$ even in the out-of equilibrium state. However, SPT models lying in other symmetry classes not having particle-hole symmetry will generically experience a dynamical symmetry breaking while out of equilibrium and hence, the topological invariant need no longer be quantized. Remarkably, as we have seen previously, a dynamically varying winding number is also experimentally relevant and manifests as a flow of polarization current (see Eq.~\eqref{w_j}),
\begin{equation}
j(t)=\frac{d\nu(t)}{dt}
\end{equation}
through the bulk system. Thus, even though the effective Hamiltonian respects the equilibrium symmetries, a dynamical symmetry breaking can indeed be probed by a bulk polarization current in such systems.\\

\subsection{Non-unitary dynamics of topological states}

\subsubsection{Lindblad master equation and steady state topology}\label{atomic_wire}

{In the previous sections we have primarily focused on the fate of symmetry protected topological phases of matter under unitary dynamical and a possible topological classification of out-of equilibrium quantum states. We now proceed to discuss the behavior of topologically non-trivial states under non-unitary dynamics which naturally arise when the system is in interaction with the environment which is difficult to avoid in thermodynamically large systems. Dissipative quantum systems, particularly those undergoing a Markovian evolution always tends to a steady state at asymptotic long times. In such open quantum systems, it instructive to observe the reduced system of interest by tracing out the degrees of freedom of the bath. The Markovian evolution of quantum systems can described by a Master equation generating the time evolution of the reduced density matrix describing only the system degrees of freedom\ct{breuer,sourav},
\begin{equation}\label{lindblad}
\partial_t\rho=\mathcal{L}(\rho)=-i[H,\rho]+\mathcal{D}(\rho), 
\end{equation}
where as usual, the Hamiltonian generates the unitary processes within the system itself and the {\it dissipator} functional $\mathcal{D}(.)$ takes care of the bath induced processes due to the system-bath coupling. The dissipator can be shown to take the form
\begin{equation}
\mathcal{D}(\rho)=\sum\limits_{i}\kappa_i\left( 2L_i\rho L_i^{\dagger}-\{
L_i^{\dagger}L_i,\rho \} \right),
\end{equation}
where the operators $L_i$ are known as {\it jump operators} containing the action of the bath on the system through the coupling strengths $\kappa_i$. For example, if the bath acts as a particle sink at a site $i$ of a fermionic system, one might set $L_i=c_i$ where $c_i$ is the annihilation operator of the system at that site. We will see the effect of such a bath when coupled to a topological system. Also, a significant property of a Markovian evolution is that the steady state is completely independent of the initial conditions when the steady state is non-degenerate. First, we proceed to discuss the possibility of the dynamical generation of topological steady states through careful enough designing of the bath. For any choice of te Lindblad jump operators, the system reaches an asymptotic steady state given by
\begin{equation} \partial_t\rho=\mathcal{L}(\rho)=0. \end{equation}
It is then possible to engineer the steady state of the system to host non-trivial properties by constructing the bath accordingly. The exact route and time taken by such a Markovian system to reach equilibrium has been extensively studied and is also an interest of current research (see Ref.~\onlinecite{vernier20}).

In fact, it has been shown\ct{zoller11} that choosing the Lindblad operators such that it destroys the 
Bogoliubov quasiparticles in a $p$-wave Kitaev chain (see Sec.~\ref{sec:kitaev_chain}), one can reach a steady state subspace in which the topological end Majorana correlations are preserved against dissipation.\\

\subsubsection{Slow quenches in a dissipative environment}\label{slow_defect}

Artificially engineered Lindblad operators has been shown\ct{zoller11} to cool a system to specific targeted pure steady states. However, in such cases the Lindblad operators are highly non-local and do not generically describe real dissipative scenarios. This is because mostly environmental interactions has an opposite effect on quantum states, i.e., the destruction of quantum coherence. For example, the bath $L_i=c_i$, where $c_i$ are the annihilation operator on site $i$, accounts for the loss of Dirac fermions from every site of the system. It can be seen that in a purely dissipative evolution, such a system will asymptotically reach the featureless and boring fermionic vacuum. In this section we attempt to understand the time and energy scales involved when such a Kitaev chain coupled to such a system is driven externally in time across the quantum critical point. Although we saw that a topological state with a perfect bulk-boundary correspondence does not exist in finite-size systems, here we will specifically study finite-size systems, as we know that a finite system is not gapless even at the critical point. However, since we are solely interested in uncovering the competition between unitary and dissipative dynamics, we can simply deal with a finite-size system which can always be driven adiabatically.\\

\begin{figure*}
\centering
\subfigure[]{
\includegraphics[width=0.92\columnwidth]{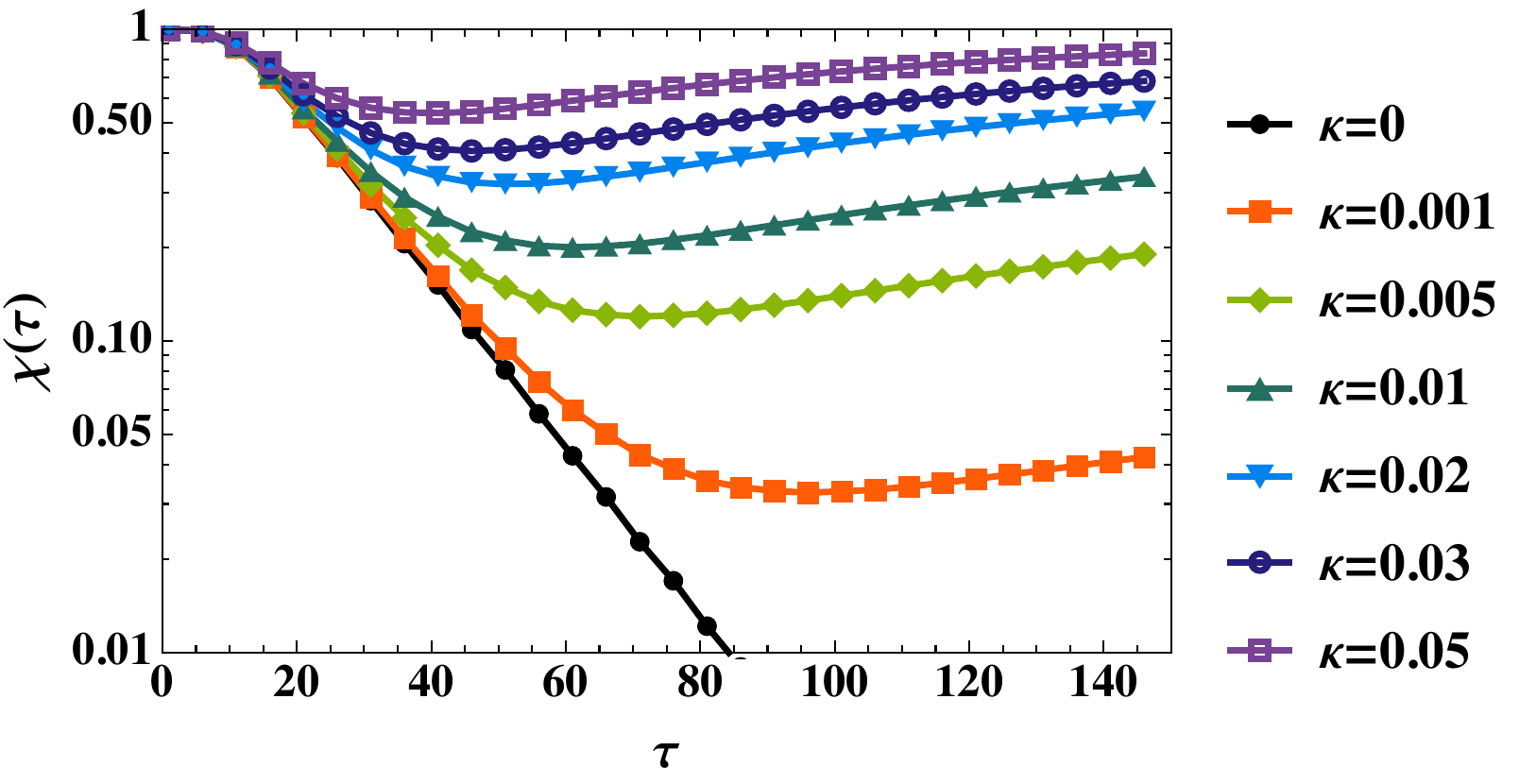}
\label{fig_3a}}\quad\quad\quad\quad
\subfigure[]{
\includegraphics[width=0.92\columnwidth]{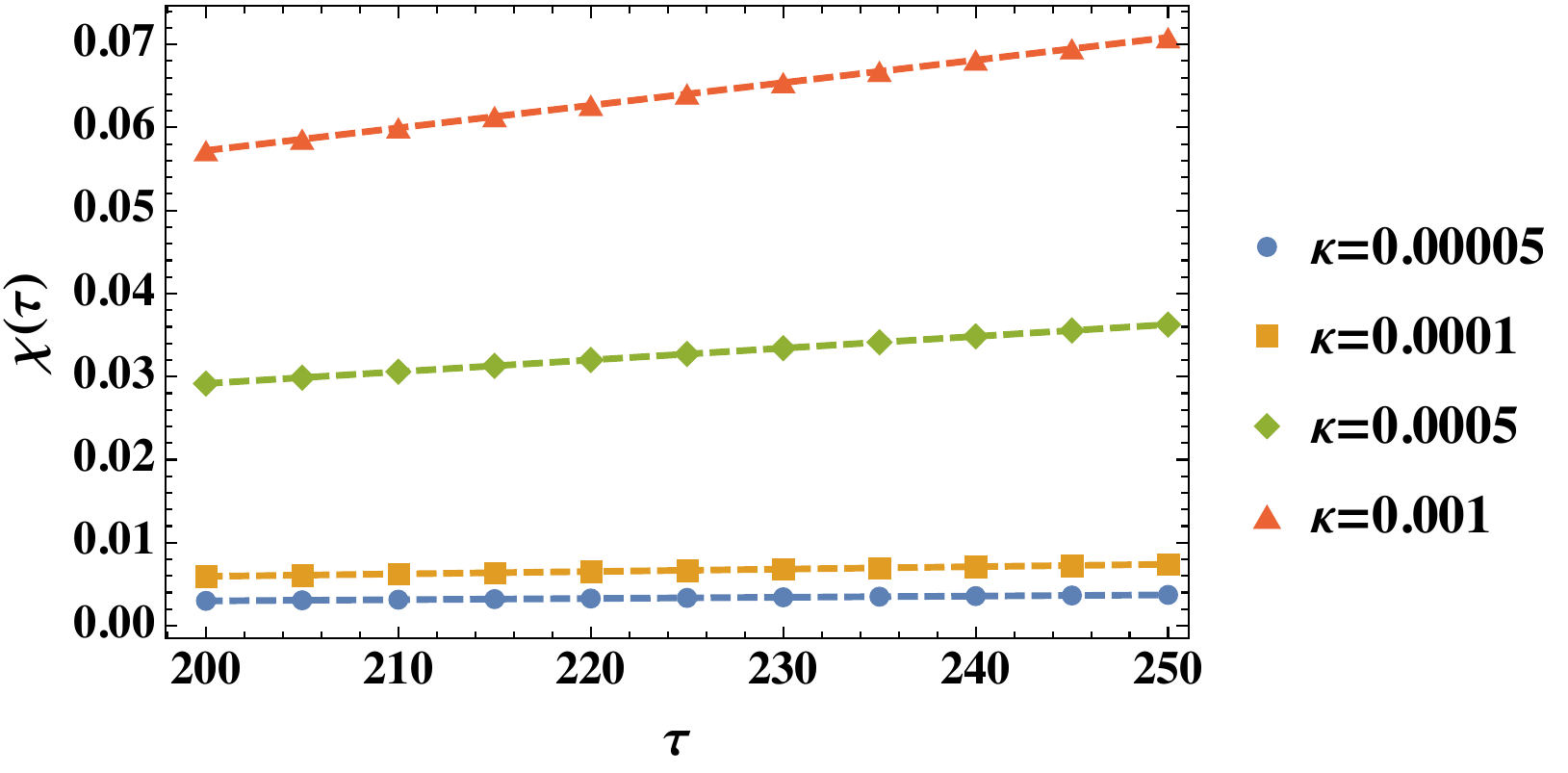}
\label{fig_3b}} 
\caption{(a) {Defects generated at the end of the ramp $\chi(\tau)$ as a function of $\tau$ for a local bath with Lindblad operators of form Eq.~\eqref{eq_lindblad_local}. For $\tau<L\lll\tau_B$, the dynamics is dominated primarily by the unitary evolution thereby resulting in lowering of defects with increasing $\tau$. For $L\ll\tau<\tau_B$, the ramp is adiabatic and induces negligible defects on its own. The defects generated is then primarily through dissipative effects resulting in more defect generation with increasing $\tau$. The interplay of the unitary and dissipative dynamics of the bath results in an optimal $\tau_o$ for which the defect generated is minimum. (b) For $L\ll\tau\ll\tau_B$, the defect scales as $\chi(\tau)\sim\kappa\tau$. The dashed lines correspond to linear fits of the data points. In both (a) and (b), the relevant parameters chosen are $L=20$, $J=1$, $\mu_i=2$ and $\mu_f=0$. (See Ref.~\onlinecite{souvik20}).}}
\end{figure*}

The presence of a QCP or vanishing gap in the bulk spectrum of a thermodynamically large chain, presents a conundrum in the context of preparing a topological state --- starting from a trivial phase of the system, it is impossible to drive the system into a non-trivial phase through a unitary dynamics. Specifically, in the problem that we consider here\ct{souvik20}, the system is initially in the ground state $\ket{\psi(0)}=\ket{\psi_i^0}$ of an initial Hamiltonian Kitaev chain $H_i: |\mu_i|>|J|$, following which the Hamiltonian is ramped across the QCP at $|\mu|=|J|$ to a final $H_f:|\mu_f|<|J|$ using the protocol,
\begin{align}\label{eq_protocol}
	\mu(t)=\left(\mu_i+(\mu_f-\mu_i)\frac{t}{\tau}\right)\Theta(\tau-t)+\mu_f\Theta(t-\tau),
\end{align}
where $\Theta(x)$ is the Heaviside step function. The protocol therefore linearly ramps the initial $\mu_i$ to a final $\mu_f$ during time $\tau$ after which $\mu$ remains frozen at the targeted $\mu_f$. In the thermodynamic limit, the quantum adiabatic theorem breaks down; the system $\ket{\psi(t)}$ therefore cannot be exclusively prepared in the ground state $\ket{\psi_f^0}$ of $H_f$. This results in generation of \textit{defects} in the correlation of the Majorana end modes which we quantify as 
\begin{align}\label{eq_defect}
	\chi(t)=\bra{\psi_f^0}\theta\ket{\psi_f^0}-\theta(t),
\end{align}
where $\theta(t)=\bra{\psi(t)}\theta\ket{\psi(t)}$.
We remark that although the ground state $\ket{\psi_f^0}$ is doubly degenerate in the thermodynamic limit, we choose the state $\ket{\psi_f^0}$ as the Bogoliubov vacuum corresponding to the final Hamiltonian. The choice of the ground state however, does not affect the results qualitatively. 
Note that for the ideal situation of a perfect unitary adiabatic preparation, the quantity $\chi(t)$ should vanish.

However, for a finite system of size $L$, the bulk spectrum is not truly gapless; the gap $\delta$ at the QCP scales as $\delta\sim 1/L$. Consequently, if the time-scale $\tau$ of the ramp is large enough such that $\tau\gg L$, the bulk of the chain (closed chain with PBC) can be prepared in the ground state of the final Hamiltonian. However, unlike the bulk, the splitting between the edge modes 
vanishes exponentially with $L$. Therefore, even for a finite system, the defect $\chi(t)$ truly vanishes only in the limit $\tau\to\infty$. Nevertheless, it is possible for $\chi(t)$ to come arbitrarily close to zero for a finite-size system even for finite $\tau \gg L$. The key point of this section is that perfect Majorana correlations cannot be generated if the ramp duration is finite with the defects scaling as $\chi(\tau)=e^{-f(L,\mu_i)\tau}$, where the coefficient $f(L,\mu_i)\geq 0$ is non-universal. \\

Now, consider the same protocol in the presence of dissipation due to a particle-loss type of bath, which effectively acts as a sink of Dirac fermions at each site of the chain. We model the dissipative effects through local Lindblad operators that act locally at each site on the Kitaev chain, barring the edge sites. Specifically, we choose
\begin{equation}\label{eq_lindblad_local}
	L_j=c_j, \quad j\in\{2,3...,N-1\}, 
\end{equation} 
where $c_j$ are the fermionic annihilation operators acting on site $j$ of the chain. In the presence of such local dissipative channels, the Majorana edge modes which were initially present in the system, are known to decay exponentially in time due to the finite overlap of the edge modes with the bulk for $\mu\neq0$. However, for $\mu=0$, the edge modes are essentially disconnected from the bulk and are therefore robust against any dissipation induced in the bulk. In our protocol, we therefore set $\mu_f=0$, so that the edge modes once created (with some defects) remain localized after the ramp, i.e., $\chi(t>\tau)=\chi(\tau)$. This allows us to focus solely on the effects of dissipation on the preparation stage of localized edge modes, thereby excluding any defects generated after the ramp is complete. 

The key observations from the numerical results are two-fold. First, as can be seen from Fig.~\ref{fig_3a}. The second and more significant result is that there exists an optimal ramp duration for which the defects generated are minimum, which, as
we will elaborate below, arises due to the competition between the unitary and the dissipative dynamics. 

To comprehend these results, it is instructive to compare the three relevant length/time scales in the dynamics -- $L$, $\tau$ and $\tau_B$. Assuming a weak coupling strength $\kappa$ ($1/\tau_B$) amounts to setting $L$, $\tau$ $\ll\tau_B$. In the adiabatic limit of the ramping protocol, i.e., $L\ll\tau\ll\tau_B$, the defects generated scale as $\chi(\tau)\sim\kappa\tau$ (see Fig.~\ref{fig_3b}). Intuitively, this monotonic rise in defect can be explained as follows --- as the edges interact with the bath indirectly through the bulk, an increase in ramp duration $\tau$ implies that the edge modes have proportionally increasing time to decay before the chemical potential is eventually ramped to $\mu_f=0$ at $t=\tau$, following which the edge modes can no longer decay. \\

Although being an expectation over localized states, this resonates perfectly with the scaling of defect density in bulk residual energy under similar dissipation and driving protocols \ct{keck17}. 
See Appendix~\ref{sec_perturb} for a perturbative expansion (with $\kappa\tau\ll 1$ as small parameter) of the solution of the dynamical equations of motion for the two-point Majorana correlations to show that any observable, which is a linear function of two-point Majorana correlations, is indeed expected to follow the same linear scaling $\sim \kappa\tau$ under the action of linear Lindblad operators.

On the other hand, {for very fast quenches $\tau<L\lll\tau_B$}, the chemical potential $\mu$ is quickly ramped to zero within a short duration $\tau$; within this duration the environment fails to induce any substantive decay in the Majorana edge correlations. The short duration of the quench however, itself results in the generation of defects. The defects scale with the ramp duration as $\chi(\tau)\sim e^{-f(L,\mu_i)}$, where the coefficient $f(L,\mu_i)\geq 0$ is a model dependent non-universal function which can not be derived within any analytical framework. The defect generation in the fast quench limit is therefore dominated by the unitary dynamics and arises due to the fast (non-adiabatic) ramping.

It follows that, there exists an \textit{optimal} ramp duration $\tau_o$ at which the defect generation {at the end of the ramp} is minimized (see Fig.~\ref{fig_3a}). From the preceding discussions, one can assume that the defect generated for $\tau\sim\tau_o$, has the form as $\chi(\tau)=\exp\left(-f(L,\mu_i)\tau\right)+\kappa\tau$. A generic expression for $\tau_o$ can be derived by minimizing the defect with respect to $\tau$ as
\begin{equation}
\frac{d}{d\tau}\left(\exp\left(-f(L,\mu_i)\tau\right)+\kappa\tau\right)|_{\tau=\tau_0}=0, \end{equation}
which gives
\begin{equation}
\tau_o=-\frac{1}{f(L,\mu_i)}\log\left(\frac{\kappa}{f(L,\mu_i)}\right).
\end{equation}
The positivity of $f(L,\mu_i)$ along with the condition of weak-coupling strength $\kappa$ implies that there exist a positive definite $\tau_o$ at which the defects are minimized. The existence of the optimal $\tau_o$ is a consequence of the competition between the unitary dynamics which demands a large $\tau$ for defect minimization and the dissipative effects (also See Ref.~\onlinecite{vicari20}) which requires short $\tau$ for the same.

We emphasize here that the above expression for the optimal ramp duration is not universal because the scaling of the defects arising from non-adiabatic effects is model dependent as reflected in the coefficient $f(L,\mu_i)$. This is unlike the universal scaling of the optimal ramp time obtained in the case of defect generation in residual energy \cite{keck17}, where the residual energy is defined as the excess energy of the time-dependent state over the instantaneous ground state and is quantitatively obtained by replacing $\theta$ in Eq.~\eqref{eq_defect} with $H(t)$ so that $H(\tau)=H_f$. {Consequently, the residual energy is an extensive (bulk) property of the system}. The universality in the scaling of residual energy stems from the fact that the contribution to the defects from the non-adiabatic excitations follow a universal Kibble-Zurek scaling. 


\subsubsection{Majorana correlations in localized phases} 

In a recent study, it was shown that the decay of the topological end-Majorana correlations $\theta(t)$ in static topological wires coupled to the particle-loss Markovian bath (modeled by Lindblad operators $L_i=c_i$ as discussed in the previous section) follow a generic exponential trend\ct{heyl15},
\begin{equation}
\theta(t)\rightarrow e^{-\gamma t}~~~\text{as}~~~t\rightarrow\infty
\end{equation} 
with a rate $\gamma$ set by the dissipative coupling $\kappa$. This exponential fallout can however be slowed down to a stretched exponential decay,
\begin{equation}
\theta(t)\rightarrow e^{-\left({\gamma t}\right)^{\alpha}}~~~\text{as}~~~t
\rightarrow\infty, \end{equation}
with an exponent $\alpha$ in many-body localized phases of the system when subjected to disorder. This has also been verified numerically for strong disorder strengths such that mostly all eigenstates are highly localized. It should now be stated that the topological Majorana correlations cannot be preserved against the action of the bath when considered to act on every site including the edge Majorana sites even with disorder based localization. However, if the edge Majorana sites are not coupled to the bath directly as in Sec.~\ref{atomic_wire} disorder based localization can indeed induce a transition into a stretched exponential decay of the end-Majorana correlations. This can be understood as follows:\\

As the end sites are not directly coupled to the bath, it can only be affected by the bath through the interaction of the edge modes with the bulk system which in turn is directly coupled to the dissipator. Therefore, in a localized phase under disorder, since all the eigenstates are now localized, the interaction of the edge Majorana modes with the bulk system is minimized which in turn reduces the interaction of the edge-Majoranas with the bulk. This slows down the decay of the Majorana edge correlations and one obtains a stretched exponential decay instead of a much faster exponential one.

\subsubsection{Mixed state topological phase transitions}
\label{sec:mixed_1d}

Although the unitary preparation of a topological state must surpass the obstacle of passing through a gapless quantum critical point, this problem can be somewhat avoided if one deals with the dissipative preparation of topological states as discussed in the previous section. However, very often in such an endeavor one ends up with mixed states rather than pure topological states. In this regard, one must identify topologically non-trivial density matrices along with the topological classification of pure states constructed so far. There has been many attempts in constructing a topological classification of mixed state density matrices. In this section we will review a recent development in mixed state topology of one-dimensional Gaussian systems which also hosts a physical manifestation through observables in mixed states. We start with the relation between the macroscopic electric polarization and the topological invariant described in Eq.~\eqref{polarization}-\eqref{XV}. As mentioned previously, one encounters a difficulty while extending the correspondence in systems with periodic boundary conditions as the many-body position operator $\hat{X}$ is not manifestly periodic. To surmount this obstacle, Resta proposed an alternative form of the relation by using translation operators instead of the simple $\hat{X}$ as\ct{resta93,bardyn18},
\begin{eqnarray}\label{resta1}
P &=& \frac{1}{2\pi}{\rm Im}\ln{\left<\hat{T}\right>}, \non \\
\hat{T} &=& e^{i\delta\hat{X}},
\end{eqnarray}
such that $\delta=2\pi/L$, $L$ being the length of the system and the expectation is to be taken over a many body state, say $\ket{\Psi}$. 
Note that the operator $\hat{T}$ is periodic by construction when one assumes periodic boundary conditions. It is then straightforward to show using the properties of determinants that for any slater determinant $\ket{\Psi}$, with periodic boundary conditions the Eq.~\eqref{resta1} reduces to,
\begin{equation}
P=\frac{1}{2\pi}\sum_{n}\int_{BZ}A^{nn}dk,
\end{equation}
where the sum is over all the occupied single-particle bands. This is nothing but the many-body version of the Berry phase (see Eq.~\eqref{Berry}) of Appendix \ref{sec_app_A}). A natural generalization to the case of mixed density matrices is now visible as,
\begin{equation}\label{mixed_winding}
\nu\rightarrow{\rm Im}\ln{{\rm Tr}\left[\hat{T}\rho\right]},
\end{equation}
$\rho$ being the density matrix of the system. Given the density matrix $\rho$ is of the Gaussian form
\begin{equation}
\rho\propto e^{-\sum\limits_{ij}\hat{\psi}_iG_{ij}\hat{\psi}_j},
\end{equation}
where $\hat{\psi}_i$ are fermionic operators, it can be shown that with periodic boundary conditions, the Eq.~\eqref{mixed_winding} simply reduces to,
\begin{equation}
\begin{split}
\nu={\rm Im}\sum\limits_{n}\left(1+\mathcal{O}_n\right),\\
\mathcal{O}_n=\prod_k e^{-g_n(k)}e^{iA_{nn}\delta},
\end{split}
\end{equation}
such that the summation is over every occupied band of the matrix $G$ (also known as {\it purity bands}) having eigenvalues $g_n(k)$ ({\it purity eigenvalues}) and the product is taken over all momenta mode $k$. Since the number of momenta mode are extensive, one can further rewrite the quantities $\mathcal{O}_n$ as,
\begin{equation}
\mathcal{O}_n=e^{-N\sum\limits_k{N^{-1}g_n(k)}}\prod_ke^{iA_{nn}\delta}.
\end{equation}
It is now clear that in the thermodynamic limit ($N\rightarrow\infty$), for a non-degenerate purity spectrum only the single band $\ket{g_0}$ having the lowest eigenvalue $g_0(k)$ gets projected out and one obtains the well-known ${\rm U(1)}$ winding number,
\begin{equation} \nu=\int_{BZ} dk ~{A}_{00}. \end{equation}
To compare with the standard definition of the winding number in pure states, here the matrix $G$ plays the role of the Hamiltonian and the topology of the Gaussian density matrix can only change when the system passes through a purity gap closing point, i.e., a point where the purity eigenvalues become degenerate. Later we will see that this topological invariant can also be extended to classify two-dimensional Chern insulating systems. 
This allows for the topological classification of finite temperature thermal density matrices of the form $\rho\sim e^{-\beta H}$ where $H$ is the Hamiltonian of the system and $\beta$ is the inverse temperature.
Furthermore, using this topological invariant in Eq.~\eqref{mixed_winding}, it can indeed be shown that under a suitable bath (which preserves the quadratic nature of the system) does undergo a topological phase transition. To exemplify, consider a one-dimensional Rice-Mele chain (see Ref.~\onlinecite{bardyn18} for an introduction of the Rice-Mele model) driven out of equilibrium in the presence of a coupling to a Markovian dissipative environment. As already introduced in Sec.~\ref{atomic_wire}, such dynamics can be simulated by a Lindblad master equation. Remarkably, although the system is non-thermal and there is no notion of temperature, a careful choice of the Lindblad jump offers guarantees a Gaussian steady in the thermodynamic limit which can be topologically non-trivial.\\

Interestingly, the topology of mixed Gaussian states can also be probed experimentally when classified by the topological invariant defined in Eq.~\eqref{mixed_winding}. This is because the interaction of such a system with electromagnetic cavity fields can be understood to take place by nothing other than the many-body position operator $\hat{X}$. Considering an incident cavity electromagnetic mode which as a gradient (such as the ${\rm TEM}_{01}$ Gauss-Hermite mode) in the direction of the chain, one can show that the emergent beam will have gained a phase directly proportional to the macroscopic electric polarization of the chain $\braket{\hat{X}}$. It is therefore seen that this topological phase can indeed be experimentally measured through a Mach-Zender interferometric setup as demonstrated in Ref.~\onlinecite{bardyn18}\\

Given the possibilities of dynamically preparing symmetry protected phases of one-dimensional quantum systems, we will now proceed to discuss an emergent direction of study, viz., the dynamical preparation of Chern topological phases which are not generically protected by restrictive symmetries.\\

\section{Generically driven two-dimensional Chern topological systems}

\subsection{The unitary no-go theorem}\label{nogo1}

Unlike symmetry protected topological phases, Chern insulators generically do not depend on a protective symmetry to host non-trivial topological states and therefore an adiabatic connection between inequivalent phases (see Fig.~\ref{figchern}) through symmetry breaking is not feasible. In fact, as we will show, it is apparently impossible to dynamically initiate topological transitions in such systems as the Chern number is invariant under unitary propagation. To see this, consider a non-interacting many-body system on a translationally invariant 
two-dimensional lattice with periodic boundary conditions and described 
by the Hamiltonian,
\begin{equation}\label{b_ham}
H=\bigoplus_{\vec{k}}H_k,
\end{equation}
where $\vec{k}$ are the conserved quasimomentum degrees of freedom. To simplify, consider the Hamiltonian in momentum space to have two bands and can therefore be written as,
\begin{equation}
H_k=\vec{B}(k) \cdot \vec{\sigma},
\end{equation}
$\vec{\sigma}\in\{\sigma^1,\sigma^2,\sigma^3\}$ being the standard Pauli matrices. Say, the system is in its ground state $\ket{\psi_0(k)}$ and the lowest energy band is completely filled. The corresponding density operator describing the state is,
\begin{equation}\label{density}
\rho_k=\ket{\psi_0(k)}\bra{\psi_0(k)}=\frac{1}{2}\left(\mathbb{I}+\vec{n}(k)
\cdot \vec{\sigma}\right),
\end{equation}
$\vec{n}(k)$ being a three dimensional vector under rotations, also known as a {\it Bloch vector} on the compact BZ $k_x,k_y\in[0,2\pi)$. Note that since the initial state is a pure state, the vector $\vec{n}(k)$ has a unit norm. In the thermodynamic limit, the topology of such systems is characterized by the Chern number (see Eq.~\eqref{chern_number} of Sec.~\ref{sec:topology}) which in terms of the density matrix $\rho(k)$ can also be written as
\begin{equation}\label{chern_rigol}
C=\frac{1}{4\pi}\int_{BZ}d^2\vec{k}~\vec{n}(k) \cdot \left[\partial_{k_x}\vec{n}(k)\times\partial_{k_y}\vec{n}(k)\right].
\end{equation}
The Chern number defined in Eq.~\eqref{chern_rigol} simply counts the winding of the Bloch vector on a Bloch sphere $\mathcal{S}^2$ as the momentum $\vec{k}$ covers the complete BZ. (This is exactly the definition of the Chern number as constructed in Sec.~\ref{sec:topology}.) A straight forward generalization can then be done to extend this topological classification to out of equilibrium states starting from the state $\rho_k(t=0)=\rho_k$. Under unitary dynamics, the time-evolved density matrix at time $t$ is given by $\rho_k(t)=U_k(t)\rho_k(0)U_k^{\dagger}(t)$ such that the propagator $U$ is generated by a time dependent Hamiltonian $H_k(t)=B_k(t).\vec{\sigma}$. Similar to the equilibrium system, the time dependent density matrix can then be described exactly as Eq.~\eqref{density} however, now through a time dependent Bloch vector $\vec{n}(k,t)$. Therefore, to classify such states topologically, one can simply define the Chern number on the time dependent Bloch vector $\vec{n}(k,t)$ as\ct{rigol15},
\begin{equation}
C(t)=\frac{1}{4\pi}\int_{BZ}d^2\vec{k}~\vec{n}(k,t).\left[\partial_{k_x}\vec{n}(k,t)\times\partial_{k_y}\vec{n}(k,t)\right],
\end{equation}
while the Schrodinger equation can be re-written as,
\begin{equation}\label{evolution}
\dot{\vec{n}}_k(t)=\vec{n}_k(t)\times\vec{B}_k(t).
\end{equation}
which also preserves the norm of the vector $\vec{n}(k)$ at all times.

However, if we differentiate the above expression with respect to time, we obtain,
\begin{eqnarray}\label{cdot}
\partial_tC &=& \frac{1}{4\pi}\int_{BZ}d^2\vec{k}~[\partial_{k_x}\vec{B}_k(t)
\cdot \partial_{k_y}\vec{n}_k(t) \non \\
&& ~~~~~~~~~~~~~~~~~-\partial_{k_y}\vec{B}_k(t) \cdot \partial_{k_x}
\vec{n}_k(t)]. \end{eqnarray}
From Eq.~\eqref{cdot} it is clear that if both $B_k(t)$ and $n_k(t)$ are smooth functions, the RHS vanishes and we see that the Chern number $C(t)$ must remain invariant under any unitary dynamics. Now, as the time evolution of each spin is dictated by a linear differential equation (see Eq.~\eqref{evolution}), one can safely deduce that if the initial spin-texture is smooth, it will remain so at all times if the magnetic field $B_k(t)$ is smooth. This invariance of the Chern number therefore also rightly hold for adiabatic time evolutions. This is because, two inequivalent Chern phases cannot be adiabatically/smoothly connected as in a thermodynamic limit, one must cross a gapless critical point which will necessarily make the process non-adiabatic. This can also be understood 
using the fact that the Hamiltonian $H_k(t)=\vec{B}_k(t) \cdot \vec{\sigma}$ acts exactly as a time dependent magnetic field on a lattice of spin-$1/2$ having Bloch vectors $\vec{n}_k(t)$ at every point on the BZ. The conservation of the Chern number then simply implies that a smooth magnetic field on the BZ which also varies smoothly in time and acting on an initial smooth spin texture must always preserve the smoothness of the spin texture thus preventing any change of the Chern number.\\

There exists a topological invariant which unlike the Chern number can topologically characterize systems with broken translational symmetry, namely the \textit{Bott index} ( see Eq.~\eqref{bott_wilson} of Sec.~\ref{sec:bott}). To define the Bott index for an equilibrium system, one considers a two-dimensional insulating lattice system with the lattice coordinates being $\{x_i,y_i\}$ and described by the Hamiltonian $H$. In the position basis, let us construct the diagonal matrices\cite{loring11,rigol15,rigol17} $X_{ij}=x_{i}\delta_{ij}$ and $Y_{ij}=y_{i}\delta_{ij}$.
	The next step is to define the unitary operators, 
	\begin{equation}
		T_{X}=e^{i\frac{2\pi}{L_x}X}~~\text{and}~~T_{Y}=e^{i\frac{2\pi}{L_y}Y},
	\end{equation}
	where $L_x$ and $L_y$ are the linear dimensions of the system along the Cartesian $x$ and $y$ axis, respectively. For simplicity, we here assume that $L_x=L_y=L$. Given that all the single-particle energy eigenstates $\ket{\psi_n}$ with energy $E_n<E_F$ are occupied, we define the projector
	\begin{equation}
		P=\sum\limits_{n; E_n<E_F}\ket{\psi_n}\bra{\psi_n},
	\end{equation}
	on the occupied subspace. The projection of the unitary operators $T_X$ and $T_Y$ into the occupied subspace is thus given by
	\begin{equation}
		\tilde{T}_X=PT_XP~~\text{and}~~\tilde{T}_Y=PT_YP.
	\end{equation}
	It is then straightforward to see that the quantity\ct{rigol15}
	\begin{equation}\label{eq_bott_main}
		\nu_{B}=\frac{1}{2\pi}{\rm Im}\left[{\rm Tr}\ln\left(\tilde{T}_X\tilde{T}_Y\tilde{T}_X^{\dagger}\tilde{T}_Y^{\dagger} \right)\right]
	\end{equation}
	is a real number which is independent of the boundary conditions imposed on the lattice. This quantity is known as the {\it Bott index}.

It has been explicitly shown\ct{rigol15} that the Bott index can indeed be changed by adiabatic evolutions and also remain quantized under open boundary conditions in a finite system as long as the corresponding thermodynamically large system has a well-defined topology. This can be demonstrated by extending the Bott index to out of equilibrium systems with physical boundaries by projecting onto the filled time-evolved state of the system rather than the equilibrium projectors, i.e., consider the time-evolved projectors,
\begin{equation}
P(t)=\sum\limits_{\alpha}U(t)\ket{\psi_{\alpha}(0)}\bra{\psi_{\alpha}(0)}U^{\dagger}(t),
\end{equation}
where $U(t)$ is the unitary propagator in time when the projector onto filled states initially is,
\begin{equation}
P(0)=\sum\limits_{\alpha}\ket{\psi_{\alpha}(0)}\bra{\psi_{\alpha}(0)}.
\end{equation}
 Using this time dependent projectors to project the translation operators, $\hat{T}_X$ and $T_{Y}$ in Eq.~\eqref{eq_bott_main} gives the Bott index of the time-evolved state. To see how this Bott index behaves in an out-of-equilibrium situation, consider a Haldane system which is exposed to high frequency circularly polarized light normally. Such a system can be described by a Hamiltonian of the form
\begin{equation}
H(t)=H_0+V(t),
\end{equation}
where $H_0$ is the bare Haldane model of graphene and $V(t)$ describes the dipole electronic interaction with the incident radiation,
\begin{equation}
V(t)=-\sum\limits_{i\alpha}c^{\dagger}_{i\alpha}c_{i\alpha}\vec{r}_{i\alpha}
\cdot \vec{E}(t), \end{equation}
where the vector $\vec{r}_{i\alpha}$ denotes the position of each lattice point with $i$ denoting the unit cell and $\alpha\in\{A,B\}$ denotes the sublattice. The operators $c_{i\alpha}$ are fermionic annihilation operators on the lattice describing the electrons and the periodically varying $\vec{E}(t)$ is the electric field vector of the incident radiation of frequency $\omega=2\pi/T$. It is well known using Floquet theory that the stroboscopic time evolution in such systems is governed by an effective Floquet Hamiltonian $H_F$. In this case, as we have seen previously in Sec.~\ref{unitary_prep} that the amplitude of the external drive $\vec{E}(t)$ can be so chosen that the Floquet Hamiltonian becomes topologically non-trivial despite the initial system being topologically trivial. However, similar to the protocol in Sec.~\ref{unitary_prep}, one must switch on the periodic driving adiabatically such that the stroboscopic system remains in the ground state of the Floquet Hamiltonian $\ket{\psi_0^F}$ such that
\begin{eqnarray} \left|\vec{E}(t)\right| &=& 0 ~~~{\rm for}~~~ t<0 \non \\
&=& \frac{t}{\tau}~~~{\rm for}~~~ 0<t<\tau \non \\
&=& 1 ~~~{\rm for}~~~ \tau<t, \end{eqnarray}
where $\tau$ is a ramping time scale such that $H_F$ is topologically non-trivial when the external field has been completely switched on at $t=\tau$. Using this driving it was exemplified that although with periodic boundary conditions, the Chern number remains invariant under arbitrary unitary dynamics, the Bott index in systems with boundaries
does indeed reflect the topological transition of the Floquet Hamiltonian and 
hence the stroboscopic state of the system by exhibiting quantized 
jumps (see Figure 3 in Ref.~\onlinecite{rigol15}).

It has also been shown in Ref.~\onlinecite{rigol17} that the lattice Chern number can capture dynamical topological phase transitions in translationally invariant finite systems which do not contain gapless points in the BZ. This is because incommensurate finite systems (systems missing gapless critical points in their BZ) allow one to maintain adiabaticity throughout the transition while the discretized lattice Chern number reflects the topology of the thermodynamic system. However, in thermodynamically large systems, such a feat is not possible in either translationally invariant systems or systems under open boundary conditions\ct{ge21} due to the necessary presence of gapless quantum critical points in the BZ. As the Bott index for finite translationally invariant systems is equivalent to the lattice Chern number (see Ref.~\onlinecite{rigol17} and Appendix~\ref{sec:bott}), it allows the dynamical variation of the Bott index to capture topological phase transitions in such systems with periodic boundary conditions. Due to the equivalence of the Bott invariant and the lattice Chern number with periodic boundary conditions in finite-size systems, the Bott index must approach the conventional Chern number as defined in Eq.~\eqref{haldane_chern} in the thermodynamic limit of such systems. Note that this does not conflict with the fact that in finite systems, the Bott index is allowed to vary dynamically while the conventional Chern number remains invariant \ct{daniel17}.

\subsection{The non-unitary no-go theorem}\label{nogo2}

In the previous section we saw that the Chern number in thermodynamically large systems with periodic boundary conditions must remain invariant under all unitary dynamics, thus being unable to describe dynamical topological phase transitions. However, one may wonder whether such a feat is possible to achieve by breaking the unitarity of the drive. In this section we will discuss precisely the problems with this approach with a simple physical picture which is nothing but a demonstration of a more general theorem\ct{goldstein19}. Let us first assume that one wishes to engineer a Chern non-trivial pure state $\ket{\Psi}$ as an asymptotic steady state of a Markovian non-unitary process. Let us then say that the target state $\ket{\Psi}$ is chosen to be the ground state of some reference Hamiltonian of non-interacting fermions on a lattice respecting a discrete translational invariance,
\begin{equation}
H^{ref}=\sum\limits_{ij\alpha\beta}h^{ref}_{\alpha\beta}(i-j)c^{\dagger}_{i\alpha}c_{j\beta}+ {\rm H.c.},
\end{equation}
where the indices $i$ refer to the unit cells while the greek indices $\alpha$ refer to the internal basis degrees of freedom, such as sublattice. The reference Hamiltonian can then be diagonalized using the translational invariance as,
\begin{equation}
H^{ref}=\sum\limits_{k\alpha}E^{ref}_{\alpha}(k)c^{\dagger}_{k\alpha}c_{k\alpha},
\end{equation}
where $k$ are the conserved quasimomenta. We now assume that the lowest energy band having energy $E_1^{ref}(k)$ of the reference Hamiltonian is dispersionless (exactly flat) and separated from the other bands by a finite gap. We might then set $E_1^{ref}(k)=0$ for simplicity without any lack of generality. We will see later as to why this assumption is necessary and how it helps to elucidate the difficulty of dissipatively preparing pure but non-trivial Chern states. Now, consider an artificially constructed initial many-body state, in which, apart from the ground state, some of the  excited states of the reference Hamiltonian are also occupied. Again, to simplify the discussion we will further assume that the Hamiltonian dynamics of the system has been suppressed while it is interacting with a Markovian bath. The bath we consider must empty all states of the system other than the ground state and simultaneously, completely fill the ground state if it is not filled initially. Thus, to produce the targeted state as a steady state, the bath must have two functions simultaneously, an evaporative effect from excited states and a condensation into the ground state of the reference Hamiltonian. First, let us design the evaporative processes in the system. To conceive such an evaporative process, there must be two different types of electron, one trapped into the lattice $c_{i\alpha}$ and one free to escape to infinity $b_{i\alpha}$ (say in the direction $\hat{z}$) such that they may never return amounting to the Markovian nature of the process and we choose the evaporation process to be described by the Hamiltonian,
\begin{eqnarray}\label{interaction}
H^{e} &=& \sum\limits_{ij\alpha\beta} (h^{ref}_{\alpha\beta}(i-j)
b^{\dagger}_{i\alpha}c_{j\beta}+ {\rm H.c.}) \non \\
&=& \sum\limits_{k\alpha} (E^{ref}_{\alpha}(k)b^{\dagger}_{k\alpha}a_{k\alpha}
+ {\rm H.c.}). \end{eqnarray}
Since the free $b$ electrons are completely free to escape in the direction $\hat{z}$, we can describe their dynamics by the Hamiltonian
\begin{equation}
H^{b}=\sum\limits_{i\alpha k_z}\left(e_z(k_z)-e_0\right)b_{i\alpha k_z}^{\dagger}b_{i\alpha},
\end{equation} 
where $e_z(k_z)$ is the energy due to motion in the free direction $z$ and $e_0$ is the left-over kinetic energy of the electron after making a transition from $a$ to $b$, i.e., the kinetic energy with which it escapes the lattice potential. Now that we have completely constructed the evaporative effects on the system, we will analyse the dynamics generated by the Hamiltonian,
\begin{equation}
H_1=H^{e}+H^{b}.
\end{equation}
Note that although the dynamics of the complete system is unitary, if one chooses to ignore the escaped electrons the dynamics of just the system of bounded electrons on the lattice is essentially non-unitary. Within the Markovian approximations, one can then trace out the degrees of freedom of the free escaped electrons belonging to the bath to obtain a dynamical Lindblad equation similar our discussion in Sec.~\ref{atomic_wire} which dictates the dynamics of the density matrix of just the system $\rho$, such that
\begin{equation}
\partial_t\rho=\mathcal{L}^e(\rho),
\end{equation}
where in this case, the Liouvillian $\mathcal{L}^e$ takes the form
\begin{equation}\label{evaporation}
\begin{split}
\mathcal{L}^e(\rho)=\sum\limits_{k\alpha}\frac{\gamma_{\alpha}^e(k)}{2}\left(2a_{k\alpha}\rho a_{k\alpha}^{\dagger}-\{\rho,a^{\dagger}_{k\alpha}a_{k\alpha}\}\right)\\
=\sum\limits_{ij\alpha\beta}\frac{\gamma_{\alpha\beta}^e(i-j)}{2}\left(2a_{i\alpha}\rho a_{j\beta}^{\dagger}-\{\rho,a^{\dagger}_{i\alpha}a_{j\beta}\}\right),
\end{split}
\end{equation}
such that the evaporation rates can be estimated by the Fermi-golden rule $\gamma^e_{\alpha}(k)\propto \left(E^{ref}_{\alpha}(k)\right)^2$ (see the interaction Hamiltonian in Eq.~\eqref{interaction}) which in real space corresponds to,
\begin{equation}\label{rate}
\gamma^e_{\alpha\beta}(i-j)\propto\sum\limits_{l\mu}h^{ref}_{\alpha\mu}(i-l)h^{ref}_{\mu\beta}(l-j),
\end{equation}
which simply takes into account virtual scattering processes with the bath such as $a\rightarrow b\rightarrow a$ and estimates the action of the bath on the system electrons bounded on the lattice. Also, note that since we have assumed that the ground state of the reference Hamiltonian is flat and that $E^{ref}_1(k)=0$ identically, the transition amplitude $\gamma^e_1(k)$ too vanishes identically. What this means is that remarkably, the evaporative process does not at all affect the lowest band of the reference Hamiltonian which is nothing but the target state. Since the other states are not dispersionless, electrons start evaporating from them as an action of the dissipative processes and eventually must all be empty in the steady state. However, the occupation of the lowest target band will not be affected at all by the evaporation. Note that this is similar to the existence of a dissipation-less subspace as we have already discussed in Sec.~\ref{atomic_wire} for the Kitaev chain. But we have not yet been completely successful with our attempt as if the ground state of the reference Hamiltonian is not completely filled initially, the bath must also fill it up completely in the asymptotic steady state. We therefore have to include a pumping/condensation action of the bath which selectively injects electrons into the different bands of the reference Hamiltonian. Such a process can be written in terms of a minimal Liouvillian,
\begin{equation}
\mathcal{L}^c(\rho)=\sum\limits_{k\alpha}\frac{\gamma_{\alpha}^c}{2}\left(2a^{\dagger}_{k\alpha}\rho a_{k\alpha}-\{\rho,a_{k\alpha}a^{\dagger}_{k\alpha}\}\right),
\end{equation}
where for the jump operators we have chosen creation operators in place of the annihilation operators in Eq.~\eqref{evaporation} and vice versa to simulate the reverse pumping into the system. One must also choose the condensation rates such that $\gamma^c_{\alpha}=0$ and only the $\gamma^c_{\alpha=1}\neq0$, which ensures that only the ground state of the reference Hamiltonian, i.e., the target state is filled completely while there is no pumping into the other excited states. The complete dissipative process can then be combined into a single dynamical process,
\begin{equation}
\partial{\rho}=\mathcal{L}^e(\rho)+\mathcal{L}^c(\rho).
\end{equation}
We have thus engineered a purely dissipative process which results in a unique and pure steady state of the system which is nothing but the ground state of the reference Hamiltonian $H^{ref}$. Now, note that the complete construction depends on the target ground state being flat which allows the steady state to be strictly pure. However, recently it has been shown that any finite range Hamiltonian cannot host topologically non-trivial perfectly flat bands. Therefore, as long as our reference Hamiltonian has a finite range, the dissipative steady state cannot be topologically non-trivial. Equivalently, from Eq.~\eqref{rate} it can be seen that no Lindblad operator which has a finite range can result in a perfectly pure and topological state as a steady state when the dynamics is purely dissipative. It can be shown that this complete discussion is simply a demonstration of a more generic no-go theorem which states that even by inclusion of the Hamiltonian dynamics in the Lindblad equation, such a feat cannot be achieved using finite-range Lindbladians such that a topologically non-trivial pure state is obtained which is also stable against external perturbations with exponential damping.

We seem to have arrived at a puzzling situation at this point. Both the no-go theorems strictly forbid the dynamical generation of non-trivial Chern topological states of matter. However, recently there have been attempts to go around the no-go theorems which might lead to successful preparation of Chern topological states even in thermodynamically large systems.

\subsection{Dissipative preparation of mixed states}

\subsubsection{Dissipators having a finite range}

As discussed in the previous section, a Markovian bath where the Lindbladians have a finite range is incapable of generating pure Chern non-trivial steady states asymptotically. Therefore, it seems unavoidable to forgo either the purity of the steady state or the locality of the bath to ensure a non-trivial Chern insulating steady state. In this section we will mainly ponder on the first possibility in which a non-trivial steady state is obtained using dissipative dynamics with local Lindblad operators\ct{diehl15}. In the process as one is bound to topologically classify mixed states as in Eq.~\eqref{density}, we simply follow the winding of the Bloch vector $\vec{n}_k$ characterizing the mixed state density matrix by defining a Chern number,
\begin{equation}
C=\frac{i}{2\pi}\int_{BZ}\frac{{\rm Tr}\{\rho_k\left[\partial_{k_x}\rho_{k},\partial_{k_y}\rho_k\right]\}}{\left(2{\rm Tr}\rho_k^2-1\right)^{3/2}},
\end{equation}
which after some algebraic manipulation reduces to the integer-quantized winding number
\begin{equation}\label{mixed_chern}
C=\frac{1}{4\pi}\int_{BZ}\hat{n}_k \cdot \left[\partial_{k_x}\hat{n}_k\times\partial_{k_y}\hat{n}_k\right],
\end{equation}
of the projection of $\vec{n}_k$ over a $\mathcal{S}^2$ Bloch sphere for the density matrix. As can be directly deduced, in analogy with the pure state Chern number, the topological invariant defined in Eq.~\eqref{mixed_chern} is protected by a finite {\it purity gap}, i.e., the value of $\left|\vec{n}_k\right|^2$ for all $k\in{\rm BZ}$ and can only change when $\left|\vec{n}_k\right|^2$ vanishes. Now, consider a purely dissipative dynamics generated by Markovian Lindbladian operators of the form of Nambu spinors,
\begin{equation}\label{lindblad_hpm}
L^C_j=\sum\limits_{i} (u^C_{i-j}\psi_{i}+v^C_{i-j}\psi_{i}^{\dagger}),
\end{equation}
where $\psi_{i}$ and $\psi_{j}^{\dagger}$ are fermionic creation and annihilation operators and $u^C_{i},v^C_{i}$ are complex numbers on a square lattice. To ensure the finite range of the Lindblad operators, let us assume that the only non-zero coefficients are $v_0^C~(=\beta ~\text{say})$, $v^C_{\pm \hat{x}}=1$ and $v^C_{\pm \hat{y}}=-1$ on the two-dimensional square lattice having $N$ lattice points with periodic boundary conditions with orthogonal directions $\hat{x}$ and $\hat{y}$, allowing for only nearest-neighbor hoppings in the Lindbladians. It is then clear from the Lindblad master equation (see Eq.~\eqref{lindblad}) dictating the dynamics of the system in which all Hamiltonian dynamics is suppressed, that the asymptotic steady state must be the ground state of the reference Hamiltonian $H^C_p=\sum\limits_{i}L_{i}^{C\dagger}L^C_i$, i.e., a vacuum of the spinors $L_i^C$. As the lindblad operators are linear in the Fermionic operators, it can be shown that in the thermodynamic limit, the steady state solution of such Lindblad equations can be written in terms of Gaussian density matrices. The density matrix of such a system can then be written in terms of decoupled quasimomentum sectors which are also quadratic in some Nambu basis $\left(\hat{a}_k,\hat{a}_{-k}^{\dagger}\right)^{T}$ as
\begin{equation}
\rho_k=\frac{1}{2}\left(\mathbb{I}-\vec{n}_k \cdot \vec{\sigma}\right),
\end{equation}
where $\sigma^i$ are Pauli matrices in the Nambu basis and the Nambu spinors in the Fourier space are given as
\begin{equation}
\hat{a}_k=\frac{1}{\sqrt{N}}\sum\limits_{j}e^{ikj}\psi_{j}.
\end{equation}
The corresponding Bloch vector $\vec{n}_k$ denotes has components along three orthogonal directions $\{\hat{e}^1,\hat{e}^2,\hat{e}^3\}$.
In the Fourier transformed reciprocal space, the Lindbladian operators $L^C_i$ can then be equivalently represented in the basis of Nambu spinors on the square lattice as,
\begin{equation}\label{nambu}
\left( \begin{array}{c}
\tilde{u}_k^C \\
\tilde{v}_k^C \\
\end{array} \right)=\left( \begin{array}{c}
2i (\sin{k_x}+i\sin{k_y}) \\
\beta+2 (\cos{k_x}+\cos{k_y}) \\
\end{array} \right), \end{equation}
which vanish together only at isolated values of the quasimomenta $k\equiv(k_x,k_y)$. The simultaneous vanishing of the functions $u_k$ and $v_k$ is intricately connected to the stability of the asymptotic fixed point of the Lindblad equation which manifests as the stability of the steady state under small perturbations. To understand this, we need to get back at the differential equation Eq.~\eqref{correlation} governing the dynamics of the correlation matrix of the system under the dissipative action of such a bath which is dictated by the two matrices $X$ and $iY$. While the matrix $X$ being real, amounts to the damping of the correlations and all the coherent oscillations are contained in the $iY$ component. It is therefore the positive semi-definite matrix $X$ which should be held responsible for any damping of the system to the steady fixed point subsequent to a small perturbation and consequently, the stability of the steady state. It is for this reason that the minimum eigenvalue of its Fourier transform $\tilde{X}(k)$ is known as the {\it damping gap} which essentially determines the rate of returning the correlations to their steady state value after a perturbative disturbance. A vanishing of the damping gap thus results in a critically damped system which is not exponentially but algebraically damped to the steady state. Note that since, the matrix $M$ in the definition of $X$ (see Eq.~\eqref{correlation}) is quadratic in the Lindblad operators, in the Nambu basis of the corresponding reciprocal space it is clear that the damping gap $\kappa=\left|\tilde{v}_k\right|^2+\left|\tilde{u}_k\right|^2$ in our situation at hand. Therefore, a simultaneous vanishing of both $\tilde{u}_k$ and $\tilde{v}_k$ results in a vanishing of the damping gap and a consequently only a critically damped stability of the steady state fixed point rather than a stable exponential decay. One such isolated point in the system is $\beta=-4$ for $k=0$, as is evident in Eq.~\eqref{nambu} which we call the {\it critical point} owing to the damping criticality the system exhibits at that point. A direct calculation then shows that at the damping criticality $\beta=-4$, the steady state density matrix Bloch vector for $k=0$, i.e., $\hat{n}_0=\hat{e}^3$ while the other components vanish and the Chern number $C$ defined through Eq.~\eqref{mixed_chern} becomes non-trivial, assuming the value $C=-1$, however with a critical stability. Away from such isolated critical points, the functions $\tilde{u}_k$ and $\tilde{v}_k$ are everywhere non-zero in the BZ and one obtains a trivial steady state in such situations having $C=0$. Therefore, even the dynamical preparation of such a critically damped Chern non-trivial state requires stringent fine-tuning as even an infinitesimal deviation from the critical point will lead to a trivial steady state. To elaborate, consider a small deviation from the critical point by $\delta$, i.e., $\beta=-4-\delta$. Under such Lindblad operators, a direct calculation shows that the third component of the Bloch vector $\hat{n}_k^3$ in the steady state reduces to $\hat{n}_k^3=\left|\tilde{u}_k\right|^2-\left|\tilde{v}_k\right|^2$, such that $\hat{n}_0^3=-\left|\tilde{v}_k\right|^2$ for a small non-zero $\left|\tilde{v}_0\right|=\delta$ which implies that $\hat{n}_0=\hat{e}^3$. However, $\hat{n}_k^3\approx\hat{e}^3$ for $\delta\l k\lll 1$ again as then $\hat{n}_k^3\approx2i(k_x+ik_y)$. Therefore, if one able to correct for this discontinuity of $\hat{n}^3_0$ for finite $\delta$, one would be able to successfully extend the topological non-trivial phase beyond the critical point and hence, also ensure exponential stability of the state. \\

To achieve this, one must notice that the quantity $(1-n_k^3)/2$ is nothing but the population $\braket{a_k^{\dagger}a_k}$ of the mode $a_k$ in the steady state. The complete attempt then reduces to making sure that in the steady state the population $\braket{a_0^{\dagger}a_0}$ vanishes for $\delta\neq0$ which in turn guarantees that $\hat{n}_0^3=\hat{e}_0^3$ leading to the steady state Chern number $C=-1$ even for a non-critical steady state having a non-vanishing $\delta$. This can be achieved by including an auxillary Lindblad operator $L^A=a_0$ along with that given in Eq.~\eqref{lindblad_hpm} which ensures the steady state must also be a vacuum of $a_0$ and corrects for the discontinuity at $k=0$, or one may also choose the auxillary Lindbladian as $L^A=\tilde{u}^A_ka_k$ such that $\tilde{u}^A_k$ is Gaussian and peaked at $k=0$, i.e., $\tilde{u}^A_k=ge^{-k^2/d^2}$ where $g$ and $d$ are controllable parameters.\\

This shows that one may indeed obtain stable topologically non-trivial Chern states as a steady state of a purely dissipative process if one forgoes the purity of the state. \\

\subsubsection{Macroscopic electric polarization and mixed state Chern number}\label{pol}

In this section we will discuss the possibility of relaxing the criteria of finiteness of range of the bath and see that it may lead to stable Chern non-trivial states as the asymptotic steady state. First, we will embark on a Chern topological classification of many-body Gaussian quantum states\ct{diss20,uni20} in two dimensions in the spirit of Sec.~\ref{sec:mixed_1d}. We will see that the macroscopic electric polarization can indeed be used to topologically classify many-body Chern states in two dimensional systems similar to its correspondence with the winding number in one dimension.\\

Consider a Gaussian density matrix of a system which can be written in terms of the real space fermionic creation and annihilation operators acting locally as,
\begin{equation}\label{eq:ss_rho}
\rho=\frac{e^{-\beta\sum\limits_{i,j}a^{\dagger}_iH^F_{ij}a_j}}{{\rm Tr}\left[e^{-\beta\sum\limits_{i,j}a^{\dagger}_iH^F_{ij}a_j}\right]},
\end{equation}
where $G$ is an effective Hamiltonian in real space.
In this state, we evaluate the macroscopic electric polarization vector of the system
in the directions of the lattice vectors spanning the two-dimensional
Bravais lattice,
\begin{equation} \vec{P}=\sum\limits_{i}~P_{\hat{i}} ~\hat{i}, \end{equation}
where $P_{\hat{i}}=\left<\hat{X}_i\right>$ (with $\hat{X}_i=\sum\limits_{j}
x_{i}^j\hat{a}^{\dagger}_j\hat{a}_j$) is the many-body position operator, and $x_i^j$ denotes the $i^{th}$ coordinate of the $j^{th}$ site. The expectation 
value is to be taken over the Gaussian density matrix as defined in Eq.~\eqref{eq:ss_rho}. The translation operator in the $i^{th}$ direction with periodic boundary conditions,
\begin{equation}
\hat{T_{i}}(\delta_i)=e^{i\delta_{i}\hat{X}_{i}},
\end{equation}
where we choose $\delta_i=2\pi/L_i$, $L_i$ being the dimension of the system in the $i^{th}$ direction. The periodicity of the exponential enforces periodic boundary conditions on the lattice. Therefore, with periodic boundary conditions and in the thermodynamic limit, the macroscopic polarization of the system assumes the following form,
\begin{equation}\label{eq:pol_def}
P_i={\rm Im}\ln\left<\hat{T}_i\right>=\sum\limits_{\alpha}p_{\alpha}\left<\hat{T}_i\right>_{\alpha},
\end{equation}
where the weighted average is taken over all the purity bands of the density matrices having purity $p_{\alpha}$.
{In the thermodynamic limit, this compactified definition of the macroscopic polarization reduces to the conventional bulk polarization of the system. This is evident from the fact that, for a many-particle pure state, $\ket{\Psi}$ (which is a slater determinant of the occupied single-particle states),
\begin{equation}
P_i={\rm Im}\ln\left<\hat{T}_i\right>={\rm Im}\ln \det{U}={\rm Im}\ln e^{{\rm Tr}\ln U},
\end{equation}
where the matrix $U$ contains all the overlap of the single-particle matrix $T_i$ between the occupied single particle states,
\begin{equation} U_{mn}=\bra{\psi_m}T_i\ket{\psi_n}, \end{equation} 
which implies 
\begin{equation} (U)_{k\alpha,k^{\prime}\beta}=\braket{\psi_{k_i+\delta_i,\alpha}|\psi_{k,\alpha}}\\
\simeq e^{-i\left(A_{i}^k\right)_{\alpha\alpha}\delta_i},
\end{equation}
where $k$ denotes the single-particle momenta while $\alpha$ and $\beta$ are the band indices and $\left(A_{i}^k\right)_{\alpha\alpha}$ is the $U(1)$ connection of the $\alpha^{th}$ occupied band. 
\begin{equation}\label{eq:pure_pol_1}
P_i=\sum\limits_{\alpha}{\rm Im}\int_{BZ}\braket{\psi_{k,\alpha}|\partial_{k_i}|\psi_{k,\alpha}}dk_{1}dk_{2},
\end{equation}
where $k$ denotes the single-particle momenta while $\alpha$ is the band indices. We extend this definition of the polarization vector for the mixed Gaussian steady state (Eq.~\eqref{eq:ss_rho}) as a weighted sum over the polarization over all Floquet eigenstates weighted by their respective populations in the stroboscopic steady state, which reduces to,
\begin{equation}\label{eq: pol_simple}
P_i={\rm Im}\sum\limits_{b}\ln\left[1+(L_i)_{bb}\right],
\end{equation}
such that components $(L_i)_{bb}$,
\begin{equation}\label{eq: pol_simple1}
(L_i)_{bb}=\prod\limits_{k}diag\{e^{-\beta E^k_{b}}\}e^{i\left(A_{i}^k\right)_{bb}\delta_i},
\end{equation}
where $\delta_i=2\pi/L_i$, $L_i$ is the dimension of the system in the 
$i^{th}$ direction, $E^k_{\alpha}$ are the Floquet quasienergies, 
and $\left(A_{i}^k\right)_{bb}$ is the $U(1)$ gauge connection over the band 
$b$ in the $i^{th}$ direction,
\begin{equation}
\left(A_{i}^k\right)_{bb}=\braket{\psi_{k,b}|\partial_{k_i}|\psi_{k,b}}.
\end{equation}

In the limit where the temperature of the bath goes to zero (i.e., $\beta\rightarrow\infty$), the many-body exponential weights predominantly selects the lowest energy band for each $k$ mode. In the $\beta\rightarrow\infty$ limit, the corresponding macroscopic polarization approaches,
\begin{equation}
P_i\simeq {\rm Im}\ln(L_i)_{gg},
\end{equation}
where $(L_i)_{gg}$ is the product of the element of the matrix $L_i$ in the lowest quasienergy state $\ket{g_k}$ over all $k$. Hence, the pure state polarization reduces simply to,
\begin{equation}\label{eq:pure_pol}
P_1=\int_{k_{02}}^{k_{02}+1}\int_{k_{01}}^{k_{01}+1}dk_{1}dk_{2}\bra{g_k}\partial_{k_1}\ket{g_k}
\end{equation}
and likewise for $L_2$, where $k_0\equiv(k_{01},k_{02})$ is the origin of the BZ over which the integration is performed. The macroscopic polarization is observed to express itself as a weighted sum of the polarization over each band of the Floquet Hamiltonian. \\

However, it is well established that the macroscopic polarization is not uniquely defined in a Chern non-trivial phase. Hence, despite its many-body nature, the macroscopic polarization is not a measurable quantity in a Chern insulator. We now proceed to define the Chern number as the change in the quantity $P_i$ {corresponding to} a shift $\delta \vec{k}_0$ in the BZ origin. This leads to the definition,
\begin{equation}\label{eq:pure_chern_I}
C\propto 
\delta P_1[\vec{k}_0]=-\delta k_{02}\int_{k_{02}}^{k_{02}+1}dk_{2}~
\partial_{k_2}\gamma(k_2),
\end{equation} 
\begin{equation}\label{eq:pure_beta}
{\rm where}~~~ \gamma(k_2)=-{\rm Im}\int_{k_{01}}^{k_{01}+1}dk_1 \bra{g_k}\partial_{k_1}\ket{g_k},
\end{equation}

The quantity $\mathcal{C}$ defined in Eq.~\eqref{eq:pure_chern_I} is invariant under a local $U(1)$ gauge transformation owing to the non-interacting nature of the systems studied in this context. 
At {\it equilibrium}, when any one of the bands is completely filled, the quantity $P_i$, reduces to the total MEP of the occupied band. In this situation, the Chern number defined in Eq.~\eqref{eq:pure_chern_I} simply detects a branch change of the function $\gamma(k_2)$ in the closed $\mathcal{S}^1$ interval $k_2\in[0,1]\equiv I$; which equivalently counts 
the winding of $\gamma(k_2)$ along $k_2$ \ct{vanderbilt09}. {This implies that the existence of a branch singularity in the map $k_2\in[0,1]\rightarrow\gamma(k_2)$, signals the Chern non-triviality of the system. In the following, we will elaborately discuss different aspects concerning the topological nature of the dynamical CN defined above.}\\

First, let us focus on the equilibrium topological characterization: the function $\gamma (k_2)$ as described in the manuscript is nothing but a uni-directional Berry phase along one of the periodic directions $\mathcal{S}_a^1\equiv k_1\in[0,1]$ and defined at each point of the $\mathcal{S}^1_b$ interval $k_2\in[0,1]$. This decomposition into two circles $\mathcal{S}^1_a$ and $\mathcal{S}^1_b$ is possible because the BZ forms a two-tori $T^2$ which is topologically equivalent to,
\begin{equation}
T^2\equiv \mathcal{S}^1_a\times \mathcal{S}^1_b.
\end{equation}
Now, the shift in the polarization is directly proportional to the Chern number which is a $\mathbb{Z}$ topological invariant. Equivalently, a branch change of the function $\gamma(k_2)$ in equilibrium at the ends of the BZ in a topological phase results in the non-uniqueness of the polarization. We however observe that it is not essential for the branch singularity to occur at the end points of the BZ. In fact, the branch singularity of $\gamma(k_2)$ at any point $k_2^*\in[0,1]$ reflects the topology of the system (as illustrated below). This is because the invariant $C$ defined in the manuscript simply provides with a homotopy classification of the map $\kappa_2\in\mathcal{S}_b^1\rightarrow\gamma(k_2)\in\mathcal{S}^1$. In fact, the invariant $C^U$ reflects the integer winding of the function $\gamma(k_2)$ as $k_2\in[0,1]$ which in turn is bound to be integer quantized as the fundamental homotopy group of the map $\mathcal{S}_b^1\rightarrow\mathcal{S}^1$ is $\pi_1(\mathcal{S}^1)\equiv\mathbb{Z}$.\\

{By fixing a gauge, such that $\gamma(k_2)$ remains continuous for all 
$I: k_2\in[0,1]$ in a topological phase, the function $\gamma(k_2)$ exhibits 
a branch change proportional to the CN, at the endpoints of the BZ, i.e.,
$C\propto\gamma(k_{02})-\beta(k_{02}+1)$ as in Ref.~\onlinecite{vanderbilt09}.}

To elaborate, choosing a smooth gauge in $I$ ensures that the derivative 
$d\gamma(k_2)/ dk_2$ is well-defined in the interval and its integration over the $\mathcal{S}^1\equiv I:k_2\in[0,1]$,
\begin{equation}\label{eq:1}
\frac{1}{2\pi}\int_0^{1}\frac{d\gamma(k_2)}{dk_2}dk_2=\frac{1}{2\pi}\left[\gamma(1)-\gamma(0)\right]=-\Delta
\end{equation}
simplify reduces to the difference between the $\gamma$ function evaluated at the "end points" of the interval $I$. Owing to the single-valuedness of the wave function at $k_2=0$ and $k_2=1$, this jump '$-\Delta$' is simply the integer quantized Chern number. Hence, a non-zero Chern index in this case implies a branch change of the map $B:\kappa_2\rightarrow\gamma(k_2)$ after a complete rotation in $k_2\in \mathcal{S}^1$. \\

Now, since, the interval $I$ forms a complete circle $\mathcal{S}^1$, the occurrence of the branch change at any other point $k_2^*$ can also be included into the same equivalence class. This can be equivalently understood as since the topological invariant counts the winding of the fibre $\gamma(k_2)$ over the base space $\mathcal{S}^1\equiv k_2\in[0,1]$ and merely changing the position of the topological kink does not change the homotopy class of the map. However, if a smooth gauge is not chosen and $\gamma$ becomes discontinuous (and hence non-differentiable) at an inner point $k_2^*\in I$, caution must be taken while evaluating the integral in Eq.\eqref{eq:pure_chern_I},
\begin{eqnarray}\label{eq:2}
&~&\frac{1}{2\pi}\int_0^{1}\frac{d\gamma(k_2)}{dk_2}dk_2= \nonumber\\
&~&\frac{1}{2\pi}\lim\limits_{\epsilon\rightarrow 0^+}\left(\int_{0}^{k_2^*-\epsilon}+\int_{k_2^*+\epsilon}^{1}\right)\frac{d\gamma(k_2)}{dk_2}dk_2,
\end{eqnarray}
where we have tactically removed the isolated point $k_2^*$ where $\gamma$ is not differentiable. This in a way is again equivalent to the destruction of simply-connectedness of the base manifold $I$ with respect to the map $B$, hence allowing for a non-trivial homotopy classification. By 
evaluating the integrals on the RHS of Eq.~\eqref{eq:2} we obtain,
\begin{eqnarray}
&~&\frac{1}{2\pi}\int_0^{1}\frac{d\gamma(k_2)}{dk_2}dk_2 \nonumber \\
&=&\frac{1}{2\pi}\lim\limits_{\epsilon\rightarrow 0^+}\left[\beta(k_2^*-\epsilon)-\gamma(k_2^*+\epsilon)\right]\nonumber\\
&=&-\Delta,
\end{eqnarray}
which is exactly the jump in $\gamma(k_2)$ and therefore may be interpreted as a signature of topological non-triviality of the equilibrium system. Also, the jump $\Delta$ is a gauge invariant quantity and must be integer multiples of $2\pi$ owing to the single-valuedness of the wave functions at every point of $I$, i.e., 

\begin{equation}\label{eq:delta}
\Delta=2\pi\mathcal{C},~~\mathcal{C}\in\mathcal{I}.
\end{equation} 

We note, in general $\gamma$ may exhibit multiple isolated discontinuities, in which case, applying a similar protocol one obtains,
\begin{equation}
\begin{split}
\mathcal{C}=\frac{1}{2\pi}\sum\limits_{\nu}\Delta_{\nu},
\end{split}
\end{equation}
where the sum is taken over all the isolated jump discontinuities of $\gamma$.\\

\subsubsection{Long-range dissipators and Floquet Chern insulators}

We have previously seen that in periodically driven systems, the stroboscopic Floquet Hamiltonian can be topologically non-trivial despite the bare undriven system being a trivial one. In this section we will see that dynamically preparing the system in its topological Floquet state is indeed possible for two-dimensional Chern insulators dissipatively. We start with a quadratic fermionic system in
two dimensions which has a sublattice structure (particularly the Haldane model of {graphene} )
subjected to a generic temporal drive generated by the Hamiltonian $H_S(t)$. The system is assumed to be coupled to a free fermionic bath $H_R$ (the {\it reservoir}) through a bilinear coupling $H_I$. The time evolution of the complete system is then described by the Hamiltonian\ct{diss20},
\begin{equation}
H(t)=H_S(t)+H_R+H_I,
\end{equation} 
such that
\begin{eqnarray} H_S(t) &=& \sum_{\alpha,\beta,n,m} [H_{\alpha,\beta,n,m}(t) 
a_{m,\alpha}^{\dagger}a_{n,\beta}+ {\rm H.c.}], \non \\
H_R &=& \sum_{i,m}\epsilon_{i}A_{m,i}^{\dagger}A_{m,i}, \non \\
H_I &=& \sum_{i,n,\alpha}[\lambda_{i,\alpha} A_{n,i}^{\dagger}a_{n,\alpha}+
{\rm H.c.}], \end{eqnarray}
where $a_{n,\alpha}$ and $A_{i}$ satisfies fermionic anti-commutation relations independently. The indices $n$ and $\alpha$ on $a_{n,\alpha},A_{n,\alpha}$
denote the sublattice and intra-sublattice index respectively, i.e., $\alpha\in\{A,B\}$ and $n\equiv\{n_1,n_2\}$ is the position of a site in the Bravais lattice having a two-point basis. We focus on a situation in which the dissipative Haldane model is driven periodically in time, i.e.,
\begin{equation}
H_{\alpha,\beta,n,m}(t)=H_{\alpha,\beta,n,m}^0+V_{\alpha,\beta,n,m}(t),
\end{equation} 
where $H^0(M,t_1,t_2,\phi)$ is the bare Haldane model. Invoking the discrete translational invariance of the Hamiltonian one can employ discrete Fourier transform to decouple the Hamiltonian $H(t)$ in momentum space.
Preparing the model initially at time $t = 0$ in the topologically trivial phase having Chern number $\mathcal{C}=0$, we subject it to a periodic driving $V(t+T)=V(t)$ with driving of frequency $\omega=2\pi/T$ such that it solely acts on the nearest-neighbor hopping amplitudes.
The corresponding single-particle Hamiltonian therefore decoupled for each momenta mode,
\begin{equation}\label{Hk}
\begin{split}
H^k_{full}(t)=\bigoplus_{k}\sum_{\alpha,\beta} H_{\alpha,\beta}^k(t)a_{\alpha}^{k\dagger}a_{\beta}^k+\sum_{i,k}\epsilon_i^kA_{i}^{k\dagger}A_{i}^k\\
+\sum_{i,\alpha}\lambda_{i,\alpha}A_{i}^{k\dagger}a_{\alpha}^k+{\rm H.c.},
\end{split}
\end{equation} 
where $k$ denotes the ordered pair $\left(k_1,k_2\right)$ and
$H^k(t)$ is the quasimomentum resolved bare Haldane Hamiltonian subjected to the periodic perturbation given as,
\begin{equation}\label{eq:driven H}
H^k(t)=H^0(k)+V(t)=H^0(k)+V_0\left[\sigma^x\cos{\omega t}+\sigma^y\sin{\omega t}\right],
\end{equation}
where $H^0(k)$ is the bare Haldane Hamiltonian in momentum space.\\

We recall the purely unitary evolution of the Haldane model dictated by the Hamiltonian in Eq.~\eqref{eq:driven H}, starting from a non-topological ground state. The
		corresponding Floquet Hamiltonian $H_F(k)$ generating
		the unitary stroboscopic evolution for $t>0$, in the high frequency limit of driving assumes the form \ct{kitagawa11},}
\begin{eqnarray}
H^F(k) &\simeq& H^0(k)+\frac{1}{\omega}\left[H_{-1}(k),H_{+1}(k)\right] \non \\
&=& H^0(k)-\frac{V_0^2}{2\omega}\sigma^z,
\end{eqnarray}
where the parameters have been chosen such that the Floquet Hamiltonian is topologically non-trivial (Fig.~\ref{fig_1a_res}) despite the bare Haldane model being in the topologically trivial phase (Fig.~\ref{fig_1a_res}).
We assume that the bath to be in an equilibrium Fermi-dirac distribution $f_{FD}(\mu,\beta)$ at all times having a temperature of $T=1/\beta$ and a chemical potential $\mu$ lying in the gap of the Floquet Hamiltonian which can be set equal to zero. In the weak coupling and high frequency limit, it can then be safely assumed that the bath cannot initiate direct electron scatterings between the system energy eigenstates. This results in the steady state stroboscopic behavior of the system to be generically governed by the Floquet Hamiltonian in equilibrium with the bath temperature and chemical potential. It is then straight forward to show by tracing out the degrees of freedom of the fermionic bath that the asymptotic stroboscopic correlations of the system reaches the time independent distribution,
\begin{equation}\label{eq:wei}
\left<f_{b}^{k\dagger}f_{b}^k\right>=\frac{\sum_n W^{knn}_{b b}(E^{k(n)}_{b})f_{FD}(E^{k(n)}_{b})}{\sum_n W^{knn}_{b b}(E^{k(n)}_{b})},
\end{equation}
which is a weighted average of $f_{FD}(E^k_{b},\mu)$ over all the photon sectors. The weights quantify the occupation of each photon sector to the asymptotic population of the Floquet bands and physically is proportional to the total number of intra-band scatterings induced by the bath. As a function of energy, the weights also reflect the energy dependence contribution of the higher photon sectors in the steady occupation 
of the effective Hamiltonian bands.
This suggests that even if the temperature of the bath is near absolute zero, the occupation of a Floquet band have significant contribution from all the photon sectors.\\

\begin{figure}
\subfigure[]{
\includegraphics[width=0.95\columnwidth]{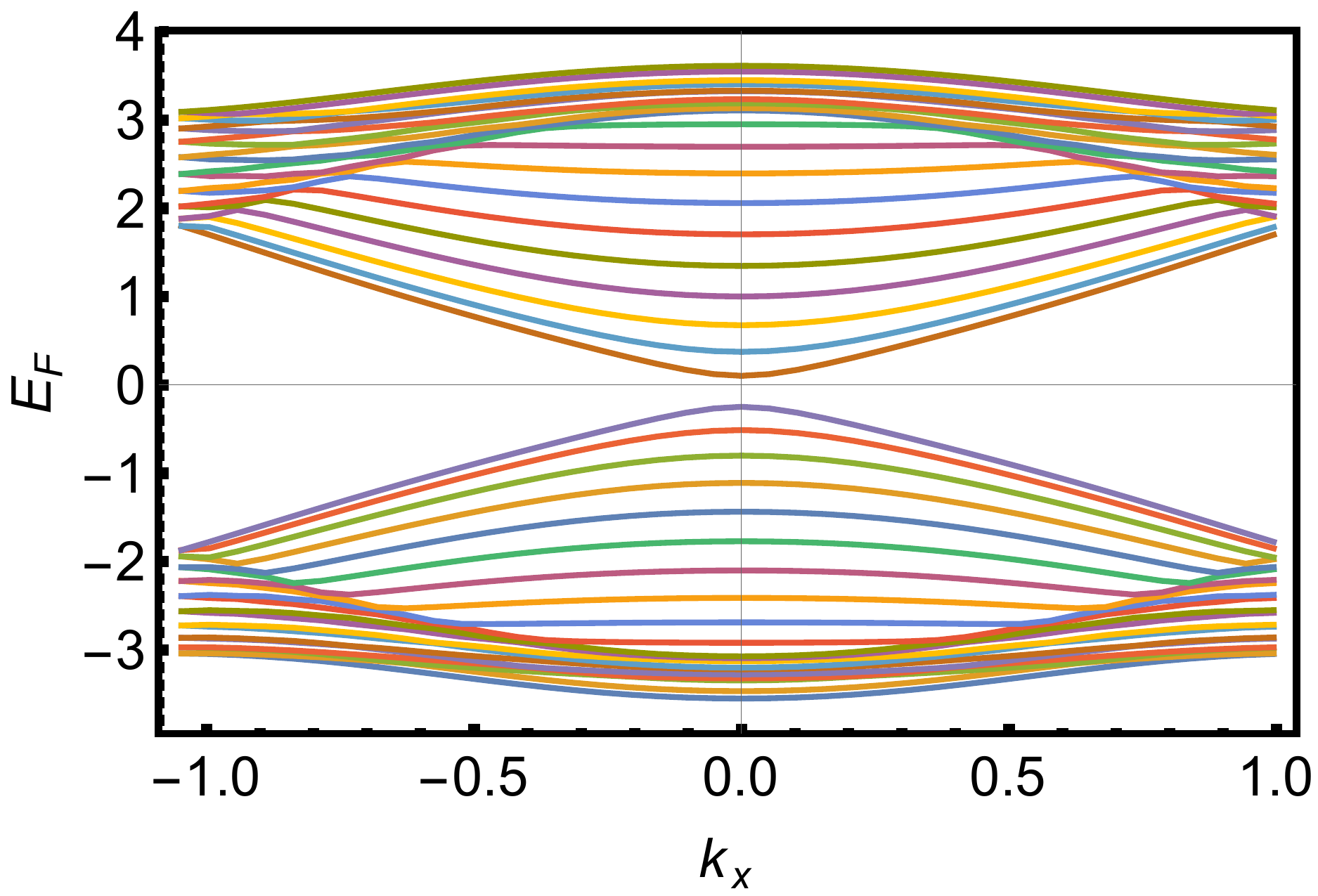}
\label{fig_1a_res}} 
\subfigure[]{		
\includegraphics[width=0.95\columnwidth]{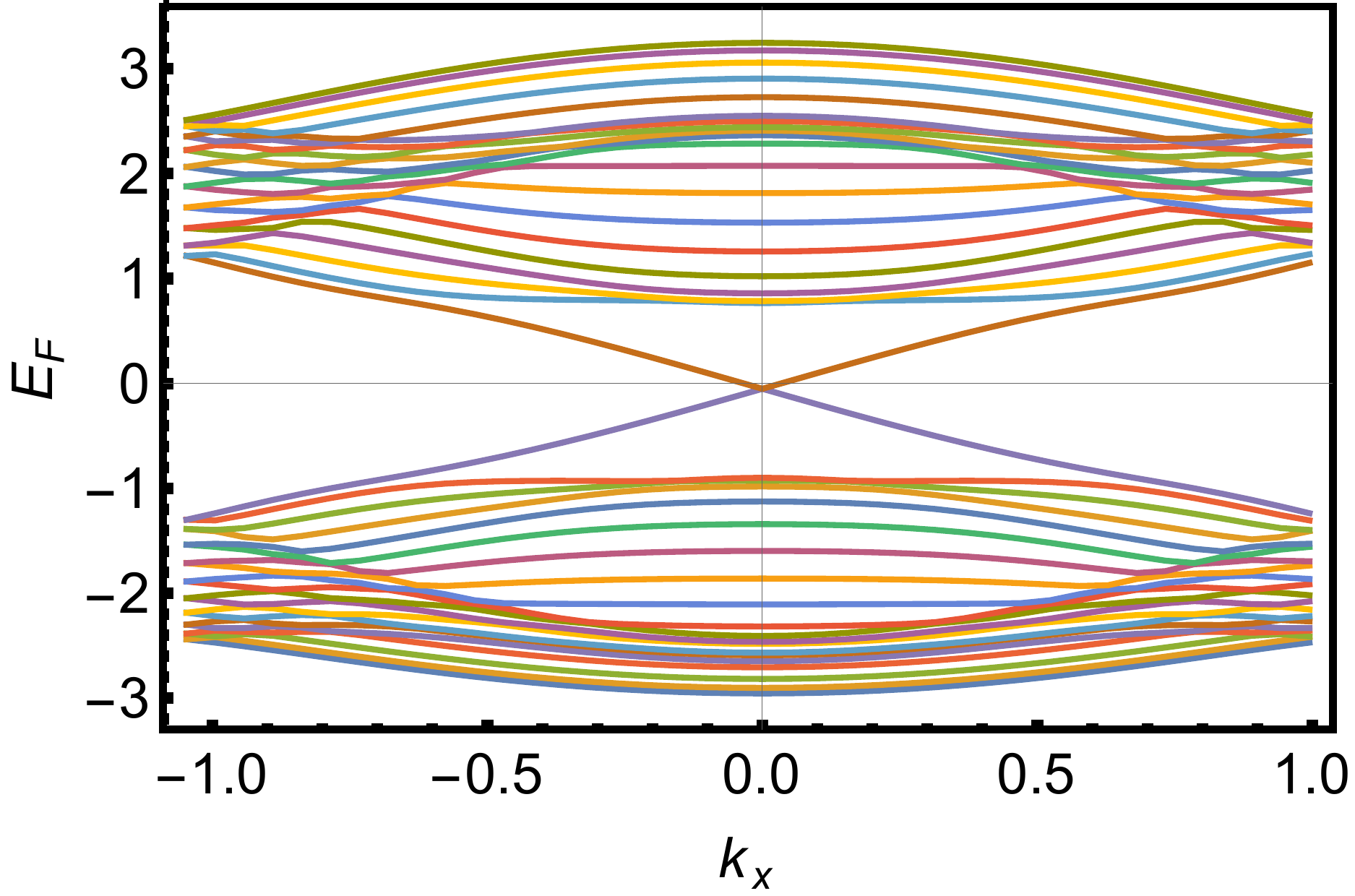}
\label{fig_1b_res}}
\caption{(a) The spectrum of the initial topologically trivial static Hamiltonian in a cylindrical geometry which is a trivial insulator with a bulk and edge gap. (b) The Floquet spectrum after the periodic driving is switched on as in Eq.~\eqref{eq:driven H}. The Floquet spectrum represents a bulk insulator and the presence of conducting edge states. (See Ref.~\onlinecite{diss20}).}
\end{figure}

However, if an energy cutoff $\Omega_c$ is introduced in the dissipative coupling {such that}
the contribution of the higher photon sectors reduce significantly in Eq.~\eqref{eq:wei}, i.e., if
\begin{equation}\label{cut}
\left|\frac{W^k(E_1)}{W^k(E_2)}\right|\rightarrow 0,
\end{equation}
for all $|E_1|\gg \Omega_c$ and $|E_2|\ll \Omega_c$, the only Floquet sectors that contribute to the sum in Eq.~\eqref{eq:wei} are such that $|E^{k(n)}_{\alpha}|<\Omega_c$. {Such a cut-off in the system bath coupling ensures that the energy window of interaction between the system and reservoir is finite.}
If the driving frequency is high enough $(\omega\gg\Omega_c)$, solely the zero photon sector contributes significantly in Eq.~\eqref{eq:wei} and the system thermalizes into an effective Fermi-Dirac distribution.
Thus, the steady state stroboscopic density matrix is a Gibbs state in the Floquet Hamiltonian $H^F(k)$,
\begin{equation}\label{eq: density_matrix}
\rho(NT)=\bigotimes_k {\cal N}_ke^{-\beta\sum\limits_{\alpha,\beta}{a^{k\dagger}_{\alpha}H^F_{\alpha\beta}(k)a^k_{\beta}}},
\end{equation}
and decoupled for each $k\in{\rm BZ}$. As the stroboscopic steady state of the system is a Gibbs state with a temperature of that equal to the bath, its purity is completely determined by the reservoir temperature. Also, one can directly classify such steady states using the Chern number devised in Sec.~\ref{pol} at all temperatures. Consequently, it has been shown that if the Floquet Hamiltonian is topologically non-trivial, so is the stroboscopic steady state at the temperature of the bath. Note that taking the temperature of the bath to zero, one gets infinitely close to a pure stroboscopic steady state which is Chern non-trivial.\\

In this set-up, the applicability of the no-go theorem (see Sec.~\ref{nogo2}) can be understood by suitably choosing the distribution of the system-bath coupling over the system energy scales. Say, the energy cut-off introduced in the system-bath coupling through the Eq.~\eqref{cut} is Gaussian. This suggests that the system-bath interaction is not strictly a finite-ranged matrix. On this we elaborate below exemplifying through a much simpler toy system similar to that in Eq.~\eqref{Hk}. Consider the complete system to be described by a Hamiltonian,
\begin{eqnarray}
H &=& H_S+H_B+H_I, \non \\
H_S &=& \sum\limits_{m,n} [\Lambda_{mn}\hat{a}_m^{\dagger}\hat{a}_n+{\rm H.c.}]
= \sum \limits_{j}E_j\hat{d}_j^{\dagger}\hat{d}_j, \non \\
H_B &=& \sum\limits_{i}\epsilon_i\hat{b}_i^{\dagger}\hat{b}_i, \non \\
H_I &=& \sum\limits_{m,n}[ \lambda_{in}\hat{b}_i^{\dagger}\hat{a}_n+{\rm H.c.}]
=\sum\limits_{ij} [\gamma_{ij}\hat{b}_i^{\dagger}\hat{d}_j+{\rm H.c.}], \non \\
&& \label{eq:ham} \end{eqnarray}
where $H_S$, $H_B$ and $H_I$ describe the system, bath and the system-bath interaction respectively. The operators $\hat{d}_i$ diagonalize the system Hamiltonian. The scattering matrix elements with energy $\Omega$ can then be approximated using the Fermi-golden rule as
\begin{equation}
W_{ij}(\Omega)=\gamma_{\mu i}^*\gamma_{\mu j} ~~~\text{such that}~~~
\epsilon_{\mu}=\Omega, \end{equation}
and we have assumed that the bath has a non-degenerate but continuous energy spectrum. In this situation, we enforce a Gaussian cutoff in the system bath coupling,
\begin{eqnarray}\label{eq:W}
W_{ij}(\Omega)&\sim&\delta_{ij}e^{-\frac{\left(\Omega-m\right)^2}{2\sigma^2}}
\non \\
{\rm implying ~~that}~~~ \left|\gamma_{\mu j}\right|^2 &\sim& e^{-\frac{\left
(\epsilon_{\mu}-m\right)^2}{2\sigma^2}}, \end{eqnarray}
for some $m$.

As both $\hat{b}_k$ and $\hat{d}_k$ represent diagonal modes of the bath and the system respectively and since both the system and bath energies have been chosen to be non-degenerate, Eq.~\eqref{eq:W} shows that the system-bath interaction is such that every mode of the system is coupled to all bath modes with varying strengths. Equivalently, if the bath is thought of as a localized lattice of varying potential at each site, every mode of the system couples to every site of te bath lattice. Thus, the system-bath interaction matrix $\gamma$ employed does not have a finite range and therefore, the applicability of the no-go theorem is unclear for extensive system-bath interactions, even if a time-local Lindbladian dynamics can be framed to describe the system.

\subsection{Going around the unitary no-go theorem: bilayer Chern systems}

As we saw in the previous section that there have been recent attempts at 
circumpassing the non-unitary no-go theorem in the dynamical preparation of Chern topological states. In this section we will discuss the possibility of unitarily preparing Chern non-trivial states while maintaining adiabaticity throughout. This has been exactly seen to be impossible in Sec.~\ref{nogo1} due to the quantum critical point separating Chern inequivalent phases. However, as we will see that such a feat can indeed be achieved by considering multiply stacked coupled Chern insulating layers\ct{peng_cheng19,barbarino_20,sen20}. \\

\begin{figure}[ht]
\ig[width=0.95\columnwidth]{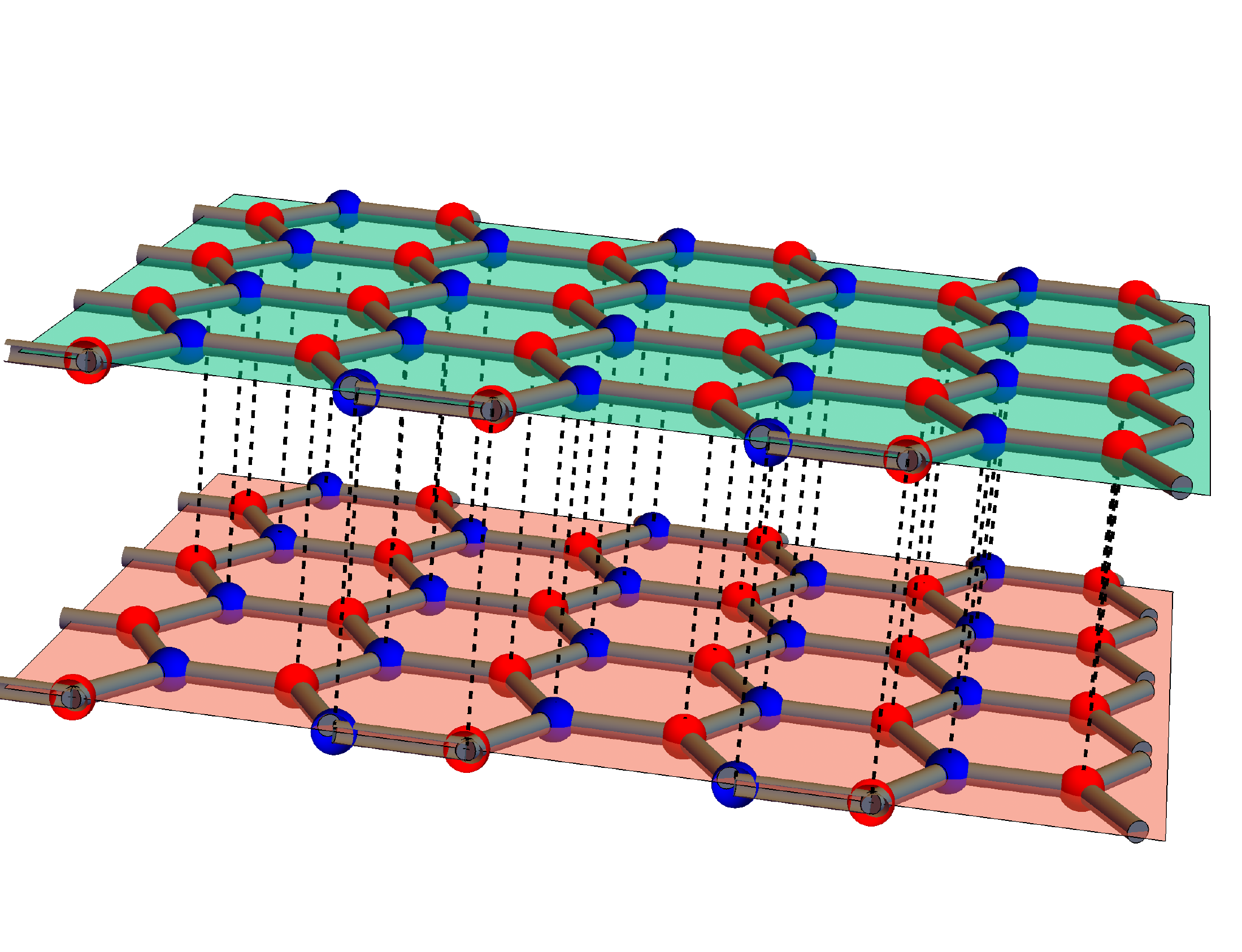}	
\caption[]{The bilayer Haldane model realized by two vertically stacked and perfectly aligned honeycomb lattices. The red and blue spheres correspond to the two sublattices and the black dashed lines indicate that each lattice point is coupled only with the one directly above or below it. (See Ref.~\onlinecite{sen20}).} \label{bilayer} 
\end{figure} 

Consider a bilayer Haldane system (see Fig.~\ref{bilayer}) in which two Haldane layers are kept in close proximity with each other with zero twist angle\ct{sen20}. This allows electrons to tunnel between the layers and the bilayers form a composite system at non-zero coupling. Assuming periodic boundary conditions for the bulk, the Hamiltonian is decoupled as $H=\bigoplus_{\bf k} c_{\bf k}^\dagger H({\bf k})c^{}_{\bf k}$, where $c_{\bf k} = \left(c_{{\bf k},\mathrm{A}}^l, c_{{\bf k},\mathrm{B}}^l, c_{{\bf k},\mathrm{A}}^u, c_{{\bf k},\mathrm{B}}^u\right)$ is a vector of the annihilation operators with $\{A,B\}$ and $\{l,u\}$ being the sublattice and layer indices, respectively. The single-particle Hamiltonian $H({\bf k})$ assumes the form 
\begin{equation}\label{eq_hamil}
H({\bf k})=\bigoplus_{\bf k}\begin{pmatrix}
H_l({\bf k}) & \Gamma\\
\Gamma^\dagger & H_u({\bf k})
\end{pmatrix},
\end{equation}
where $H_{l(u)}$ is the Haldane Hamiltonian corresponding to the lower (upper) layer and $\Gamma$ denotes the interaction potential between the layers. We recall the Bloch form of the Haldane Hamiltonians (see Eq.~\eqref{eq_hamil_k_ap}), $H_{l(u)}({\bf k})={\vec d}^{l(u)} ({\bf k})\cdot {\vec \sigma}$, where 
${\vec d}^{l(u)}=\{d_x,d_y,d_z\}^{l(u)}$ and $\vec \sigma$ is a vector of pseudo-spin operators. Note that only $ d_z^{l(u)}$ depends on the complex phase and is therefore annotated with distinct superscripts for each layer. In what follows, we consider a staggered interlayer coupling of the form $\Gamma=\gamma\tau_z$, where $\tau_z$ is another pseudo-spin operator. \\

 Considering the ground state of such a system, where multiple single-particle states are occupied, the Chern invariant characterizing the topological phases is calculated from the $U(2)$ Berry curvature which is non-Abelian (see Eq.~\eqref{nab} of Appendix \ref{sec:topology}). It can then be shown that by tuning the inter-layer coupling the spectral gap between the occupied single-particle energy states and the vacant ones can be kept finite while the parameters of each layers such as the semenoff mass $M$ and the magnetic fluxes $\phi$ are altered. Also, since, now due to a finite coupling between the Chern insulators, their individual Chern numbers may become undefined at points where the occupied single-particle states become degenerate. However, the ${\rm U(2)}$ Chern number is well-defined even at all such points and should remain invariant under any unitary dynamics of the composite system. The ${\rm U(2)}$ Chern number can be shown to be equivalent to the total Chern number calculated by summing up the Abelian curvatures of the individual bands. \\
 
 However, it has been shown that starting from a decoupled system in which both the layers are in their respective ground states and have well-defined Chern numbers, one can adiabatically switch on an inter-layer coupling $\Gamma$. In the presence of the finite inter-layer coupling one may then modify the parameters of the individual layers while always maintaining a finite gap between the occupied and the unoccupied single-particle states, thus maintaining adiabaticity. Then if the inter-layer coupling is switched off adiabatically, the individual ${\rm U(1)}$ symmetries of the decoupled layers are now completely restored and the individual layers regain well-defined ${\rm U(1)}$ Chern numbers. Through this protocol, one can show that the individual Chern numbers of both the layers can indeed be changed adiabatically such that the ${\rm U(2)}$ Chern number of the composite system is unchanged.\\
 
 Using a similar approach but with a Chern insulator on a square lattice (having a total Chern number $C_{tot}=0$), it has been shown\ct{barbarino_20} that starting with two decoupled layers of trivial Chern layers, one can adiabatically transform the system to two decoupled non-trivial topological layers with non-zero Chern numbers which are equal and opposite (Thus, preserving the $C_{tot}$).
 

\section{Majorana modes in one-dimensional systems and transport}
\label{sec:mmtr}

The end modes of the Kitaev chain, discussed in Sec.~\ref{sec:kitaev_chain}, 
have a profound effect on electronic transport through the system, namely, 
there are peaks in the differential conductance when the voltage 
bias applied to the leads (these are normal metal regions which are connected 
to the two ends of the system) is equal to the energy of end modes.
We will now examine what happens if the wire is not very long;
in that case, the modes at the two ends will hybridize with each other and
thereby shift their energies away from zero energy.
Such a system was studied in Ref.~\onlinecite{thakurathi15} as we now discuss.

Let us denote the second quantized wave function as $\psi= (c,~ d)^T$, where
$c(x,t), ~d(x,t)$ are the electron (particle) and hole components respectively
The Hamiltonian can then be written as
\bea H &=& \int_{-\infty}^\infty dx ~[ c^\dagger (- \frac{\hbar^2 
\partial_x^2}{2m} -\mu ) c ~-~ d^\dagger (- \frac{\hbar^2 \partial_x^2}{2m} 
- \mu ) d \nonumber \\
&& ~~~~~~~~~- \frac{i\Delta (x)}{k_F} ~( c^\dagger \partial_x d + d^\dagger 
\partial_x c ) ], \label{ham1_ds} \eea
where $\mu$ is the chemical potential, $k_F = \sqrt{2 m \mu}/\hbar$ is the
Fermi wave number, and $\Delta (x)$ is the $p$-wave pairing 
amplitude which will be assumed to be real everywhere. The Fermi velocity is 
$v_F = \hbar k_F/m$. (We will henceforth set $\hbar = 1$, except in places 
where it is required for clarity). Solving the Heisenberg equations of motion 
$i \partial_t c = - [H,c]$ and $i \partial_t d = - [H,d]$, we find that
for a wave function which varies as $e^{i(kx - Et)}$, the energy is given by 
$E = \pm (k^2/(2m) - \mu)$ if $\Delta = 0$, and by $\pm \sqrt{(k^2/(2m) - 
\mu)^2 + \Delta^2 (k/k_F)^2}$ if $\Delta \ne 0$. Thus the energy 
spectrum in a SC region with constant $\Delta$ has a gap equal to $2\Delta$ 
at $k= \pm k_F$. We will study the differential conductance for energies 
lying in the gap.

We now consider a system consisting of a SC wire in a region given by
$0 < x < L$ (we take $\Delta$ to be a non-zero constant in this region)
which is connected to semi-infinite leads on the left
and right given by $x < 0$ and $x > L$ respectively. In the leads we set the
pairing $\Delta = 0$. Suppose that an electron is incident from the left 
lead with an energy $E$ lying in the gap $[-\Delta, \Delta]$. The wave
functions in the left lead, SC wire and right lead are then respectively
given by $e^{-iEt}$ times $\psi_1$, $\psi_2$ and $\psi_3$, where
\bea \hspace*{-0.8cm} \psi_1 &=& e^{ik x} \left( \begin{array}{c}
1 \\
0 \end{array} \right)
+ r_n e^{-ik x} \left(\begin{array}{c}
1 \\
0 \end{array} \right)
+ r_a e^{ik x} \left(\begin{array}{c}
0 \\
1 \end{array} \right), \nonumber \\
\hspace*{-0.8cm} \psi_2 &=& t_1e^{ik_1x} \left( \begin{array}{c}
1 \\
e^{i\phi} \end{array} \right)
+ t_2 e^{ik_2x} \left(\begin{array}{c}
1 \\
-e^{-i\phi} \end{array} \right) \nonumber \\
& & +~ t_3 e^{ik_3x} \left(\begin{array}{c}
1 \\
e^{-i\phi} \end{array}\right)
+ t_4 e^{ik_4x} \left(\begin{array}{c}
1 \\
-e^{i\phi} \end{array} \right), \nonumber \\
\hspace*{-0.8cm} \psi_3 &=& t_ne^{ik x} \left( \begin{array}{c}
1 \\
0 \end{array} \right)
+ t_a e^{-ik x} \left(\begin{array}{c}
0 \\
1 \end{array} \label{3eq}\right), \label{waves} \eea
and $e^{i\phi}= (E -i\sqrt{\Delta^2 -E^2})/\Delta$. In $\psi_1$ (left lead),
$r_n$ and $r_a$ denote the normal and Andreev (hole) reflection amplitudes, 
and in $\psi_3$ (right lead), $t_n$ and $t_a$ denote the normal and Andreev 
(hole) transmission amplitudes. A schematic picture of this normal metal - 
superconductor - normal metal (NSN) system along with the four amplitudes 
is shown in Fig.~\ref{fig:nsn1}. A calculation involving matching the wave 
functions and their derivatives at $x=0$ and $L$ shows that these 
amplitudes satisfy the conservation relation $|r_n|^2 + |r_a|^2 + |t_n|^2 + 
|t_a|^2 = 1$. The total probability for an electron charge
to be transmitted from the left lead to the right lead is given by 
$|t_n|^2 - |t_a|^2$ which leads to the differential conductance
\beq G_N ~=~ \frac{e^2}{h} ~(|t_n|^2 ~-~ |t_a|^2), \label{gn} \eeq 
where $h = 2 \pi \hbar$ and the electron charge is $-e$. The total probability 
for electrons to be transmitted into the SC in the form of Cooper pairs is 
$2 (|r_a|^2 + |t_a|^2)$ which leads to the differential conductance
\beq G_C ~=~ \frac{2 e^2}{h} ~(|r_a|^2 ~+~ |t_a|^2). \label{gc} \eeq

\begin{figure}[htb]
\begin{center}
\includegraphics[width=3.0in]{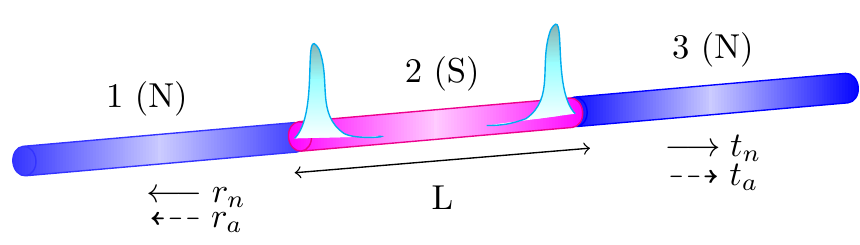}
\caption{Schematic picture of a NSN system. The middle part
(2) with length $L$ is the $p$-wave superconductor, while the left and right
parts (1 and 3) are normal metal leads. Four amplitudes are shown: $r_n, ~r_a$
are normal and Andreev reflections in the left lead, and $t_n, ~t_a$ are normal
and Andreev transmissions in the right lead. Majorana modes at the two ends of
the SC region are shown. (See Ref.~\onlinecite{thakurathi15}).} 
\label{fig:nsn1} \end{center} \end{figure}

A numerical study of the conductances shows the following features 
depending on the ratio of the length of the SC wire $L$ and the length scale 
associated with the SC gap, $\eta = \hbar v_F /\Delta$~\cite{thakurathi15}. 
When $L \ll \eta$, the wire behaves like a superconducting box and it 
contains a state with energy lying in the gap (called a subgap state) only 
when the box quantization condition is satisfied, $k_F L = n \pi$, where $n$ 
is an integer. As $k_F L$ is varied across one of these quantized values, the 
energy $E$ of the subgap states changes rapidly between $+\Delta$ and $-\Delta$
(see Fig.~\ref{fig:en} (a)). We then find that $G_N$ has peaks at these 
energies and is almost zero otherwise, while $G_C$ is almost zero for all 
energies. When $L \sim \eta$, the modes at the two ends of the wire hybridize 
strongly with each other, thereby shifting their energies away from zero; the 
energies of these modes oscillate with $L$ as $\pm \sin (k_F L)$ 
(Fig.~\ref{fig:en}
(b)). Both $G_N$ and $G_C$ then have peaks at these energies and are almost 
zero otherwise. When $L \gg \eta$, the end modes do not hybridize and they
have zero energy (Fig.~\ref{fig:en} (c)). $G_C$ then shows a
peak at zero bias while $G_N$ is almost zero for all energies.

Figure~\ref{fig:en} shows the energy (in units of $\Delta$) of the subgap 
states as a function of the
length $L$ of a SC wire with $k_F=1, ~m =1$ and $\Delta=0.01$ (so that $v_F = 
1$ and $\eta = 100$), for three ranges of $L$, namely, (a) $L \ll \eta$ (note
that the energies changes rapidly nearly $k_F L = 2 \pi$ and $3 \pi$), (b) 
$L \sim \eta$ (the energies cross zero near $k_F L = 37 \pi$ and $38 \pi$), 
and (c) $L \gg \eta$. The results are in agreement with the discussion 
presented above.

\begin{widetext}
\begin{center} \begin{figure}[htb]
\subfigure[]{\includegraphics[width=2.29in]{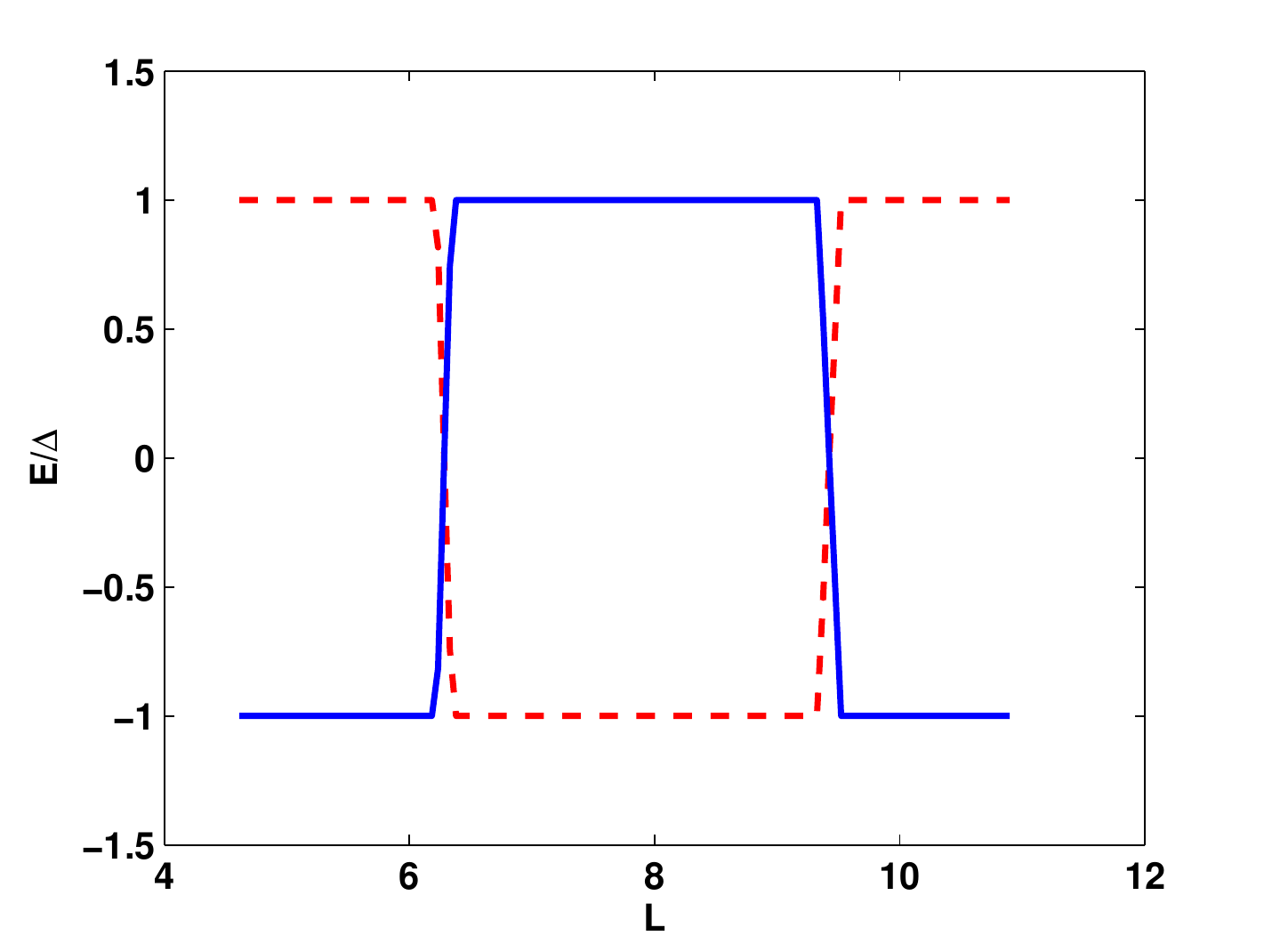}}
\subfigure[]{\includegraphics[width=2.29in]{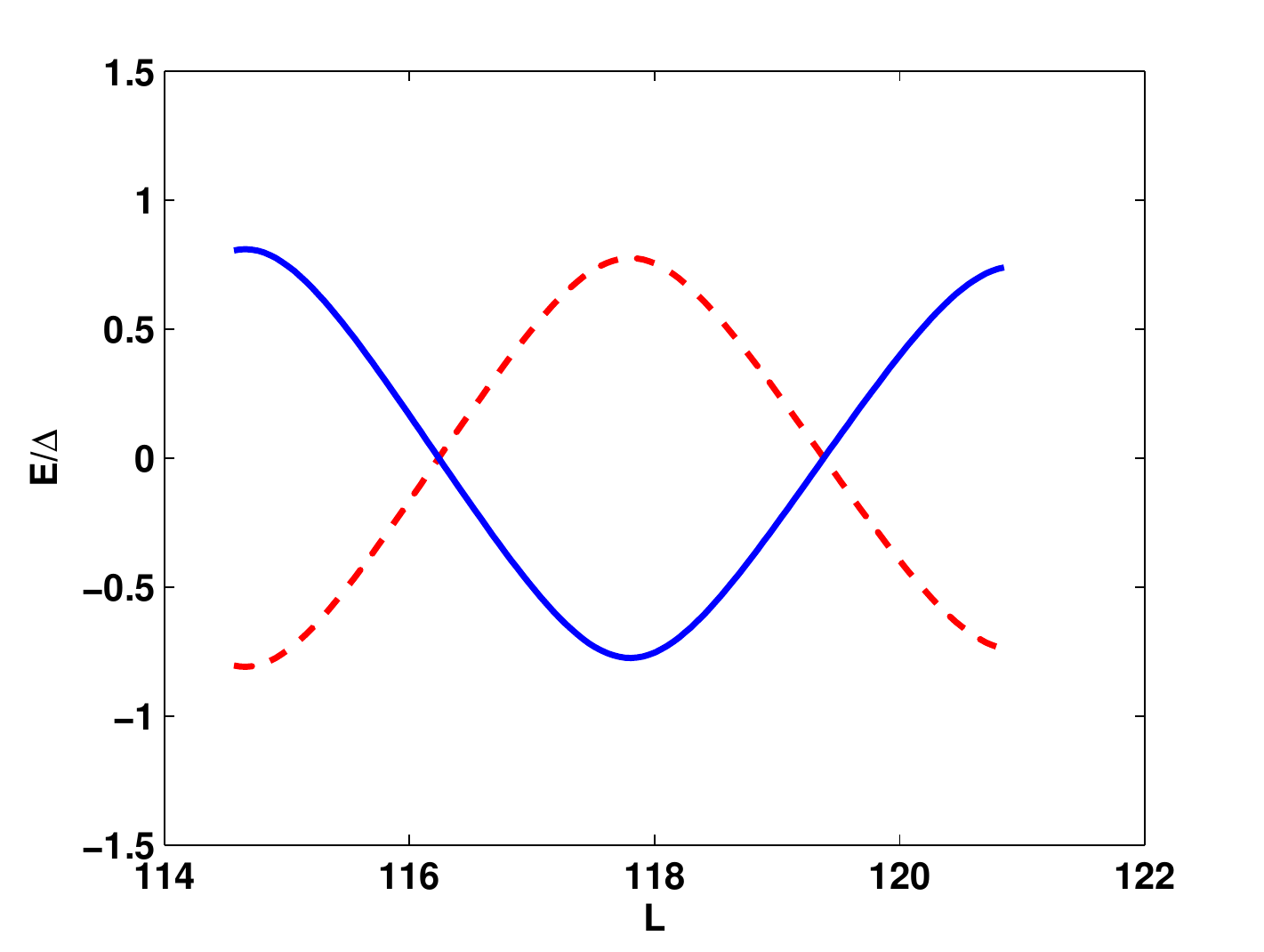}}
\subfigure[]{\includegraphics[width=2.29in]{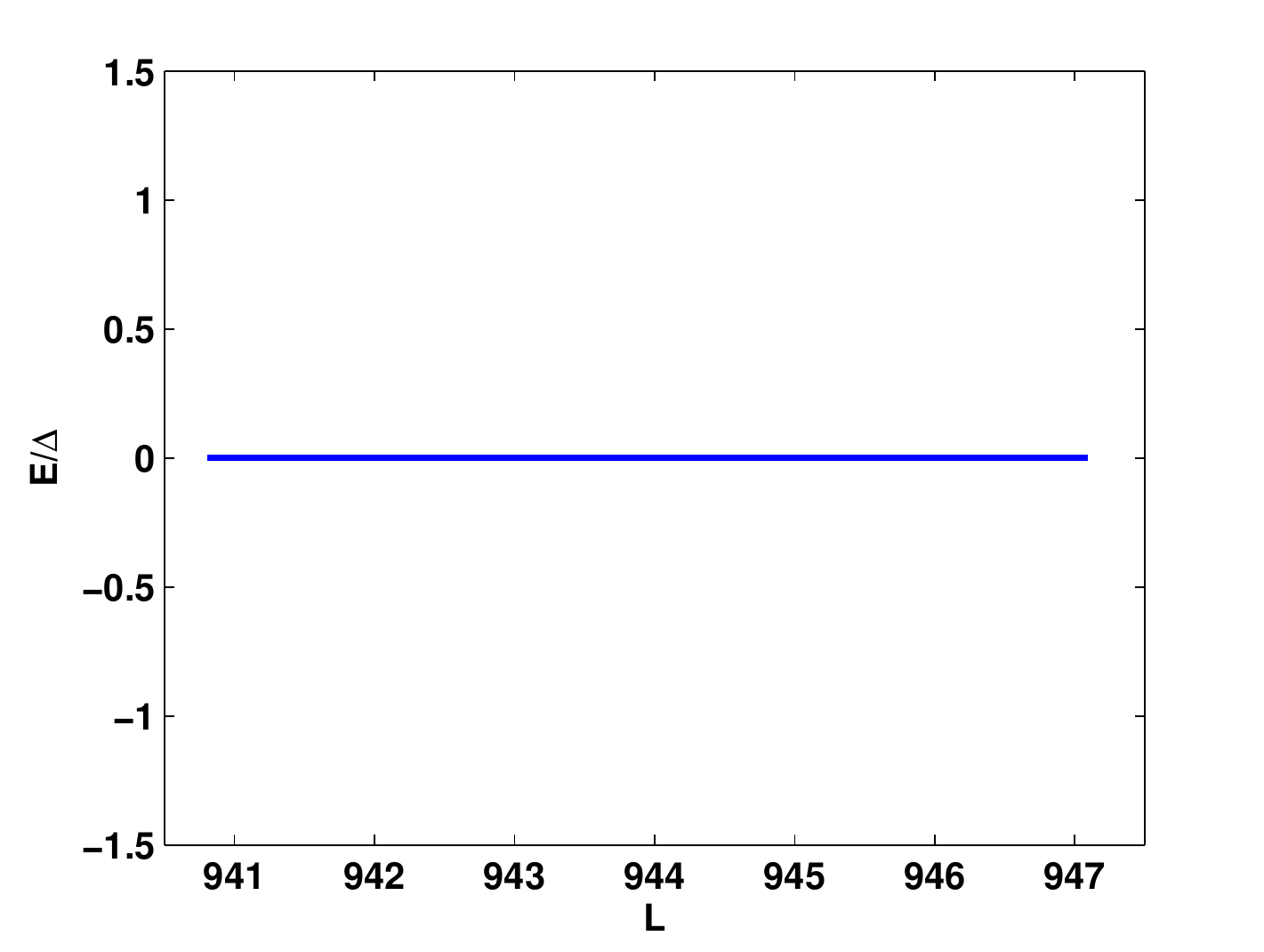}}
\caption{Energies $E/\Delta$ of the Majorana modes versus $L$ for 
(a) $L \ll \eta$, (b) $L \sim \eta$, and (c) $L \gg \eta$, where $\eta = 
100$. (See Ref.~\onlinecite{thakurathi15}).} \label{fig:en} \end{figure} 
\end{center} \end{widetext}

The system discussed above had zero energy Majorana modes only at the ends.
It would be interesting to look at systems which can have such modes
in the bulk and to study their effects on the peaks in the differential
conductance. To this end,
let us consider a junction of three $p$-wave SC wires each of which is 
connected to normal metal leads~\cite{deb16}. The wires are labeled as 
1, 2 and 3. For simplicity we will take the pairings $\Delta_i$ 
in the different wires to be real and equal in magnitude, but we will allow
their signs to be different from each other. We will see that this generally 
gives rise to additional Majorana modes at the junction which can significantly affect the differential conductances. 
The Hamiltonian for this system again has the form given in Eq.~\eqref{ham1_ds} 
in each of the three wires which we label as $j$ where $j=1,2,3$. However,
since each wire is semi-infinite, the limits of the integral over $x_j$ will 
be taken to go from 0 at the junction to $L_j$ at the point where the SC part 
ends and the lead part begins, and then to $\infty$. In the SC part ($0 < x_j 
< L_j$), $\Delta (x_j)$ will be taken to be a non-zero constant denoted as 
$\Delta_j$, while in the lead ($L_j < x < \infty$) we set $\Delta (x_j) = 0$.
To complete the description of the system, we also have to specify the
boundary conditions on the operators $c_j (x_j), ~d_j (x_j)$ and their
derivatives with respect to $x_j$ when we approach the three-wire junction, 
i.e., as $x_j \to 0$. The boundary condition can be described by a $3 \times 3$
matrix $M$ which linearly relates the derivatives of $c_j, ~d_j$ to 
$c_j, ~d_j$. For convenience, we will assume the most symmetric possible 
matrix $M$ all of whose elements are equal to each other~\cite{deb16}.

Consider now what happens when an electron is incident from lead 1 with an 
energy which lies in the SC gap. This can give rise to normal and Andreev 
reflections with amplitudes denoted as $r_{n1}$ and $r_{a1}$ and normal
and Andreev transmissions to wires 2 and 3 which are denoted as $t_{n2}$ 
and $t_{a2}$ and $t_{n3}$ and $t_{a3}$ respectively (see Fig.~\ref{fig:3sc1}).
These satisfy the conservation relation $|r_{n1}|^2 + |r_{a1}|^2 + |t_{n2}|^2 
+ |t_{a2}|^2 + |t_{n3}|^2 + |t_{a3}|^2 = 1$.
We find that there are three possible differential conductances
corresponding to transmission to leads 2 and 3 and transmission into the SC
wires in the form of Cooper pairs. These are respectively given by
\bea G_{N2} &=& \frac{e^2}{h} ~(|t_{n2}|^2 ~-~ |t_{a2}|^2), \nonumber \\
G_{N3} &=& \frac{e^2}{h} ~(|t_{n3}|^2 ~-~ |t_{a3}|^2), \nonumber \\
G_C &=& \frac{2e^2}{h} ~(|r_{a1}|^2 ~+~ |t_{a2}|^2 ~+~ |t_{a3}|^2). 
\label{g3} \eea

\begin{figure}[htb]
\begin{center} 
\includegraphics[width=3in]{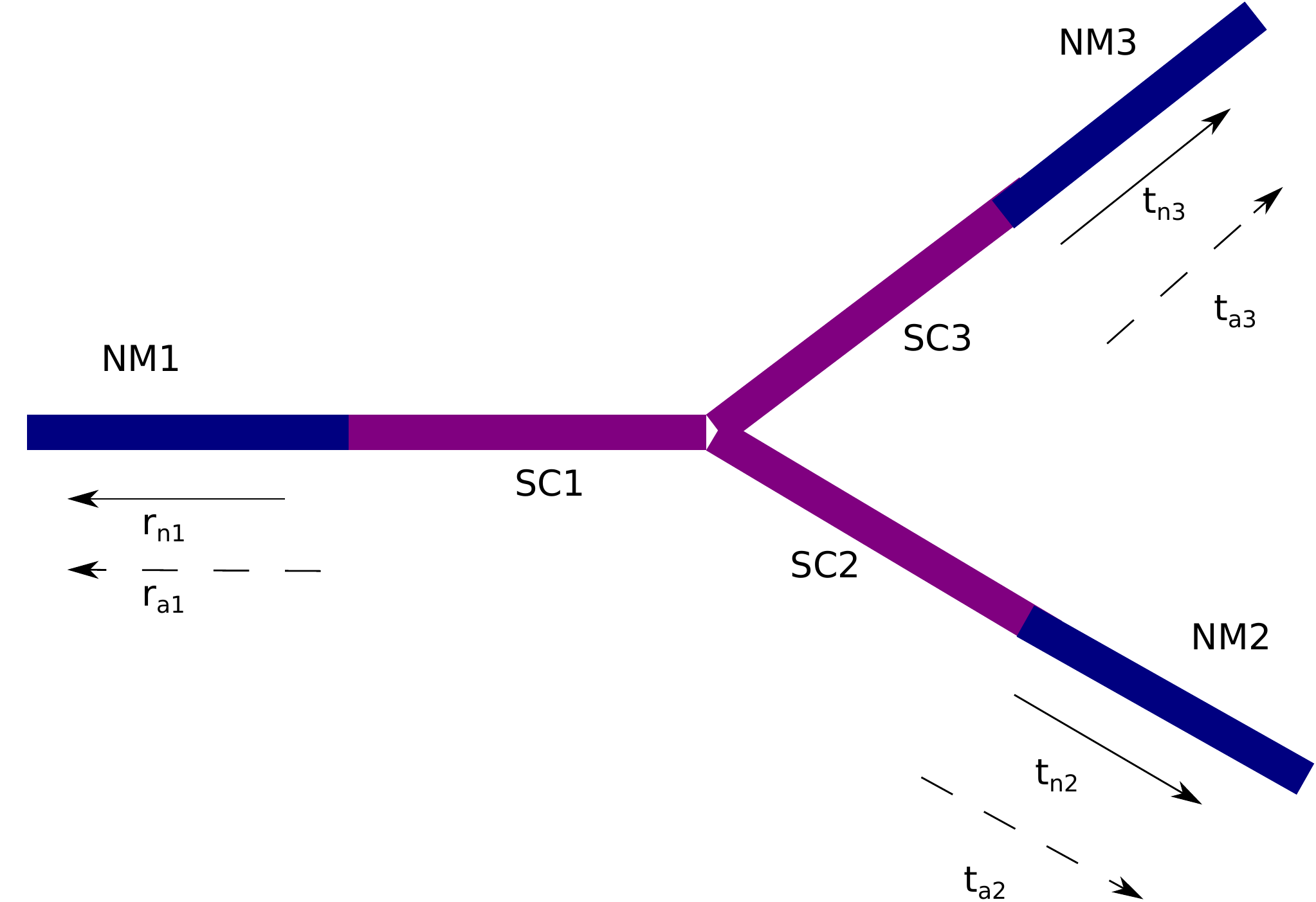} 
\caption{Schematic picture of a $Y$-junction of three wires labeled as 1, 2, 3.
The inner (lighter) regions of the wires are $p$-wave SCs while the outer 
(darker) regions are normal metal (NM) leads. For an electron incident from
NM1, six amplitudes are shown in the leads: $r_{n1}, ~r_{a1}$ are normal and
Andreev reflections in NM1, $t_{n2}, ~t_{a2}$ are normal and Andreev
transmissions in NM2, and $t_{n3}, ~t_{a3}$ are normal and Andreev
transmissions in NM3. (See Ref.~\onlinecite{deb16}).} 
\label{fig:3sc1} \end{center} \end{figure}

We now examine two cases: (a) $\Delta_1 = \Delta_2 = \Delta_3$ and 
(b) $- \Delta_1 = \Delta_2 = \Delta_3$. In both cases, we find one zero
energy Majorana mode at each of the three SC - lead junctions, i.e., at $x_j = 
L_j$, assuming that all the $L_j$'s are much larger than $\eta$. However, 
there is a remarkable difference at the junction $x_j = 0$ in the two cases. 
When all the $\Delta_j$'s have the same sign, we find {\it three} Majorana 
modes at that junction; they do not hybridize with each other and therefore 
all stay at zero energy due to a symmetry~\cite{deb16}. The system therefore 
has a total of six zero energy Majorana modes. However, when one of the 
$\Delta_j$'s has the opposite sign to the other two, the symmetry is absent 
and only one of the three modes at the three-wire junction stays at zero 
energy. As a result the system has only four zero energy Majorana modes.
Since the peaks in the differential conductances lie at the energies of the 
Majorana modes, the pattern of peaks is therefore looks different 
in cases (a) and (b). Numerically we find that when all the $L_j$'s are less 
than or of the order of $\eta$, conductance peaks are found at six different 
subgap energies in case (a) and at four different energies in case (b);
these energies are non-zero since $L_j \sim \eta$ allows hybridization between
all the Majorana modes. When all the $L_j$'s are much larger than $\eta$, 
however, there is no hybridization and we only find conductance peaks at 
zero energy.

Next, we discuss how cases (a) and (b) can be experimentally realized. We 
begin with a microscopic model consisting of a wire with a Rashba spin-orbit 
coupling of the form $\alpha_R p_r \sigma^y$, where $p_r$ denotes the momentum 
of the electrons in the wire (assumed to be along either the $+ \hat x$ or 
$- \hat x$
direction) and $\sigma^y$ is one of the Pauli spin matrices~\cite{ganga11}.
We then place the wire in a magnetic field in the $\hat z$ direction (with a
Zeeman coupling $\Delta_Z$ to the spin of the electrons) and in proximity to a
bulk $s$-wave SC with pairing $\Delta_S$. Ref.~\onlinecite{ganga11} then shows 
that this system is equivalent, for some range of parameters, to a $p$-wave SC 
of the form shown in Eq.~\eqref{ham1_ds} where the $p$-wave pairing is given by 
$\Delta = \pm \alpha_R k_F \Delta_S /\Delta_Z$ where the $\pm$ sign depends 
on whether the electron momentum is along $+ \hat x$ or $- \hat x$. Now 
suppose that the three wires are arranged as in Fig.~\eqref{fig:exp} (a) so 
that electrons in all of them move away from the three-wire junction along the 
$+ \hat x$ direction. Then all the three $\Delta_j$'s will have the same sign 
as in case (a). However, if the wires are arranged as in Fig.~\eqref{fig:exp} 
(b) so that electrons in one of the wires move away from the junction in a
direction opposite to electrons in the other two wires, then $\Delta_j$
in that wire will have the opposite sign to the $\Delta_j$'s in the other two
wires, thus realizing case (b). We note that the vertical region shown in 
Figs.~\ref{fig:exp} (a) and (b) denotes the three-wire junction; it is clear 
that this region is likely to scatter the electrons between the different 
wires. We have modeled this scattering using the symmetric matrix $M$ 
mentioned above. Further, the conductance
$G_C$ (which was described earlier in terms of Cooper pairs going into the 
SC wires) can now be understood at a microscopic level in terms of processes 
in which Cooper pairs go between the bulk $s$-wave SC and the three SC wires.

\begin{figure}[htb]
\subfigure[]{\includegraphics[width=0.99in]{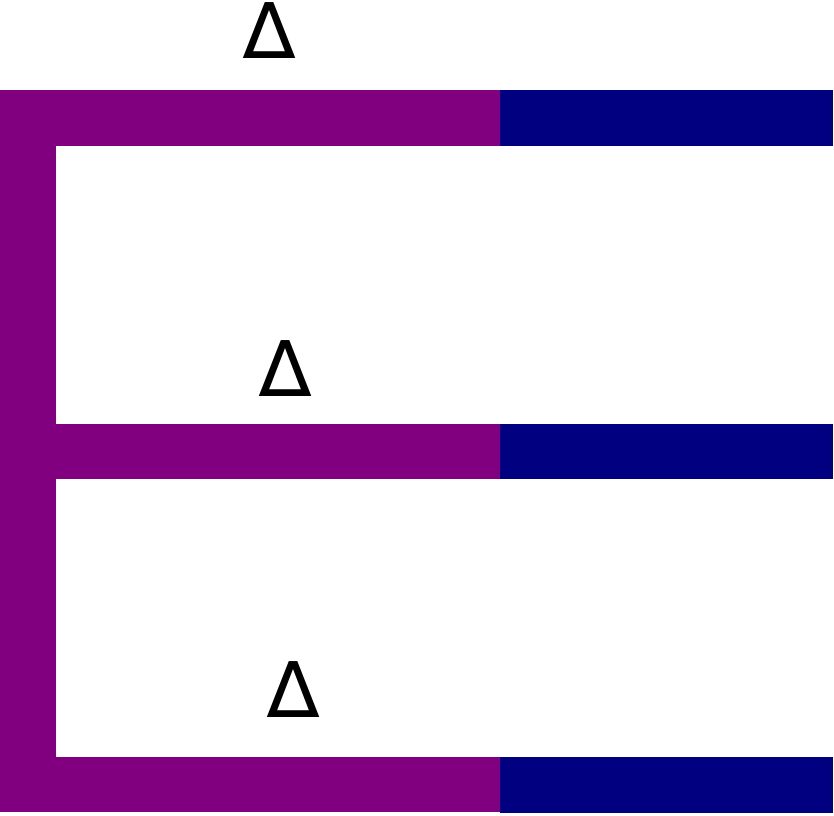}} 
\hspace*{.35cm} 
\subfigure[]{\includegraphics[width=2.04in]{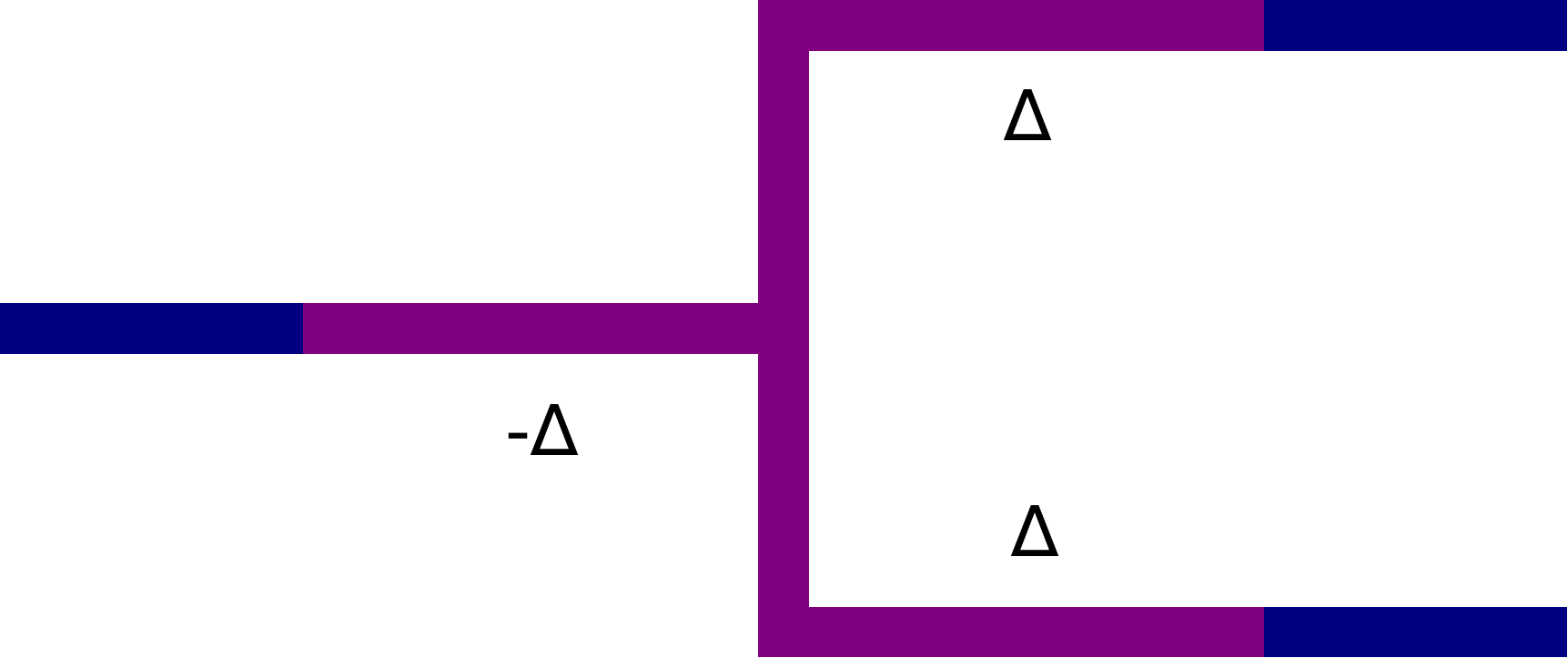}}
\caption[]{Three SC wires with (a) the sign of $\Delta_j$ being the same in 
all the wires, (b) the sign of $\Delta_j$ being different in the wire on the 
left compared to the other two wires. The SC wires (NM leads) are shown in a 
lighter (darker) shade respectively. (See Ref.~\onlinecite{deb16}).} 
\label{fig:exp} \end{figure}

\section{Transport on surfaces of topological insulators}
\label{sec:trti}

Topological insulators in three dimensions are characterized by bulk 
states which are gapped and gapless surface states~\cite{hasan10,qi11}. At low
temperatures which are much smaller than the bulk gap (about $0.56$ eV
in Bi$_2$Se$_3$), electronic transport in such systems is dominated by the
surface states. The surface states are governed by Hamiltonians which have
the form of massless Dirac equations (i.e., derivatives along the two 
coordinates which define the surface multiplied by some traceless and
anticommuting matrices); the precise form depends on the orientation of the 
surface~\cite{zhang12,deb14}. In particular, the Hamiltonians for electrons
on the top and bottom surfaces (with coordinates $x, ~y$) are given by
\bea H_{top} &=& -~i \hbar v ~(\sigma^x \frac{\partial}{\partial y} ~-~ 
\sigma^y \frac{\partial}{\partial x}), \nonumber \\
H_{bottom} &=& ~i \hbar v ~(\sigma^x \frac{\partial}{\partial y} ~-~
\sigma^y \frac{\partial}{\partial x}), \label{hamtb} \eea
where $\sigma^{x,y}$ are Pauli spin matrices, and $v$ is the velocity (for 
Bi$_2$Se$_3$, $\hbar v = 0.333$ eV nm). Note that $H_{top}$ and $H_{bottom}$ 
have opposite signs. 

If the topological insulator is sufficiently thick (namely, its thickness is much
larger than the decay length of the surface states), the tunneling between 
the top and bottom surfaces can be ignored and transport on the two 
surfaces can be studied independently of each other. An interesting problem 
arises in the case of thin topological insulators where the tunneling between 
the two surfaces cannot be neglected. Such a situation has been studied in 
Ref.~\onlinecite{udupa18}. In this work, the effect of potential and magnetic 
barriers placed on only the top surface was investigated; a schematic picture 
of the system is shown in Fig.~\ref{fig:bar1}.

\begin{figure}[htb]
\centering
\hspace*{-.4cm} \includegraphics[scale=0.57]{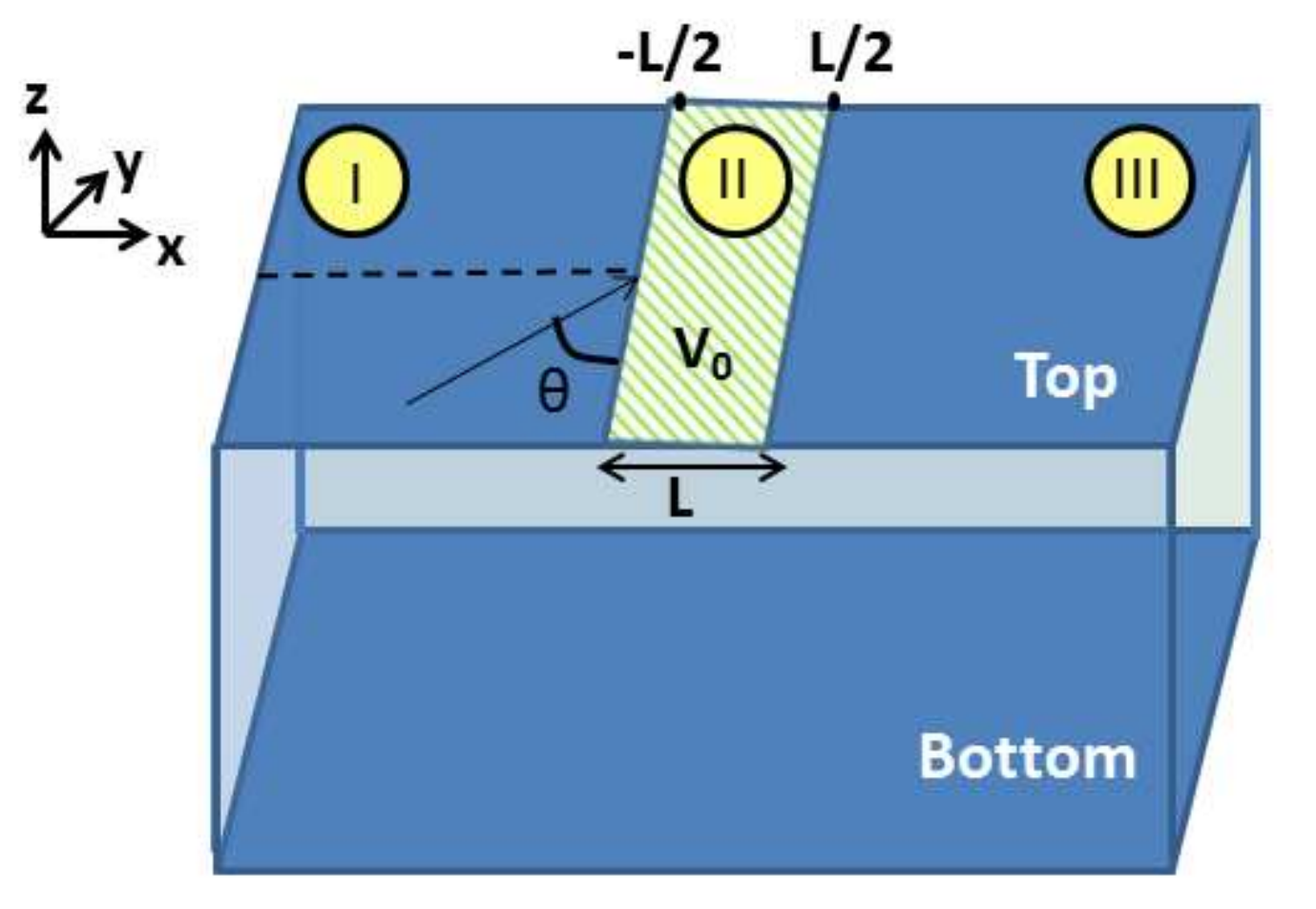}
\caption{Schematic picture showing the top and bottom surfaces of a 
topological insulator, a potential barrier with strength $V_0$ and width
$L$ on the top surface (region $II$), and a wave coming in from region $I$
with an angle of incidence $\theta$. (See Ref.~\onlinecite{udupa18}).} 
\label{fig:bar1} \end{figure}

Since there are two surfaces, the Hamiltonian of the system is a $4 \times 4$
matrix given by
\beq H_{0} = \begin{pmatrix}
H_{top} ~+~ V (x) I_2 & \lambda I_2 \\
\lambda I_2 & H_{bottom} \end{pmatrix}, \label{ham2_ds} \eeq
where $H_{top}$ and $H_{bottom}$ are the Hamiltonians for the top and bottom 
surfaces given in Eq.~\eqref{hamtb}, $\lambda$ is the tunneling amplitude 
between the two surfaces, $V(x)$ denotes a barrier placed on the top surface, 
and $I_2$ denotes a two-dimensional identity matrix. We will consider 
finite-width barriers which can be a potential barrier with $V(x) = V_0$ for 
$-L/2 < x < L/2$ or a magnetic barrier with $V(x) = V_0 \sigma^x$ for 
$-L/2 < x < L/2$. A magnetic barrier can be experimentally realized by placing 
a ferromagnetic strip on the top surface; the magnetization in the strip will 
have a Zeeman coupling to the electron spin, and we are assuming that the
magnetization points in the $\hat x$ direction.

Far away from the barrier (i.e., for $|x| \gg L/2$), the energy of an electron
with momentum ${\vec k} = (k_x,k_y)$ is given by $E = \pm \sqrt{\hbar^2 v^2 
{\vec k}^2 + \lambda^2}$, each with a double degeneracy (due to the presence 
of two surfaces). The gap between the positive and negative energy bands is 
$2 |\lambda|$.

The differential conductance of this system is calculated as follows. Suppose
that an electron comes from the left in Fig.~\ref{fig:bar1}) with a momentum 
${\vec k} = (k_x,k_y)$ (where $k_x > 0$), a positive energy $E$, and an angle 
of incidence $\theta = \tan^{-1} (k_y/k_x)$. (We are following the convention 
shown in the figure that $\theta$ lies in the range $[0,\pi]$, with $\theta = 
\pi/2$ denoting normal incidence). Given ${\vec k}$ and $E, k_x > 0$, there 
are two possible incident waves due to the double degeneracy; we call these 
$\psi_1 (k_x,k_y)$ and $\psi_2 (k_x,k_y)$. An electron incident with wave 
function $\psi_m (k_x,k_y)$ ($m$ can be 1 or 2) can be reflected back to the 
left from the barrier with the wave function $\psi_n (-k_x,k_y)$ and amplitude 
$r_{nm}$ or transmitted to the right with the wave function $\psi_n (k_x,k_y)$ 
and amplitude $t_{nm}$ ($n$ can be 1 or 2 in either case). Current conservation
implies that $|r_{1m}|^2 + |r_{2m}|^2 + |t_{1m}|^2 + |t_{2m}|^2 = 1$ for 
$m =1, ~2$). The wave function in the barrier region ($-L/2 < x < L/2$) will
have the same values of $E, ~k_y$ as the incident wave but can have momenta 
$\pm k'_x$ (each with two possible wave functions) where $k'_x$ is fixed by 
the conservation of $E$ and $k_y$ and the barrier strength $V_0$. We can then 
find the reflection and transmission amplitudes by matching the wave functions 
at $x = \pm L/2$. The differential conductance $G$ is calculated by adding up
the total transmission probability for the two possible incident wave 
functions and integrating over the angle $\theta$. More precisely, we have
\beq G ~=~ \frac{e^2 W \sqrt{E^2 - \lambda^2}}{v(2\pi \hbar)^2} ~\int_0^{\pi} 
d\theta ~\sin \theta \sum_{m,n=1,2} ~|t_{nm}|^2, \label{condG1} \eeq
where $W$ is the width of the system in the $\hat y$ direction, and $E$ is the 
chemical potential in the leads. In addition to the charge
current which enters the expression for $G$, we can also calculate the 
transmitted spin current.

The main results obtained for this system are as follows~\cite{udupa18}.
For a potential barrier, the conductance $G$ is an oscillatory function
of the barrier strength $V_0$. This is a consequence of the massless Dirac
nature of the electrons; the electrons can propagate through the barrier 
(as in Klein tunneling) but there are interference effects in that 
region which depend on the ratio $V_0 L/(\hbar v)$. The conductance at 
the peaks of these oscillations reaches almost unity for specific values 
of the barrier potential $V_0$ thus demonstrating near-perfect
transmission resonances. The transmitted spin current
(with spin component along the $\hat y$ direction) is observed to be
always negative (positive) at the top (bottom) surface, but their
sum is always positive. This is due to the opposite forms of
spin-momentum locking on the two surfaces as can be seen from 
Eq.~\eqref{hamtb}; an electron with positive energy and moving in the
$+ \hat x$ direction on the top (bottom) surface has a spin pointing
in the $- \hat y$ ($+ \hat y$) direction, respectively. (Hence this
system can be used to split an incoming current into two separate
spin currents with opposite polarizations. These spin currents can be 
detected by attaching spin-polarized metallic leads to the two surfaces).
For a magnetic barrier in which the magnetization points along the $\hat x$
direction, the conductance $G$ does not oscillate but decreases
and reaches a constant value as the barrier strength $V_{0}$ increases, in
contrast to the case of a potential barrier. Interestingly, even when
$V_0$ is very large, there is always a non-zero current due to the presence of
the bottom surface; an electron incident from the left along the top surface 
can avoid the magnetic barrier by tunneling to the bottom surface and then
transmitting to the right. The transmitted spin currents again have opposite 
signs on the top and bottom surfaces due to the spin-momentum locking as 
discussed above.

While most experimental studies of topological insulators have looked at 
crystalline systems, there has been a recent study of a granular 
topological insulator~\cite{banerjee17}. These are 
thin films consisting of an assembly of tunnel coupled nanocrystals of 
Bi$_2$Se$_3$. Magnetoconductance measurements show that there is a range of 
temperature and film thickness in which the system has decoupled states at 
the top and bottom surfaces which have unusually large penetration depths 
which can range from 10 nm to 30 nm; remarkably, this is much larger than the 
vertical size of 2-3 nm for a single grain. The penetration depth depends on 
the ratio of inter-grain to intra-grain tunnelings which can be tuned by 
varying the temperature. We would also like to mention a theoretical 
study where it was shown that an amorphous system with appropriate symmetries 
of the Hamiltonian can behave like a topological insulator with robust surface 
states which exhibit quantized charge conductances~\cite{agarwala17}.

We have already seen in Sec.~\ref{sec_periodic} that periodic driving can have significant effects on topological systems. Further, periodic driving can also have interesting effects on electronic transport in such systems~\cite{mondal19,udupa20}. One way to 
produce such driving is to apply electromagnetic radiation of a particular 
frequency on the system. This can produce peaks in the differential 
conductance, and this can be used to detect the radiation and measure its 
frequency. This idea has been proposed for detecting radiation with 
frequencies in the range of 1 to 10 terahertz using a double barrier structure 
placed on the surface of a three-dimensional
topological insulator~\cite{mondal19}. As the barrier strength and the 
driving parameters (the amplitude and frequency $\omega$ of the radiation) 
are varied, the conductance shows pronounced peaks at both the incident energy 
$E$ of the electron (which is determined by the voltage bias applied to the 
device) and at the subband energies given by $E + n \hbar \omega$, where $n$ 
takes integer values.

\begin{figure}[htb]
\centering
\includegraphics[scale=0.36]{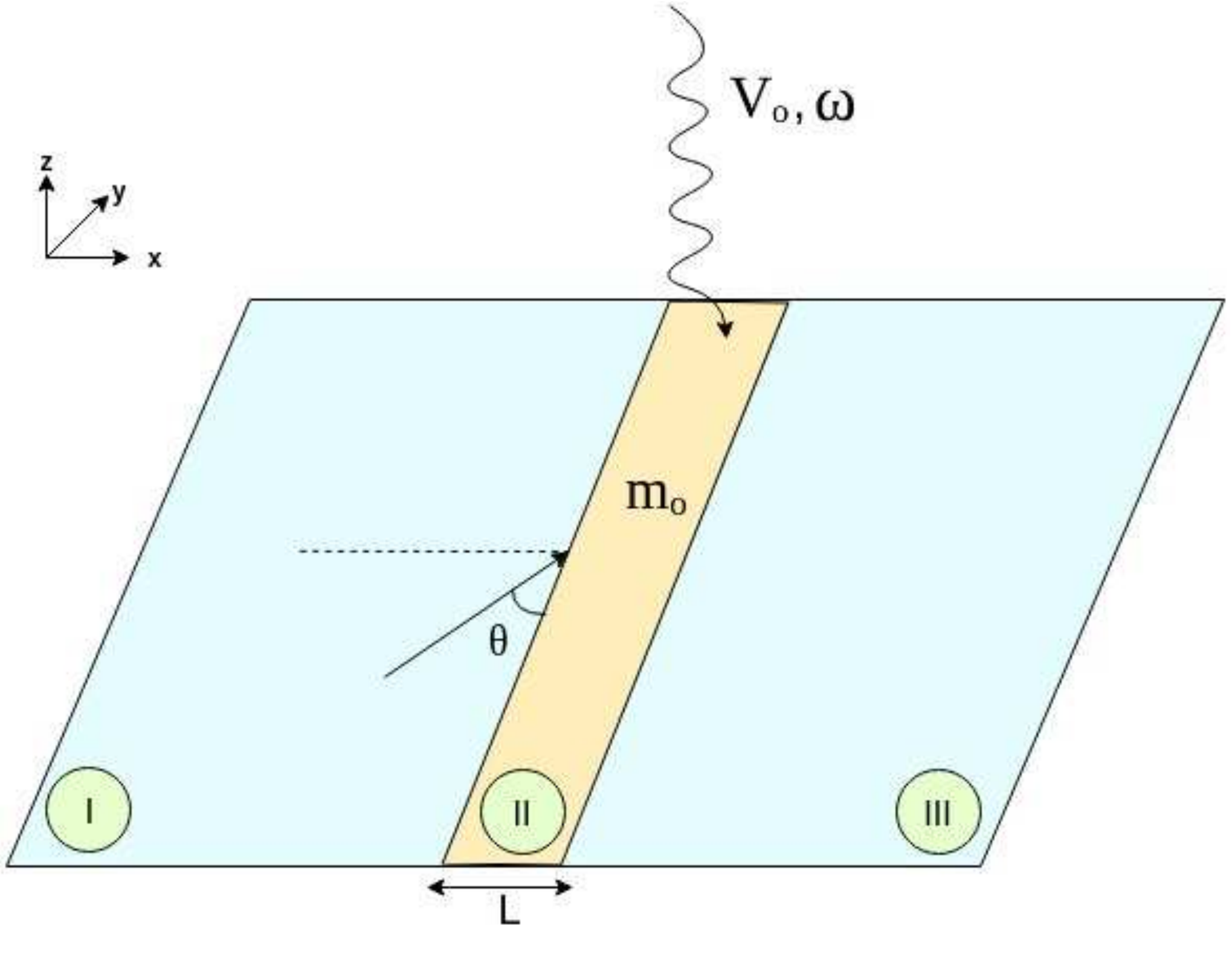}
\caption{Schematic picture of the system showing the top surface of a 
topological insulator, with 
a static magnetic barrier of strength $m_0$ and electromagnetic radiation with 
frequency $\omega$ and amplitude $V_0$ which is incident on the barrier. The 
barrier has a width $L$ (region $II$ where $0 < x < L$), and an electron 
comes in from region $I$ (where $x < 0$) with an angle of incidence $\theta$. 
(See Ref.~\onlinecite{udupa20}).} \label{fig:bar2} \end{figure}

We now discuss a specific model for studying ballistic transport on the 
surface of a topological insulator in the presence 
of a time-dependent barrier. The barrier is 
formed by applying a magnetic field (leading to a static Zeeman coupling
to the electron spin) as well as linearly polarized radiation on a strip-like 
region on the top surface of a topological insulator 
as shown in Fig.~\ref{fig:bar2}. Assuming 
that the magnetic field points in the $\hat x$ direction while the electric 
field and therefore the vector potential of the radiation points in the 
$\hat y$ direction, the Hamiltonian governing a surface electron, with 
momentum $\hbar k_y$ in the $\hat y$ direction (this momentum is a conserved 
quantity since the barrier is independent of the $y$ coordinate), has the form
\beq H_{II} ~=~ \hbar v \bigl[k_y \sigma^x ~+~ i \sigma^y \frac{\partial}{
\partial x} ~+~ [m_0 ~+~ V_0 \cos (\omega t)] \sigma^x \bigr] 
\label{hamII} \eeq
in the barrier (region II in Fig.~\ref{fig:bar2}), and $\hbar v [ k_y 
\sigma^x + i \sigma^y \partial /\partial x]$ outside the barrier (regions I 
and III). 

Now consider an electron approaching the barrier from the left (region I) 
with a positive energy $E_0$ and a momentum $\hbar (k_{x,0},k_y)$ (so that 
$E_0 = \hbar v \sqrt{k_{x,0}^2 + k_y^2}$). It can be reflected back with an 
energy $E_n$ and momentum $\hbar (-k_{x,n}, k_y)$, where $E_n = E_0 + n \hbar
\omega$ (this is called the $n$-th side band) and $k_{x,n} = \sqrt{(E_n/\hbar 
v)^2 - k_y^2}$ is either real and positive or imaginary with a positive
imaginary part so that the corresponding wave function $e^{-i k_{x,n} x}$ goes
to zero as $x \to - \infty$. The wave function in region I is therefore
\bea \psi_I &=& e^{i (k_{x,0} x + k_y y - E_0 t /\hbar} \nonumber \\
&& +~ \sum_{n=-\infty}^\infty r_n \psi_n (-k_{x,n}, k_y, E_n) e^{i (-k_{x,n} 
x + k_y y - E_n t /\hbar)}, \nonumber \\
\label{psiI} \eea
where $\psi_n (-k_{x,n}, k_y, E_n)$ is a normalized two-component eigenstate 
of the Hamiltonian in region I and $r_n$ is the reflection amplitude. 
Similarly, the wave function in region III (corresponding to a transmitted 
electron) is given by
\beq \psi_{III} ~=~ \sum_{n=-\infty}^\infty t_n \psi_n (k_{x,n}, k_y, E_n) 
e^{i (k_{x,n} x + k_y y - E_n t /\hbar)}, \label{psiIII} \eeq
where $t_n$ is the transmission amplitude. In region II (the barrier), a
particular wave function is given by
\beq \psi_{II,j} ~=~ \sum_n ~\psi'_{j,n} (k'_{x,j}, k_y, 
E_n) e^{i (k'_{x,j} x + k_y y - E_n t /\hbar)}, \label{psiIIj} \eeq
where the meaning of the label $j$, the momenta $k'_{x,j}$ and the normalized
wave functions $\psi'_{j,n} (k'_{x,j}, k_y, 
E_n)$ are as follows. To do numerical calculations, we must truncate 
the total number of bands to some large but finite number $N+1$; we choose
$N$ to be an even integer so that the subband label $n$ goes symmetrically from
$-N/2$ to $N/2$. Then the Schr\"odinger equation $i \hbar \partial \psi_{II}
/\partial t = H_{II} \psi_{II}$ gives the coupled set of equations
\bea E_n \psi'_{j,n} &=& \hbar v ~[ (k_y + m_0) \sigma^x \psi'_{j,n} ~-~ 
k'_{x,j} \sigma^y \psi'_{j,n} \nonumber \\
&& ~~~~~+~ \frac{V_0}{2} \sigma^x ~(\psi'_{j,n+1} ~+~ \psi'_{j,n-1}) ],
\label{eqpsiII} \eea
where the label $n$ in $\psi'_{j,n}$ only goes from $-N/2$ to $N/2$.
Since $n$ takes $N+1$ values and $\psi'_{j,n}$ are two-component wave 
functions, we can take the term involving $k'_{x,j}$ in Eq.~\eqref{eqpsiII} to 
one side and then multiply both sides by $\sigma^y$ to obtain an eigenvalue 
equation for a $2(N+1)$-dimensional matrix, where $k'_{x,j}$ is the eigenvalue
and the eigenvector is $\psi'_{j,n}$ written as a single column with 
all values of $n$, thus giving a total of $2(N+1)$ entries. We then obtain 
$2(N+1)$ possible values of $k'_{x,j}$ which are labeled by $j$. The total 
wave function in region II will be a superposition of all these eigenvectors 
with some amplitudes $C_j$,
\beq \psi_{II} ~=~ \sum_{j=1}^{2(N+1)} C_j ~\psi_{II,j}, \label{psiII} \eeq
where $\psi_{II,j}$ is given by Eq.~\eqref{psiIIj}. 
We now have a total of $4(N+1)$ amplitudes: $N+1$ of both $r_n$ and $t_n$ and 
$2(N+1)$ of the $C_j$'s. These are determined by matching the wave functions 
$\psi_I$ and $\psi_{II}$ at $x=0$ and $\psi_{II}$ and $\psi_{III}$ at $x=L$; 
each of these gives $2(N+1)$ conditions. Current conservation implies that
\beq \frac{k_{x,0}}{E_0} ~=~ \sum_n ~(|t_n|^2 ~+~ |r_n|^2) ~
\frac{k_{x,n}}{E_n}, \label{curcon} \eeq
where the sum only runs over values of $n$ for which $k_{x,n}$ is real and 
positive; bands for which $k_{x,n}$ is imaginary do not contribute to the 
current since those wave functions are localized. 

The transmitted particle current in the $\hat x$ direction in region III is 
given by
\beq J_x ~=~ \sum_n ~|t_n|^2 \left( \frac{\hbar v^2 k_{x,n}}{E_n} \right),
\label{jx} \eeq
where the sum again runs only over values of $n$ for which $k_{x,n}$ is 
real. The differential conductance is obtained
by integrating over the angle of incidence $\theta = \tan^{-1} (k_{x,0}/k_y)$,
\beq G ~=~ \frac{e^2 W E_0}{(2 \pi \hbar v)^2} ~\int_0^\pi d \theta ~J_x 
(\theta), \label{condG2} \eeq
where $W$ is the width of the system, and the chemical potential in the leads 
is $E_0$. The maximum value of $G$ is obtained when there is no barrier
(i.e., $m_0 = V_0 = 0$) giving $J_x = v \sin \theta$. Then the conductance
is given by $G_0 = e^2 W E_0 /[2 v (\pi \hbar)^2]$.

Particularly interesting results are found in this system when $\omega$ and
$V_0$ are large compared to $E_0$ and $m_0$~\cite{udupa20}. Setting $m_0 = 0$ 
for simplicity, we find that the behavior of the conductance $G$ is quite 
different depending on whether $\omega /(v V_0)$ is larger than or smaller than 
$0.83$ (this is equal to $2/z_0$ where $z_0 \simeq 2.405$ is the first zero 
of the Bessel function $J_0 (z)$). When $\omega /(v V_0) \gtrsim 0.83$, the 
conductance is large along certain curves 
in the $(V_0,\omega)$ plane; these curves correspond to resonances, and their 
spacing is equal to $\pi v /L$. When $\omega /(v V_0) \lesssim 0.83$, the
conductance is generally small; however it is particularly small along 
certain lines whose slopes are related to the successive zeros of $J_0 (z)$.
Figs.~\ref{fig:surf} (a) and (b) shows these results as a surface plot of 
$G/G_0$ as a function of $V_0$ and $\omega$. In that figure,
$E_0 = 2$ has been taken in units of $0.01$ eV (so that it is 
much smaller than the bulk gap of $0.56$ eV in Bi$_2$Se$_3$, ensuring that 
there is no contribution to the current from the bulk states), the barrier 
width $L$ is in units of $\hbar v / (0.02$ eV) $\simeq 17$ nm, $\omega$ is in 
units of $0.01$ eV/$\hbar \simeq 15.2$ THz, and $V_0$ is in units of 
$0.01$ eV/$(\hbar v) \simeq 0.03$ nm$^{-1}$.

The resonance curves and lines of very small conductance in Fig.~\ref{fig:surf}
can both be analytically understood using a Floquet perturbation 
theory~\cite{soori10} to study the solution inside the barrier (region II).
Taking $\psi_{II} = e^{i(k'_x x + k_y y)} f(t)$, we have to solve the equation
$i \hbar \partial f/\partial t= H f$, where 
\beq H ~=~ \hbar v[ k_y \sigma^x - k'_x \sigma^y + V_0 \cos (\omega t) 
\sigma^x]. \eeq
(This follows from Eq.~\eqref{hamII} with $m_0 = 0$). To develop the Floquet
perturbation theory, we now write $H(t) = H_0 (t)+ V (t)$ where $H_0$ is much 
larger than $V$. Depending on the various parameters, we will decompose $H$ 
in two different ways.

\begin{figure}[htb]
\centering
\subfigure[]{\includegraphics[scale=0.24]{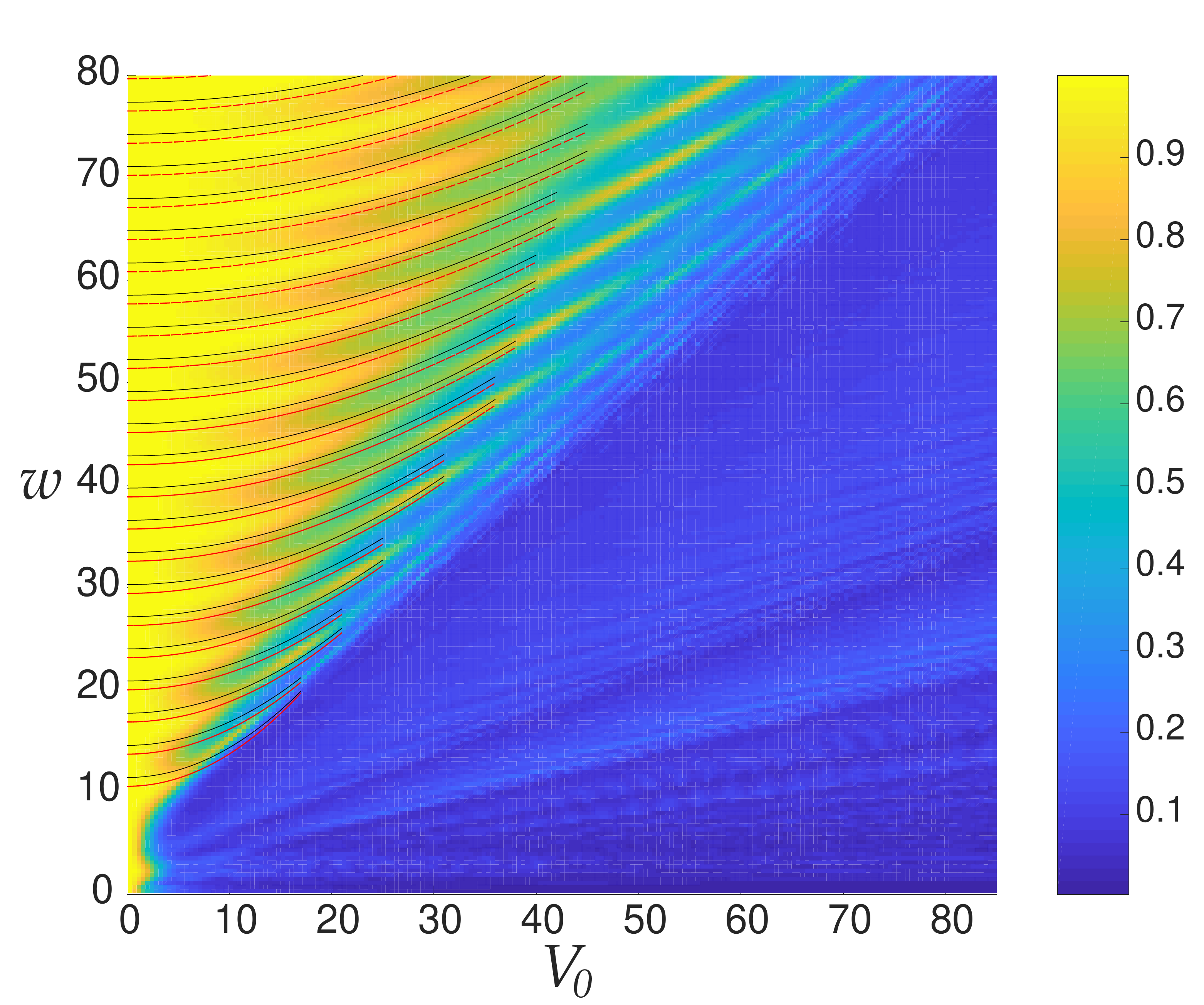}}
\subfigure[]{\includegraphics[scale=0.25]{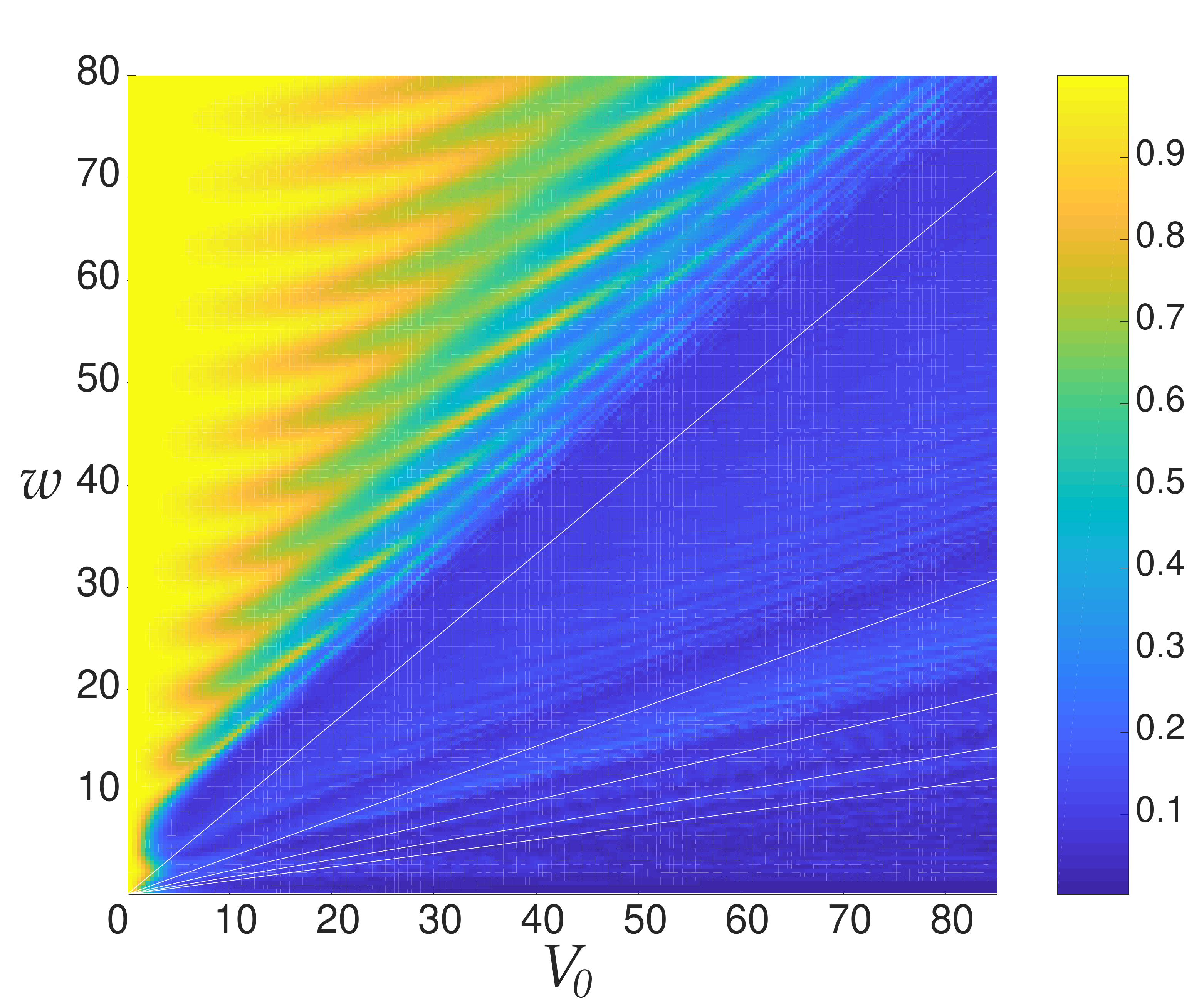}}
\caption{Surface plot of $G/G_0$ as a function of $V_0$ and $\omega$, for 
$E_0 = 2$, $m_0 = 0$, and $L=1$. The same plot appears in the two figures but 
some lines have been shown to highlight some features in the regions with 
small and large conductances respectively. In (a), black and red lines given 
by Eq.~\eqref{om1} are shown, where the second term is $\pm E_0/\hbar$ 
respectively, and the integer $p$ increases as we go up from the bottom to the 
top. In (b), some white lines corresponding to $J_0 (2V_0/\omega) = 0$ are 
shown; from top to bottom, these lines have slopes given by $0.832, 0.362, 
0.231, 0.170$ and $0.134$. (See Ref.~\onlinecite{udupa20}).} 
\label{fig:surf} \end{figure}

\noindent (i) For $k'_x \gg k_y, V_0$, we take
\bea H_0 &=& - \hbar v k'_x ~\sigma^y, \nonumber \\
V &=& \hbar v ~[k_y ~+~ V_0 \cos (\omega t)] ~\sigma^x. \label{h0v2} \eea
Floquet perturbation theory then leads to the expression~\cite{udupa20}
\beq k'_x ~=~ \frac{\omega}{v} ~\pm~ \frac{E_0}{\hbar v} ~-~
\frac{v k_y^2}{2 \omega} ~-~ \frac{v V_0^2}{3 \omega}. \label{kpx2} \eeq
Since the barrier length is $L$, we expect to get transmission resonances
when $k'_x = p \pi/L$, where $p = 1, 2, 3, \cdots$. Then in the regime 
$\omega \gg E_0/\hbar,~ vV_0$, Eq.~\eqref{kpx2} gives the approximate relation
\beq \omega ~=~ \frac{\pi p v}{L} ~\pm~ \frac{E_0}{\hbar} ~+~ \frac{vL
V_0^2}{3 \pi p}. \label{om1} \eeq 
This explains why in Fig.~\ref{fig:surf} (a), the resonance regions have a 
spacing given by $\pi v /L$ when $V_0 = 0$ and curve up as $V_0^2$ as $V_0$ 
increases. The black and red lines in that figure are given by Eq.~\eqref{om1},
where the second term is $\pm E_0 /\hbar$ respectively, and the integer $p$ 
increases as we go up from the bottom to the top. 

\noindent (ii) $V_0 \gg k_y, k'_x$, we take
\bea H_0 &=& \hbar v V_0 \cos (\omega t) ~\sigma^x, \nonumber \\
V &=& \hbar v ~(k_y ~\sigma^x ~-~ k'_x ~\sigma^y). \label{h0v1} \eea
In this case, Floquet perturbation theory leads to the expression
\beq k'_x ~=~ \pm ~\frac{\sqrt{(E_0/\hbar v)^2 ~-~ k_y^2}}{J_0(2vV_0 /\omega)}.
\label{kpx1} \eeq
This implies that if $J_0(2vV_0 /\omega) = 0$, there is no allowed value of 
$k'_x$
in region II. Hence, it is not possible for an electron to transmit through 
region $II$ if the parameters $(V_0,\omega)$ lie on one of the lines where 
$J_0 (2vV_0 /\omega) = 0$. The first five zeros of $J_0(2vV_0 /\omega)$ 
correspond to $\omega/(vV_0) = 0.832, 0.362, 0.231, 0.170$ and $0.134$. We see
in Fig.~\ref{fig:surf} (b) that the conductance is indeed particularly small
on the white lines whose slopes correspond to the above values.

\section{Other topological systems}
\label{sec:otherti}

We would like to mention two unusual examples of systems whose properties 
exhibit some topological features~\cite{seshadri18,deb18}. The first system
involves a spin system in which the elementary excitations, called magnons,
are bosons. The second system is a Josephson junction of several 
superconducting wires and the objects of interest are the Andreev bound states.

The spin system is defined on a kagome lattice and it has both $XXZ$ anisotropy
(i.e., coupling $J$ between the $xx$ and $yy$ components and $J \Delta$
between the $zz$ components of the two spins) and a Dzyaloshinskii-Moriya (DM)
interaction $D$ between nearest-neighbor spins~\cite{seshadri18}. As a 
function of $\Delta$ and $D/J$, the system has a large number of phases some 
of which have a topological character. In these topological phases, the spins 
have a simple ferromagnetic order in the ground state. However, 
the three magnon (spin-wave) bands have 
an interesting structure: each band has a Berry curvature which is a function 
of the wave number ${\vec k} = (k_x,k_y)$, and integrating the Berry curvature
over the BZ gives a Chern number which is non-zero in some of the 
bands. A strip-like system which is infinitely long in one direction and
has a finite width in the other direction has modes at the two edges; the
energies of these modes lie in the gaps between the bands, and the number of
edge modes is related to the Chern numbers. Further, the system has a non-zero
thermal Hall conductivity (ratio of heat current in one direction to a 
temperature gradient applied in the perpendicular direction) which is given 
by the sum over the bands $j$ (where $j=1,2,3$) of the integral over the 
BZ of a product of the Berry curvature and a function of the 
Bose-Einstein distribution $1/[\exp [E_j ({\vec k})/(k_B T] - 1]$.
This is a function of the system parameters $J$, $\Delta$, $D$, and the 
temperature $T$. Unlike a fermionic topological system, the thermal Hall 
conductivity is not quantized and it varies smoothly with the temperature.

In the Josephson system, we consider a junction of three superconducting wires,
each of which may have $s$-wave or $p$-wave pairing~\cite{deb18}. We denote 
the phases of the pairings in the three wires as $\phi_j$, where $j=1,2,3$. 
The junction can host three 
Andreev bound states whose energies lie within the superconducting gap and 
are functions of the two independent differences of pairing phases, namely, 
$\phi_{13} = \phi_1 - \phi_3$ and $\phi_{23} = \phi_2 - \phi_3$; these play
the same role as the momenta $(k_x,k_y)$ of a two-dimensional topological 
system. The three bands of Andreev bound states have Berry curvatures which
depend on the parameters of the system, in particular, on the scattering
matrix at the junction which determines the amplitude for an electron or hole 
incident on the junction from one of the wires to either get reflected back 
to the same wire or transmitted to the other two wires. The integral of the
Berry curvature over $(\psi_{13},\phi_{23})$ gives the Chern number of the 
band. Some of the transport properties of the system turn out to be quantized
at zero temperature. For instance, when a constant voltage bias $V_1$ is 
applied to wire 1 (keeping $V_2 = V_3 = 0$), the phase $\phi_1$ changes 
linearly with time while the phases $\phi_2, \phi_3$ remain constant in time. 
We then find that there is a time-averaged current 
$I_2 = - I_3$ in the other two wires (the time-averaged current in the same
wire, $I_1$, turns out to be zero). The ratio $G_{21} = I_2/V_1$ is called
a transconductance. We then find that the integral of $G_{21}$ over $\phi_2$
is quantized in an integer multiple of $4e^2/h$, where the integer is related
to the Chern numbers of the filled bands of Andreev bound states.

\section{Summary and future directions}

We will now summarize the contents of this review and provide some recent progress in the field which might lead to exciting future studies. We began by 
introducing four prototypical systems, namely, the SSH model,
the Kitaev chain, the Kitaev honeycomb model, and the Haldane model, and
the different topological invariants which distinguish between the topological
and non-topological phases in these systems. We then discussed what happens 
when some of the parameters in the Hamiltonians of the Kitaev chain and the 
Kitaev honeycomb model are driven periodically in time. It is found that 
periodic driving can generate
boundary modes and there are some topological invariants
which can predict the number of such modes. 

We then proceed to discuss the possibility of dynamically preparing topologically non-trivial states. In particular, we review the protocols for adiabatically connecting topologically inequivalent phases by manipulating the protective symmetry in SPT insulators. To elaborate, we shed light on the two possibilities of explicit and dynamical symmetry breaking in preparation of SPT phases. In teo-dimensional Chern insulators, we discuss the unitary and non-unitary no-go theorems which represent the problems associated with the dynamical preparation of non-trivial Chern states. Subsequently, we discuss some recent attempts in going around the no-go theorems in dynamical unitary and non-unitary settings. Particularly, we discuss the existence of a many-body Chern number, constructed out of the Resta polarization, which successfully captures the topology of Gaussian states at all temperatures and show that it indeed characterizes mixed topological states in periodically driven systems. We also discuss how the coupling of multiple two-dimensional Chern insulating layers may lead to interesting dynamical protocols which adiabatically connect inequivalent Chern states of the individual layers.

Next, we discussed transport through
a Kitaev chain and through a junction of three Kitaev chain. These systems
have Majorana modes with different energies at the ends or at the junction, 
and these modes lead to peaks in the differential conductance at those 
energies. We then discussed transport through the surface states of 
three-dimensional topological insulators. Transport through a thin
topological insulator, where the top and bottom surfaces hybridize
significantly with each other, shows a number of interesting features such
as an oscillatory dependence of the conductance on the strength of a 
barrier placed on the top surface. Transport in a granular topological 
insulator was discussed briefly, and it is found that such a system
can behave like a single crystal of a topological insulator. The 
effect of periodic driving (produced by applying electromagnetic
radiation to a strip-like region on the top surface) was discussed. 
Depending on the amplitude and frequency of the driving, the system
can exhibit large or small differential conductance. Thus the system
can behave like an optical switch whose properties can be tuned by
the parameters of the radiation. Finally, in the Appendices we have included a pedagogical discussion regarding adiabatic transformations and topology in quantum many-body systems.

Recently, there has been many intriguing attempts to topologically classify non-equilibrium quantum many-body states which at the same time, also sheds light on novel out of equilibrium phase transitions. Particularly, in Ref.~\onlinecite{roy17}, it was shown that the expectation values of some local observables belonging to the bulk system indeed exhibit emergent non-analyticities due to the presence of topological critical points, even when the system is probed in out of equilibrium situations. It has also been established recently\ct{perfetto20} that, shining laser pulses on a non-topological p-wave excitonic insulator can induce a non-equilibrium topological phase transition in the system. As far as engineering topological phases is concerned, a promising prospect which has garnered great interest is the construction of optimal adiabatic paths connecting inequivalent mono-layer Chern phases through counter-diabatic\ct{claeys19} driving protocols. A related and exciting development in this regard has been the advent of so called \textit{dynamical quantum phase transitions} (dqpts) \cite{heyl13,sharma15,budich16,sharma16,utso17,utso17_2,utso_mixed17,heyl_mixed17,sedlmayr18,bhattacharjee18,dora18,bandyopadhyay18,halimeh20,bandyopadhyay20}, where the return amplitude of the wave function of the system exhibits non-analyticities with time, following a perturbation of the system. Remarkably, a dynamical topological order parameter \cite{budich16} can be identified which assumes only integer values and can change only at those instants of time when the return amplitude becomes non-analytic, thus mimicking equilibrium topological phase transitions. In recent times, there has also been an emerging trend of greater interest in the topology of interacting quantum many-body systems \cite{rachel18} which holds significant experimental relevance. Currently, a few studies\cite{tarnowski19,xinchen20} are also opening up an approach to characterize the Chern topology of non-equilibrium systems by studying the linking invariant. A complete understanding of the rigorous connection between the linking invariant and the lattice Chern number or the Bott index requires further studies.


\vspace*{.8cm}

\centerline{\bf Acknowledgments}
\vspace*{.4cm}

The authors thank Amit Dutta for numerous discussions.
Souvik Bandyopadhyay acknowledges financial support from PMRF, MHRD, India.
Sourav Bhattacharjee acknowledges CSIR, India for financial support.
Diptiman Sen thanks DST, India for Project No. SR/S2/JCB-44/2010 for 
financial support.

\appendix

\section{A pedagogical introduction to the geometry and topology of quantum states}\label{sec_app_A}

In this appendix, we elaborate on the intrinsic connection between topology and the dynamics of physical non-interacting systems through the generation of adiabatic time-evolution of classical and quantum systems. We then proceed to identify a deeper geometric origin of both topological properties and adiabatic transformations which allow us to quantitatively construct topological invariants, characterizing inequivalent topological phases of matter and dynamics. 

\subsection{Adiabatic transformations}\label{sec:adiabatic}

Consider a fermionic non-interacting quantum system whose dynamics is generated {by a Hamiltonian $H(\lambda(t))$}, where $\lambda(t)$ is a time-dependent parameter. As for example, $\lambda$ can be an external time-dependent magnetic field vector acting on a spin-$1/2$ particle. Generically, the system can be dependent on a family of {time-dependent parameters}.
For simplicity, we assume that the Hamiltonian $H(\lambda)$ has non-degenerate eigenvalues which are positive semi-definite and therefore, its eigenvectors,
\begin{eqnarray}\label{character}
	H(\lambda)\ket{n(\lambda)}&=&E_n(\lambda)\ket{n(\lambda)},\nonumber\\
\end{eqnarray}
form a complete basis. 

{A quantum mechanical state $\ket{\phi(\lambda)}$ is said to evolve adiabatically in time if the solution of the time-dependent Schrodinger equation,
\begin{equation}\label{eq_schrodinger}
i\hbar\frac{\partial}{\partial t}\ket{\phi(\lambda)}=H(\lambda)
\ket{\phi(\lambda)}, \end{equation}
satisfies the condition
\begin{equation}\label{eq_cond_adiab} 
\left|\braket{n(\lambda)|\phi(\lambda)}\right|^2=\mathrm{constant}
\end{equation}
	for all $\ket{n(\lambda)}$. It is important to note that an adiabatic transformation is not necessarily always restricted to time evolution. As such, in subsequent discussions, we will associate $\lambda$ with any \textit{controllable parameter} which defines a family of Hamiltonians $H(\lambda)$. 
}

The Hamiltonian $H(\lambda)$ can always be diagonalized by a unitary transformation,
\begin{equation}\label{diagonal}
	\tilde{H}(\lambda)=U^{\dagger}(\lambda)H(\lambda)U(\lambda),
\end{equation}
where $U(\lambda)$ is a unitary operation that {diagonalizes $H(\lambda)$ for all values of $\lambda$}. {The unitary transformation thus connects a $\lambda$-independent basis $\{\ket{b_n}\}_n$ on the vector space $\mathcal{V}$ with the eigenbasis of $H(\lambda)$, i.e.},
\begin{equation}
	\ket{n(\lambda)}=U(\lambda)\ket{b_n}.
\end{equation}
{It follows then},
\begin{eqnarray}\label{agp}
	i\partial_{\lambda}\ket{n(\lambda)}&=&\mathcal{A}_{\lambda}\ket{n(\lambda)},\nonumber\\
	\mathcal{A}_{\lambda}&=&i\left(\partial_{\lambda}U(\lambda)\right)U^{\dagger}(\lambda),
\end{eqnarray}
where $\mathcal{A}_{\lambda}$ is called the {\it generator of adiabatic transformations}\ct{anatoli17,anatoliMbukov} for the system in the parameter $\lambda$ and {essentially generates eigenvectors $\ket{n(\lambda)}$ of $H(\lambda)$}. {The nomenclature follows from the fact that the above equation is identical to the Schrodinger equation in Eq.~\eqref{eq_schrodinger} and the states $\ket{n(\lambda)}$ trivially satisfy the adiabatic condition, given in Eq.~\eqref{eq_cond_adiab}.}

An alternate form of the generator $\mathcal{A}_\lambda$ can be derived as follows. Differentiating Eq.~\eqref{character} with respect to the parameter $\lambda$, it is straightforward to show that
\begin{equation}\label{eq:g}
\partial_{\lambda}H-i\left[\mathcal{A}_{\lambda},H\right]=
\sum_{n}\partial_{\lambda}E_n(\lambda)\ket{n(\lambda)} \bra{n(\lambda)},
\end{equation}
where we have omitted the $\lambda$ dependence of $H$ for simplicity in notation. It then follows that
\begin{equation}\label{agp_2}
	\left[\partial_{\lambda}H-i\left[\mathcal{A}_{\lambda},H\right],H\right]=0.
\end{equation} 
Solving this equation does not explicitly require the knowledge of the eigenstates of the Hamiltonian $H(\lambda)$. It can in fact be mapped to a minimization problem which also allows one to construct approximate adiabatic transformations with considerable ease. Thus, the eigenstates can now be generated for all $\lambda$ using the generator $\mathcal{A}_\lambda$, without explicitly diagonalizing $H(\lambda)$. An added advantage of the form in Eq.~\eqref{agp_2} is that it is readily extended to classical Hamiltonian system \ct{anatoli17} as
\begin{equation}\label{agp_classical}
	\{\partial_{\lambda}H-\{\mathcal{A}_{\lambda},H\},H\}=0,
\end{equation}
where $\{.\}$ denotes the Poisson bracket with respect to the generalized coordinates. This is not surprising as adiabatic transformations are also valid canonical transformations in classical Hamiltonian systems.\\

\subsubsection{Topology of adiabatically connected states}


{To understand the deep-seated connection between adiabaticity and topology, let us consider an adiabatic transformation along a trajectory in the one-dimensional parameter space of $\lambda$. We will be primarily concerned with the topological aspects of non-interacting fermionic many-body systems. Such systems permit an underlying (quasi) particle description, where importantly, the (quasi) particles are non-interacting and their number remains conserved under any unitary transformation. As such, it often suffices to consider only the single-particle sector of the many-body Hamiltonian. In general, there may exist multiple single-particle eigenstates for a given ${\lambda}$ and the many-body ground state corresponds to the situation in which half of the lowest energy single particle states are completely filled. We consider the simple case in which there exists two single-particle states for each $\lambda$; the ground state $\ket{n_0(\lambda_0)}$ of $H(\lambda_0)$ therefore corresponds to one occupied single-particle state. We also assume that the ground state is separated from other excited states by a finite spectral gap. At any arbitrary point along the trajectory, the adiabatic evolution is generated by the action of $\mathcal{A_\lambda}$ on the state $\ket{n_0(\lambda)}$. Taking an inner product of $\mathcal{A_\lambda}\ket{n_0(\lambda)}$ with the initial state, we find from Eq.~\eqref{agp}, 

	\begin{equation}\label{loop}
		i\partial_{\lambda}\braket{n_0(\lambda_0)|n_0(\lambda)}=i\braket{n_0(\lambda)|\partial_{\lambda}|n_0(\lambda)}\braket{n_0(\lambda_0)|n_0(\lambda)}.
	\end{equation}
For passage along a closed loop, such that $H(\lambda_0+\lambda_T) = H(\lambda_0)$ for some constant $\lambda_T$, the solution of the Eq.~\eqref{loop} can be written as,
\begin{eqnarray}
\braket{n_0(\lambda_0)|\tilde{n}_0(\lambda_0)} &=& e^{-i\oint \braket{n_0
(\lambda)|\partial_{\lambda}|n_0(\lambda)}d\lambda} \non \\
&=& e^{-i\oint \mathcal{A}_{\lambda}^{n_0n_0}d\lambda}, \end{eqnarray}
where $\ket{\tilde{n}_0(\lambda_0)}$ is simply the initial state after being adiabatically transported along a closed parametric loop. One therefore obtains for the phase\ct{berry84,zela12}
	\begin{equation}\label{Berry}
		\Phi_B^{n_0}=\oint \mathcal{A}^{n_0n_0}_{\lambda}d\lambda=c_{n_0},
\end{equation}}
popularly known as the Berry phase.
\begin{figure}[ht]
\ig[width=0.95\columnwidth]{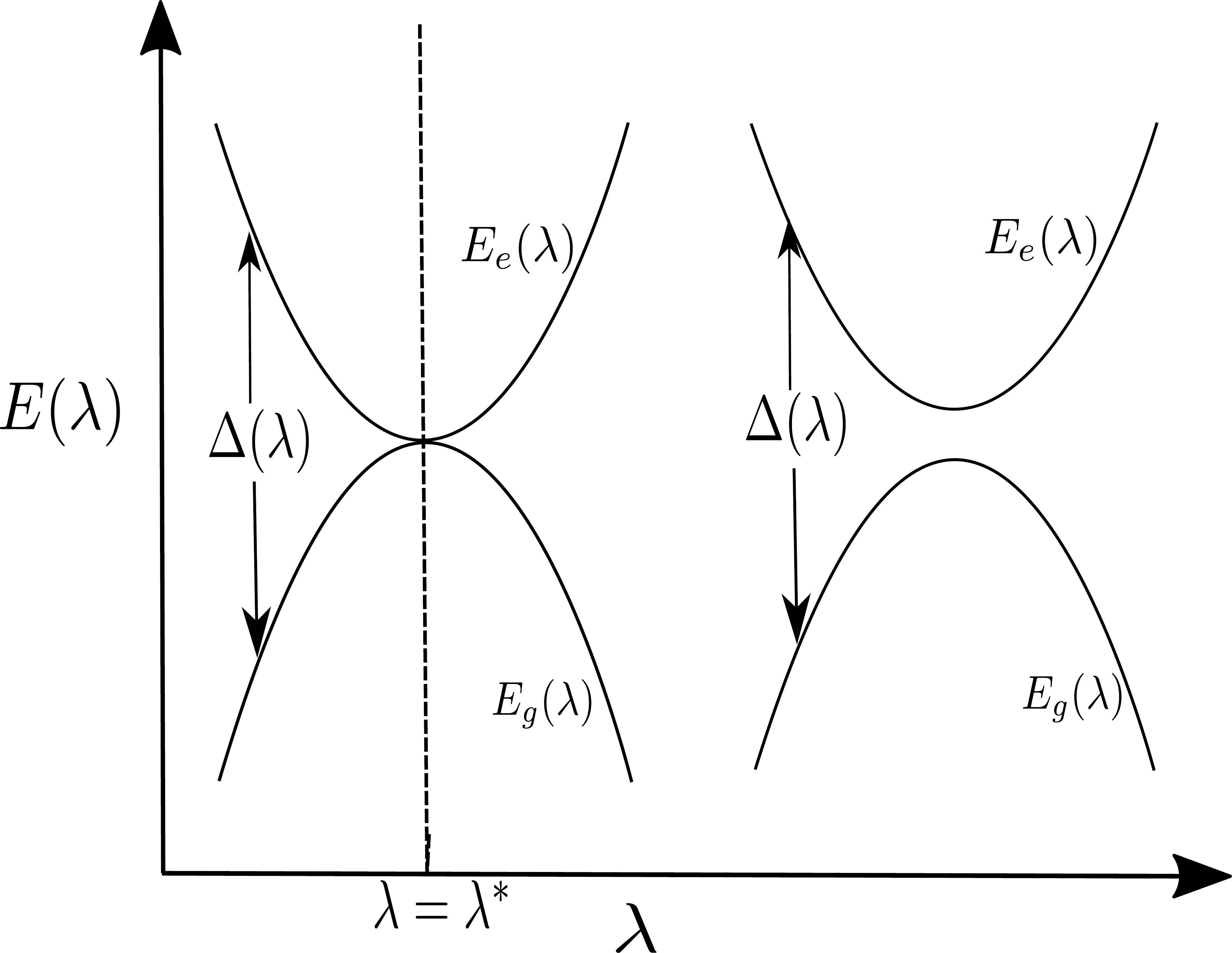}	
\caption[]{Schematic diagram of the spectrum of a two-level system with $E_{g}(\lambda)$ and $E_{e}(\lambda)$ as the ground state and excited state energies respectively depending on a parameter $\lambda$. The spectrum on the left shows a band degeneracy at the critical point $\lambda=\lambda^*$ destroying adiabaticity as one tunes the parameter $\lambda$. The spectrum on the right however shows two eigenstates which are always separated by a finite gap $\Delta$. \label{fig:adiabatic} }
\end{figure} 
It is important to note that the Berry phase must be an adiabatic invariant of the system. 
This is a crucial consequence of the fact that the integral in Eq.~\eqref{Berry} is performed over a closed loop. Every point on the trajectory is thus adiabatically connected to every other point and the integral is thus independent of $\lambda_0$.
However, the situation changes if two or more eigenstates become degenerate for some $\lambda^*$; this leads to a breakdown of the adiabaticity condition as $\ket{n_0(\lambda^*)}$ is not uniquely defined. It is only across these points that the Berry phase can change. As for example, if $\lambda^*$ is a \textit{band-crossing} point between two eigenstates (see Fig.~\ref{fig:adiabatic}), then the Berry phases of the ground state and the excited states can interchange at this point. 


At this point, we would like to state that the requirement of a closed loop in Eq.~\eqref{Berry} has a much deeper origin than stated above. To elucidate, we first recall that the quantum states describing a physical system are not uniquely specified with respect to physical observables $\mathcal{O}(\lambda)$. 
In particular, if the states transform as,
\begin{equation}\label{gauge}
	\ket{\psi(\lambda)}\rightarrow e^{i\chi\left(\lambda\right)}\ket{\psi(\lambda)},
\end{equation}
the observable expectations $\braket{\mathcal{O}(\lambda)}=\braket{\psi(\lambda)|\mathcal{O}(\lambda)|\psi(\lambda)}$ remain invariant. Therefore, there is a whole collection of quantum states which describe the same physical phenomena. This internal degree of freedom of quantum states is known as a {\it gauge freedom/symmetry} and internal transformations (as in Eq.~\eqref{gauge}) relating such states are called {\it gauge transformations}. It is thus expected that all observable quantities must remain invariant under such transformations. The transformations in Eq.~\eqref{gauge} belong in the ${\rm U(1)}$ group and it can therefore be said that generic quantum states have an inherent ${\rm U(1)}$ gauge symmetry. Under such a ${\rm U(1)}$ gauge transformation of the eigenstates $\ket{n(\lambda)}$, the generator $\mathcal{A}_{\lambda}$ transforms as,
\begin{equation}\label{vector_potential}
	\mathcal{A}_{\lambda}\rightarrow\mathcal{A}_{\lambda}+\partial_{\lambda}\chi,
\end{equation}
and is therefore not a gauge invariant quantity. As in the previous exercise, if we transport the state $\ket{n_0(\lambda_0)}$ along an open loop in the parameter space, to a point say $\lambda=\lambda_f$, we observe the accumulation of a phase,
\begin{equation}
	\ket{n_0(\lambda_f)}=e^{-i\int_{\lambda_0}^{\lambda_f}\mathcal{A}_{\lambda}^{n_0n_0}d\lambda}\ket{n_0(\lambda_0)}.
\end{equation}
However, under gauge transformation (Eq.~\eqref{gauge}), the complete protocol transforms as,
\begin{eqnarray}\label{gauge_choice}
\ket{n_0(\lambda_f)}^{\prime}&=&e^{-i[ \int_{\lambda_0}^{\lambda_f}\mathcal{A}^{n_0n_0}_{\lambda}d\lambda + \chi(\lambda_f)-\chi(\lambda_0)]}
\ket{n_0(\lambda_0)}^{\prime}\nonumber\\
\text{where~}\ket{n_0(\lambda)}^{\prime}&\rightarrow& e^{i\chi(\lambda)}\ket{n_0(\lambda)}.
\end{eqnarray}
From Eq.~\eqref{gauge_choice} it is easy to see that a gauge choice
\begin{equation}
	\chi(\lambda_f)-\chi(\lambda_0)=-\int_{\lambda_0}^{\lambda_f}
\mathcal{A}^{n_0n_0}_{\lambda}d\lambda ~\mod 2\pi, \end{equation}
completely does away with the phase accumulation due to the adiabatic transport. Hence, it cannot be a physically relevant quantity. However, this is taken care of if one considers closed loops in the parameter space. For closed loops there is an additional restriction on the gauge choice due to single-valuedness of the wave vectors, 
\begin{equation} \chi(\lambda_f)-\chi(\lambda_0)=2\pi l, \end{equation}
where $l\in\mathcal{Z}$. Thus, the accumulated phase due to an adiabatic transformation about a closed loop in parameter space is {\it gauge invariant modulo $2\pi$}, i.e., under a gauge transformation,
\begin{equation}
\Phi^{n_0}_B\rightarrow\Phi_B^{n_0}+2\pi l,~~~l\in\mathcal{Z}.
\end{equation}
Its gauge invariance therefore makes it impossible to get rid of the phase $\Phi_B^n$ by a simple gauge transformation. 
Thus, a closed trajectory ensures not only that the accumulated phase is an adiabatic invariant but is also gauge invariant modulo $2\pi$.
For the special $N=d=1$ case, the phase $\Phi_B^n$ is also called the {\it Zak phase/winding number} of the filled $\ket{n(\lambda)}$ state.\\

In the language of topology, the different eigenstates having differing indices $c_n$ are said to belong to different {\it topological classes/phases} which, as we have seen, cannot be adiabatically/smoothly deformed into each other. The number $c_n$ is termed a {\it topological invariant} (which in this case is an adiabatic invariant) characterizing the different topological phases separated by gapless critical points. In conventional terminology, a phase in which the system has a non-zero value of the topological invariant is called a {\it non-trivial phase} and it is otherwise called a {\it trivial phase}. We have thus uncovered a topological structure of quantum states based on adiabatic transformations. {It is also remarkable to note that the generator of adiabatic transformations $\mathcal{A}_{\lambda}$ under a gauge transformation, transforms exactly as a ${\rm U(1)}$ gauge field similar to the magnetic vector potential in electromagnetic theory (see Eq.~\eqref{vector_potential}).
	This subtle insight already hints that the topological classification constructed in this section is deeply rooted in an underlying geometric structure of quantum states which bridges electromagnetism, topology and many other apparently disconnected areas of physics through the premise of adiabatic transformations.} 

\subsection{The underlying geometry of quantum states}\label{sec:geometry}

{To comprehend the underlying geometric structure behind adiabatic transformations, it is essential to understand that quantum states form a bundle of vector spaces, with a complete local vector space attached to each point of the parameter space (a manifold) $\mathcal{B}$. Each local vector space $\mathcal{V}_{{\lambda}}$ is equipped with a complete basis. This local basis may be taken to be the eigenbasis $\{\ket{n({\lambda})}\}_n$ of the local Hamiltonian $H({\lambda})$, where ${\lambda}$ is the position of the local vector space in the parameter-space. For the sake of generality, in this section we deal with a parameter-manifold of arbitrary dimension $d$ and $N$ number of occupied single-particle states, where the latter equals the dimension of the local vector space. Such a geometry in the field of mathematics is known as a {\it vector bundle} on $\mathcal{B}$. We must also note that similar to the previous discussion, each $\mathcal{V}_{{\lambda}}$ now hosts a local non-Abelian ${\rm U(N)}$ gauge freedom.}\\

To proceed with the analysis of the quantum vector bundle \ct{frankel11,zanardi10,budich13,wang19}, one must first redefine the actions of algebra and analysis on the generic parameter-manifold. Say the local vector spaces are equipped with the $N$-dimensional basis $\{\ket{e_i({\lambda})}\}_i$. Let us now define a family of states $\ket{\phi(\lambda)}$ for each $\lambda\in\mathcal{B}$, following some general rule. As for example, if $\lambda$ is chosen as a time-dependent parameter, then the family of states $\ket{\phi(\lambda)}$ correspond to solution of the time-dependent Schrodinger equation. The vector $\ket{\phi(\lambda)}$ can be expanded in the eigenbasis of the local vector space as,
\begin{equation}
	\ket{\phi(\lambda)}=\phi_i(\lambda)\ket{e_i(\lambda)},
\end{equation}
where henceforth we impose that repeated indices are being summed over (unless explicitly specified) and $\phi_i(\lambda)=\braket{e_i(\lambda)|\phi(\lambda))}$. Taking a derivative of the vector along the direction $\lambda_{\mu}$ in the parameter space, we obtain,
\begin{equation}
	\frac{d\ket{\phi(\lambda)}}{d\lambda_{\mu}}=\frac{\partial\phi(\lambda)_i}{\partial\lambda_{\mu}}\ket{e_i(\lambda)}+\phi_i(\lambda)\frac{\partial\ket{e_i(\lambda)}}{\partial\lambda_{\mu}},
\end{equation}
where the second term on the R.H.S. signifies the change in the basis vectors as one moves along the parameter manifold $\mathcal{B}$. One may thus rewrite the derivative of a vector in terms of the corresponding action on its components $\phi_i(\lambda)$,
\begin{equation}\label{cov_der}
	D_{\mu}\phi_i=\partial_{\mu}\phi_i-iA^{\mu}_{ij}\phi_j,
\end{equation}
where we must set,
\begin{equation}\label{connection}
D_{\mu}\phi_i\equiv\bra{e_i}\frac{d}{d\lambda_{\mu}}\ket{\phi} ~~~\text{and}~~~
A^{\mu}_{ij}=i\braket{e_i|\partial_{\mu}|e_j}. \end{equation}
The operator $A_{ij}^{\mu}$ is called a {\it connection} on the bundle because it in a sense `connects' different points on the manifold by describing how the local basis changes as one travels around the parameter manifold.
The differential operator $D_{\mu}$ is popularly referred to as the {\it covariant derivative} on the parameter manifold and It is for this reason that the connection is often referred to as the {\it gauge potential}. Also, it is straightforward to check that the connection transforms under a gauge transformation as \ct{frankel11,zanardi10},
\begin{equation}\label{Agauge}
A^{\mu}\rightarrow A^{\mu\prime} = UA^{\mu}U^{\dagger} + i U\partial_{\mu}
U^{\dagger}, \end{equation}
{The reason behind its nomenclature is readily seen if one probes the way it transforms under a gauge transformation. The generic local gauge transformation in this scenario,
\begin{equation*}\label{gauge_vec}
\ket{\phi(\lambda)}\rightarrow \ket{\phi(\lambda)}^{\prime} =
U(\lambda)\ket{\phi(\lambda)}, \end{equation*}
	actively rotates the local vectors. This same action amounts to rotating the basis vectors passively,
	\begin{equation*}
		\ket{e_i}\rightarrow \left[U^{\dagger}(\lambda)\right]_{ji}\ket{e_j}.
	\end{equation*}
	Using Eq.~\eqref{cov_der} and Eq.~\eqref{Agauge}, it is straight forward to check that the covariant derivative of a vector transform under a gauge transformation covariantly,
\begin{equation}
\non D_{\mu}\phi_i\rightarrow U_{ji}\left(D_{\mu}\phi_j\right),
\end{equation}
	i.e., exactly similar to the transformation of a vector itself. However, there is an interesting interpretation of both the connection and the covariant derivative in terms of the flatness and curvature of the underlying parameter-manifold. Taking derivatives of vectors basically involves adding/subtracting two vectors at different points on a manifold after transporting them parallely to the same point. This is required because adding two vectors requires both the vectors to belong to the same vector space, i.e.,
\begin{equation}
+:\mathcal{V}_{{\lambda}}\times\mathcal{V}_{{\lambda}}\rightarrow\mathcal{V}_{{\lambda}}.
\end{equation}
Transporting a vector $\ket{\phi}$ on a flat manifold is intuitively simple to understand as the basis remains the same. Thus, the net change in the vector vanishes if 
\begin{equation} \partial_{\mu}\phi_i=0~~~{\rm for ~all}~~i, \end{equation}
	as the connection $A_{\mu}$ vanishes identically.
	However, on a curved surface, the basis itself is different at all points and thus, the condition of parallel transport requires
\begin{equation}\label{parallel}
D_{\mu}\phi_i=\partial_{\mu}\phi_i-iA^{\mu}_{ij}\phi_j=0~~~
{\rm for ~all}~~i. \end{equation}
	This can be equivalently written in terms of a column vector $\vec{\Phi}$ the vectorial form as,
	\begin{equation}\label{parallel_transport}
		i\partial_{\mu}\vec{\Phi}=-A^{\mu}\vec{\Phi}.
	\end{equation}
	Note the exact similarity between the connection $A^{\mu}$ and the generator of adiabatic transformations $\mathcal{A}_{\mu}$ defined in Eq.~\eqref{agp}. Therefore, the same connection which defines a parallel transport on a curved manifold, also generates adiabatic transformations on the eigenstates of $H(\lambda)$. This leads us to conclude that adiabatic transformations are transformations that parallely transport the eigenvectors of $H(\lambda)$ on a generic parameter-space. 
	
	At this point, it is instructive to find a measure of the curvature of the underlying manifold. A defining characteristic of a curved manifold (like the surface of the Earth) is that, unlike in flat-space, if a vector is parallely transported in a closed loop, the transported vector is different with respected to the initial one. The magnitude of the difference is directly proportional to the curvature of the manifold. Consider two orthogonal directions $\mu$ and $\nu$ on the curved parameter space $\mathcal{B}$. We first transport the vector $\ket{\phi}$ along the direction $\mu$ infinitesimally, followed by an infinitesimal transport along the direction $\nu$. The final vector takes the form
\begin{eqnarray}
\ket{\phi}^{\prime} &\simeq& \ket{\phi}+D_{\mu}\ket{\phi}d\lambda^{\mu}+D_{\nu}\ket{\phi}d\lambda^{\nu} \non \\
&& +D_{\mu}^2\ket{\phi}d\lambda^{\mu}d\lambda^{\mu}+D_{\nu}^2\ket{\phi}d\lambda^{\nu}d\lambda^{\nu} \non \\
&& +D_{\nu}D_{\mu}\ket{\phi}d\lambda^{\mu}d\lambda^{\nu}.
\end{eqnarray}
	If the transport direction is then reversed, i.e., an initial translation along $\nu$ is followed by a transport along $\mu$, the vector returns to the same point but is not identical to the initial vector. The change in the vector due to this cyclic transport is found to be
	\begin{equation}
		\delta\ket{\phi}\simeq\left[D_{\mu},D_{\nu}\right]\ket{\phi}d\lambda^{\mu}d\lambda^{\nu}.
	\end{equation}
	This change $\delta\ket{\phi}$, also known as {\it holonomy} in the language of geometry, vanishes in the case of a flat manifold. We therefore define the curvature of the parameter space as
	\begin{equation}
		F_{\mu\nu}=i[D_{\mu},D_{\nu}]=\partial_{\mu}A^{\nu}-\partial_{\nu}A^{\mu}-i[A^{\mu},A^{\nu}].
	\end{equation} 
The holonomy around the infinitesimal loop then reduces to
	\begin{equation}
		\delta\ket{\phi}\simeq -iF_{\mu\nu}d\sigma^{\mu\nu}\ket{\phi},
	\end{equation}
	where $d\sigma^{\mu\nu}=d\lambda^{\mu}d\lambda^{\nu}$ is simply the area enclosed by the infinitesimal loop in parameter space. For the Abelian case of $N=1$, one simply obtains,
	\begin{equation}
		F_{\mu\nu}=\partial_{\mu}A^{\nu}-\partial_{\nu}A^{\mu},
	\end{equation}
	as the $A^{\mu}$ are now scalar functions and therefore commute. 
Thus, the curvature $F$ is nothing but the curl of the vector field $A$ in complete analogy to the magnetic field and the vector potential in Maxwell's electrodynamics. This intriguing semblance is due to the fact that the electromagnetic theory is also equipped with a local ${\rm U(1)}$ gauge symmetry which through Noether's theorem ultimately leads to the local conservation of electric charge. It is because the $N=1$ Abelian theory enjoys a similar ${\rm U(1)}$ gauge freedom in our case, we obtain these striking resemblance with electrodynamics. One can further the semblance by observing that the holonomy $\delta\ket{\phi}$ is nothing but the flux of the local ``magnetic field" (which is similar to the curvature/``{\it gauge field}" $F_{\mu\nu}$ in quantum mechanics) through an infinitesimal loop on $\mathcal{B}$.\\
	
	We note that the gauge field $F_{\mu\nu}$ is not generically gauge-invariant except for the Abelian case where it transforms under a gauge transformation as,
	\begin{equation}
		F_{\mu\nu}\rightarrow UF_{\mu\nu}U^{\dagger}.
	\end{equation} 
	To construct a gauge-invariant quantity, let us recall the definition of parallel transport of a vector $\ket{\phi}$ on a curved manifold via Eq.~\eqref{parallel}. The general solution of the Eq.~\eqref{parallel_transport} can be written in terms of a Dyson series,
	\begin{equation}\label{para_propagate}
		\vec{\Phi}_{||}({\lambda}_f)=\mathbb{P}_{||}({\lambda}_f,{\lambda_i})\vec{\Phi}({\lambda}_i),
	\end{equation}
	where $\vec{\Phi}_{||}(\vec{\lambda}_f)$ is $\vec{\Phi}(\vec{\lambda}_i)$ after being transported from the point $\vec{\lambda}_i$ to $\vec{\lambda}_j$ on $\mathcal{B}$ and,
	\begin{equation}\label{para_propagate2}
		\mathbb{P}_{||}(\vec{\lambda}_f,\vec{\lambda_i})=\mathcal{P}e^{i\int_{\vec{\lambda}_i}^{\vec{\lambda}_f}A^{\mu}d\lambda_{\mu}},
	\end{equation}
	is the parallel propagator which transports a vector parallely along a curve $C$; $\mathcal{P}$ being the path-ordering operator on $C$. The parallel propagator is sometimes also known as a Wilson line. From Eqs.~\eqref{Agauge} and~\eqref{para_propagate2}, it is apparent that the parallel propagator also suffers from the ambiguity of gauge freedom. Under a gauge transformation, the propagator transforms as,
	\begin{equation}
		\mathbb{P}_{||}(\vec{\lambda}_f,\vec{\lambda_i})\rightarrow U(\vec{\lambda}_f)\mathbb{P}_{||}(\vec{\lambda}_f,\vec{\lambda_i})U^{\dagger}(\vec{\lambda}_i).
	\end{equation}
	
	In a similar methodology employed in Sec.~\ref{sec:adiabatic}, a gauge invariant can be constructed by considering a parallel transport along a closed curve ($\vec{\lambda}_i=\vec{\lambda}_f=\vec{\lambda}$), which is realized by the parallel propagator,
	\begin{equation}
		\mathbb{P}_{o}=\mathcal{P}e^{i\oint A^{\mu}d\lambda_{\mu}},
	\end{equation}
	Under a gauge transformation, the closed loop propagator transforms as,
	\begin{equation}
		\mathbb{P}_{o}\rightarrow U(\vec{\lambda})\mathbb{P}_{o}U^{\dagger}(\vec{\lambda}).
	\end{equation}
	It can now be seen that the quantity,
	\begin{equation}\label{loop_dependent}
		W={\rm Tr}\left(\mathcal{P}e^{i\oint A^{\mu}d\lambda_{\mu}}\right),
	\end{equation}
	must be gauge invariant under a generic ${\rm U(N)}$ gauge transformation due to the cyclic property of trace. This gauge invariant quantity is popularly known as the {\it Wilson loop}\ct{wang19} holonomy on the parameter manifold $\mathcal{B}$. A more useful form is the gauge-invariant action/index which is defined as,
	\begin{equation}\label{nabcn_1}
		\mathcal{C}=\frac{1}{i}\ln{\rm Tr}\left(\mathcal{P}e^{i\oint A^{\mu}d\lambda_{\mu}}\right)=c_{\phi}.
	\end{equation}
	
	We have thus constructed a gauge invariant index $c_{\phi}$ that characterizes states $\ket{\phi({\lambda})}\in\mathcal{V}_{{\lambda}}$ on the generic parameter manifold $\mathcal{B}$. However, in practice, the parameter manifold is often discrete as opposed to a continuous one.
	To exemplify, consider a many-body Hamiltonian describing non-interacting particles moving on a periodic lattice. This Hamiltonian then respects a discrete translational symmetry and not a continuous one. Utilizing this symmetry, one might recast the complete description in Fourier space parametrized by conserved quasimomenta $(\{k_i\}_{i=1}^{i=d})$ which now form a $d$-dimensional discrete parameter space. The discrete space approaches a continuum manifold in the thermodynamic limit of the system. 
	It is thus important to extend the formalism, discussed thus far in the context of continuous manifold, to situations in which the parameter space is discrete.
	
	For the sake of simplicity, let us consider the case of a two dimensional parameter space. An extension to higher dimensions can be subsequently made in a straight forward manner. 
	Assuming a square lattice, we label each point on the discrete parameter space by the coordinates $\vec{\lambda}\in\{na\hat{\mu},ma\hat{\nu}\}$ where $a$ is the distance between neighboring lattice points and the numbers $\{n,m\}$ take integral values. The discrete space then reaches a continuum manifold in the limit $a\rightarrow 0$. We now proceed to construct Wilson loops on this discrete parameter space. It is evident that the Wilson line originating from the point $\vec{\lambda}$ and connecting the next consecutive lattice point in the direction $\hat{\mu}$, for $a\ll1$, can be approximated as,
	\begin{equation}\label{link}
		\mathcal{U}_{\mu}({\lambda})=\mathbb{P}_{||}({\lambda})\simeq e^{iA^{\mu}({\lambda})a}.
	\end{equation} 
	The group-valued quantities $\mathcal{U}_{\mu}$ on the lattice is also known as a {\it link variable} simply because it lives on the ``links" connecting lattice points while the quantum states $\ket{\phi({\lambda})}$ reside on the lattice points itself. Using the link variables, we can therefore generate a parallel transformation on a square loop on the lattice such that the enclosed area is a minimum. The rules to do so can be inferred immediately: (i) to parallely transport a vector from $\vec{\lambda}$ to
\begin{figure*}[htb]
\centering
\includegraphics[scale=0.35]{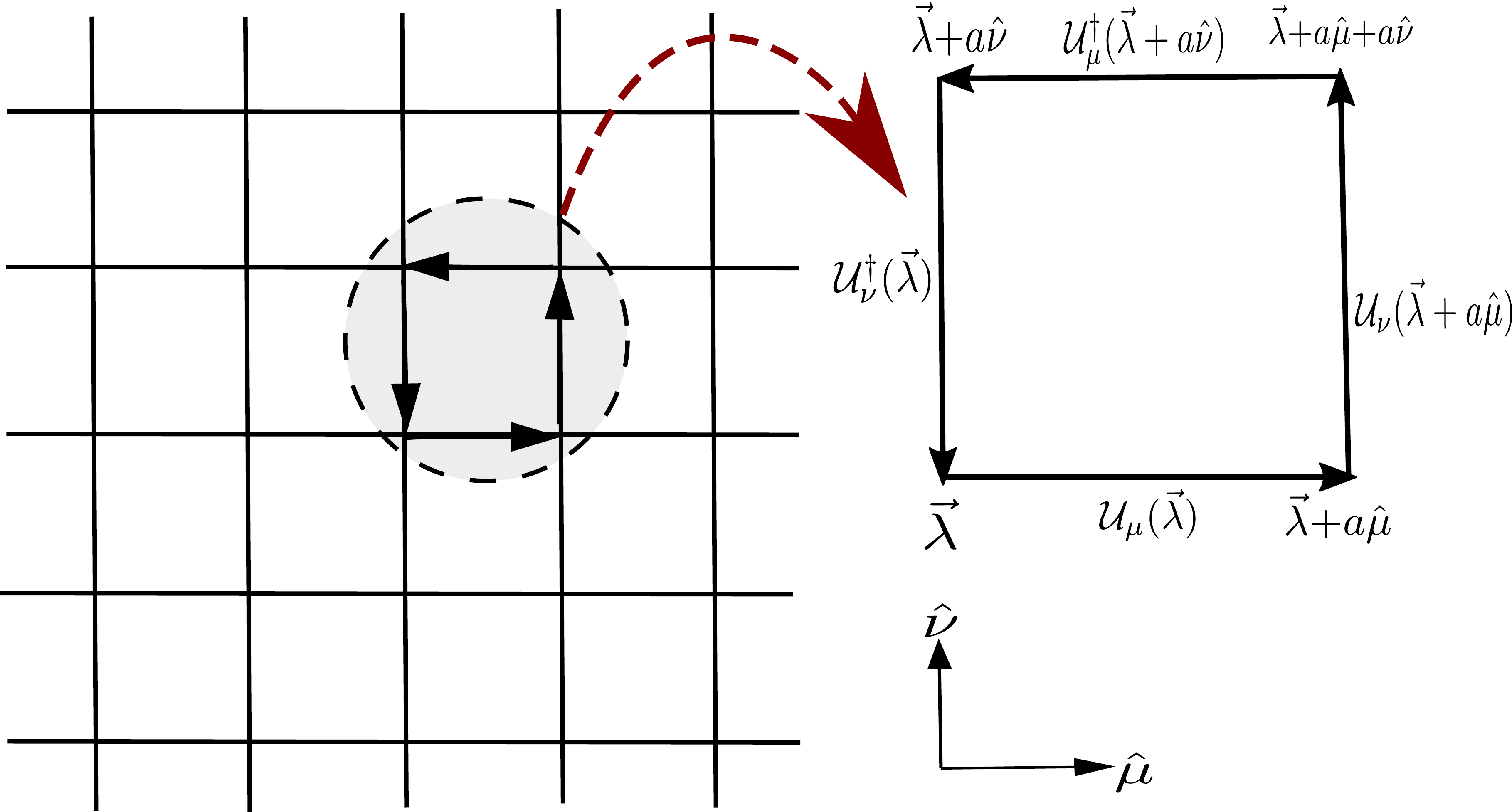}
\caption{Schematic picture of the lattice Wilson loop on a discrete square lattice denoting the link variables and path traversed around the closed loop enclosing an area of $a^2$, $a$ being the lattice parameter.} \label{fig:wilson} \end{figure*}
$(\vec{\lambda}+a\hat{\mu})$, one must operate with $\mathcal{U}_{\mu}(\vec{\lambda})$, and (ii) to induce the reverse transport from $(\vec{\lambda}+a\hat{\mu})$ to $\vec{\lambda}$, one must similarly operate with $\mathcal{U}^{\dagger}_{\mu}(\vec{\lambda}+a\hat{\mu})$. Hence, for the infinitesimal square loop over a plaquette of minimum area ($=a^2$) on the discrete lattice, the logarithm of the gauge invariant \textit{lattice Wilson loop} $W_L$ (see Fig.~\ref{fig:wilson})can be written as\ct{fukui05},
	\begin{equation}\label{fukui_1}
		W_{L}=\frac{1}{i}\ln{\rm Tr}\left[\mathcal{U}_{\nu}^{\dagger}(\vec{\lambda})\mathcal{U}^{\dagger}_{\mu}(\vec{\lambda}+a\hat{\nu})\mathcal{U}_{\nu}(\vec{\lambda}+a\hat{\mu})\mathcal{U}_{\mu}(\vec{\lambda})\right].
	\end{equation}
	Substituting Eq.~\eqref{link} for the link variables in Eq.~\eqref{fukui_1} and retaining terms up to the lowest order in $a$, one obtains,
	\begin{equation}
		W_L\simeq\frac{1}{i}\ln{\rm Tr}\left(e^{ia^2F_{\mu\nu}+....}\right),
	\end{equation}
	where we have retained terms up to quadratic in order $a$. This expression can further be simplified for $a\ll1$ as,
	\begin{eqnarray}\label{lattice_wilson}
		W_L&\simeq& a^2{\rm Tr}F_{\mu\nu}={\rm Tr}\left(F_{\mu\nu}\right)d\lambda^{\mu}d\lambda^{\nu},
	\end{eqnarray}
	where we have evaluated the logarithm in its principal branch, i.e.,
	\begin{equation}\label{branch}
		\left|{\rm Tr}\left(F_{\mu\nu}\right)d\lambda^{\mu}d\lambda^{\nu}\right|<\pi.
	\end{equation}
	
	In the continuum limit $a\rightarrow 0$, Eq.~\eqref{lattice_wilson} approaches an exact equality. It is also to be noted that the Wilson loop around an infinitesimal plaquette is simply the flux of the gauge field through the plaquette. Therefore, it is natural to expect that if the Wilson loop encloses a finite region $S$, the total flux through the loop equals the sum of the fluxes through all infinitesimal plaquettes $dS\in S$ within the loop. We can therefore write the discretized version of the index $\mathcal{C}$ defined in Eq.~\eqref{nabcn_1} as,
	\begin{eqnarray}\label{discrete_wilson}
		\mathcal{C}(C)&\simeq&\sum\limits_{dS\in S}W_L(dS)\nonumber\\
		&=&\sum\limits_{dS\in S}{\rm Tr}\left(F_{\mu\nu}\right)\Delta^{\mu\nu}+2\pi\eta(dS),
	\end{eqnarray}
	where $\Delta^{\mu\nu}$ is the area element of each plaquette. The integer valued field $\eta(dS)$ defined over each plaquette restricts the complex logarithm in Eq.~\eqref{fukui_1} in the principal branch, thus satisfying Eq.~\eqref{branch}. In the continuum limit, the relation becomes an exact equality and is also known as the {\it Stokes' theorem},
	\begin{equation}\label{eq_stokes}
		\mathcal{C}(C)=\frac{1}{i}\ln{\rm Tr}\left(\mathcal{P}e^{i\oint_C A^{\mu}d\lambda_{\mu}}\right)=\int_{S}{\rm Tr}\left(F_{\mu\nu}\right)d\sigma^{\mu\nu},
	\end{equation}
	where all logarithms are evaluated in the principal branch.
	
\subsection{Topological invariants}
\label{sec:topology}

\subsubsection{Chern invariants}
	
	We are now fully equipped with the necessary abstractions and ready to probe the topological phases of physical quantum systems. Consider a non-interacting fermionic many body system described by the Hamiltonian $H$ on a periodic lattice. As discussed previously, one can utilize the discrete spatial translation symmetry to rewrite the Hamiltonian in terms of decoupled/conserved momentum modes $\vec{k}$ in Fourier space,
	\begin{equation}
		H=\bigoplus_{\vec{k}}H(\vec{k}),
	\end{equation}
	where the vector $\vec{k}$ parametrizes the underlying manifold. For example, in $d=2$, we have $\vec{k}\equiv\{k_1,k_2\}$ where the coordinates $k_1$ and $k_2$ specify position along two basis directions in reciprocal space. As we have already seen in Sec.~\ref{sec:adiabatic}, for one-dimensional systems, the Berry phase $\Phi_B$ (for singly filled bands i.e., $N=1$) characterizes the topological classification of the eigenstates of $H(k)$. In this section, we focus primarily on the topological classification of two dimensional ($d=2$) systems and construct the relevant {\it topological invariants} which characterize such systems. 
	
	The presence of translational invariance allows us to impose periodic boundary conditions, which implies that all physical observables are uniquely defined within the first BZ in the reciprocal space. The underlying parameter-manifold is therefore \textit{compact}, i.e., the manifold closes in on itself. For 
example, the BZ in one dimension, $k\in[0,2\pi)$, has the geometry of a closed ring (known as a $\mathcal{S}^1$). Similarly, in $d=2$, the BZ assumes the form of a two-dimensional torus (denoted as a $T^2$) as both $k_1,k_2\in[0,2\pi)$.
	
	Let us first consider the Abelian case of $N=1$ where only one single-particle eigenstate of the Hamiltonian $H(\vec{k})$ is occupied for all $\vec{k}\in BZ$. Any simple closed loop $\Gamma$ on the surface of the two-dimensional BZ torus can be contracted to a point on the surface itself. The holonomy of an eigenstate $\ket{\psi}=\ket{n(k_1,k_2)}$, defined in Eq.~\eqref{nabcn_1}, can be evaluated as,
	\begin{equation}\label{circulation}
		\mathcal{C}(\Gamma)=\frac{1}{i}\ln\left(e^{i\oint_{\Gamma} A^{\mu}dk_{\mu}}\right),
	\end{equation}
	where we have omitted the path ordering operator $\mathcal{P}$ as the gauge group is Abelian. The loop $\Gamma$ then divides the surface of the parameter-space into two distinct sections. Although the area enclosed by the loop in the two sections are different, the line integral in Eq.~\eqref{circulation} is identical to the surface integral (see Eq.~\eqref{eq_stokes}) on either of the sections. For our purpose, we evaluate the surface integral over the surface $S$ which expands in area as the loop $\Gamma$ shrinks in perimeter. We thus arrive at,
	\begin{equation}\label{abelian_stokes}
		\mathcal{C}(\Gamma)=\frac{1}{i}\ln\left(e^{i\oint_{\Gamma} A^{\mu}dk_{\mu}}\right)=\int_{S}\left(F_{\mu\nu}\right)d\sigma^{\mu\nu},
	\end{equation}
	
	However, note that as the loop $\Gamma$ shrinks to a point, the line integral in Eq.~\eqref{abelian_stokes} vanishes and the surface $S$ starts covering the complete BZ. This apparently suggests that the net gauge flux through the total BZ must always vanish. However, this only holds true for {\rm topologically trivial} systems. It is seen that there also exists topologically non-trivial systems, for which it becomes impossible to uniquely define a consistent gauge for all single-particle states residing on the complete BZ. For example, consider the family of quantum systems in which the ground state in a given basis can be represented by points on a Bloch sphere,
	\begin{equation}
		\ket{\psi(\theta(\vec{k}),\phi(\vec{k}))}\equiv \begin{bmatrix}
			f(\theta(\vec{k}),\phi(\vec{k})) \\
			g(\theta(\vec{k}),\phi(\vec{k})) \\
		\end{bmatrix}=
		\begin{bmatrix}
			e^{i\phi(\vec{k})}\sin\frac{\theta(\vec{k})}{2} \\
			\cos\frac{\theta(\vec{k})}{2} \\
		\end{bmatrix},
	\end{equation}
	where $\theta(\vec{k})\in[0,\pi]$, $\phi(\vec{k})\in[0,2\pi]$ and we have chosen the gauge in which the function $g$ satisfies,
	\begin{equation}\label{gauge_fixing}
		{\rm Im}[g(\theta(\vec{k}),\phi(\vec{k}))]=0.
	\end{equation}
	The gauge-fixing rule in Eq.~\eqref{gauge_fixing} is well behaved at all points except the south pole $\theta=\pi$, where the Eq.~\eqref{gauge_fixing} is ill-defined. On the other hand, changing the gauge condition to ${\rm Im}[f(\theta(\vec{k}),\phi(\vec{k}))]=0$ does not resolve the issue as it is then ill-defined at $\theta=0$. If the ground state $\ket{\psi(\theta(\vec{k}),\phi(\vec{k}))}$ is topologically non-trivial, it is found to be mapped to both the poles of the Bloch sphere for two different values of $\vec{k}$ within the BZ. Thus evaluating the integral in Eq.~\eqref{abelian_stokes} necessitates choosing at least two different gauges on disjoint sections, $S_1$ and $S_2$, of the BZs such that the gauge-fixing condition is well-defined in their respective sections.
	Thus, now the total gauge flux through the compact BZ,
	\begin{equation}
		\mathcal{C}=\oint_{BZ}F_{\mu\nu}d\sigma^{\mu\nu}=\left(\int_{S_1}d\sigma^{\mu\nu}_1-\int_{S_2}d\sigma^{\mu\nu}_2\right)~F_{\mu\nu}.
	\end{equation}
	Also, the connection under two different gauges in $S_1$ and $S_2$ must be related by a simple ${\rm U(1)}$ gauge transformation,
	\begin{eqnarray}
		\ket{\psi}_1&=&e^{-i\gamma}\ket{\psi}_2,\nonumber\\
		A^{\mu}_1&=&A^{\mu}_2-\partial_{\mu}\gamma.
	\end{eqnarray}
	Therefore, from the Stokes' theorem separately applied on both the sections, the total flux reduces to,
\begin{equation}
\mathcal{C}=\oint_{\Gamma^*}\partial_{\mu}\gamma dk^{\mu}=2\pi\nu ~~~{\rm where}~~~ \nu\in\mathcal{Z}, \end{equation}
	where $\Gamma^*$ is the boundary between the disjoint sections. The second equality in the above equation follows from the requirement of single-valuedness of the vectors $\ket{\psi(\vec{k})}$ on the boundary loop $\Gamma^*$ and we thus obtain an integer quantized gauge-invariant topological index. The gauge fixing is however not an issue while evaluating the total flux as the gauge field $F_{\mu\nu}$ is in itself completely gauge invariant. We have therefore the gauge invariant integral quantity, also popularly known as the {\it Chern number} of a state\ct{tong16},
	\begin{equation}\label{chern_number}
		\nu=\frac{1}{2\pi}\oint_{BZ}F_{\mu\nu}d\sigma^{\mu\nu}.
	\end{equation}
	From the discussion following Eq.~\eqref{Berry} and that the Chern number can be represented in terms of a holonomy on the manifold generated by the gauge potential $A_{\mu}$ (see Eq.~\eqref{eq_stokes}), it follows that the Chern number must remain invariant under adiabatic transformations on the state $\ket{\psi(k_1,k_2)}$ over which it is being evaluated. Thus, in analogy with the Berry phase, the Chern number induces a topological classification of quantum states on a two dimensional parameter-manifold. For the generic case when $N$ single-particle states are occupied, it follows from the discussion in Sec.~\ref{sec:geometry} that the appropriate gauge-invariant quantity (see Eq.~\eqref{loop_dependent}) classifying the topological phases of the system is the ${\rm U(N)}$ Chern number defined by\ct{tong16,hughes08},
	\begin{equation}\label{nab}
		\tilde{\nu}=\frac{1}{2\pi}\oint_{BZ}{\rm Tr}\left(F_{\mu\nu}\right)d\sigma^{\mu\nu}.
	\end{equation}
	
	For a finite-size system, the parameter space or the BZ is discrete. As discussed in Sec.~\ref{sec:geometry}, one can construct Wilson loops on the discrete $d=2$ lattice with the lattice parameter chosen as $a=2\pi/L\ll1$, where $L$ is the linear dimension of the system. To calculate the total flux through the discrete parameter lattice, we simply sum over the discrete Wilson loops (see Eq.~\eqref{discrete_wilson}) over all the infinitesimal square plaquettes $dS$ of area $a^2$ making up the BZ,
	\begin{equation}\label{fukui_2}
		\nu_{L}=\frac{1}{2\pi}\sum\limits_{dS\in BZ}W_L(dS),
	\end{equation}
	where $W_L$ is given by Eq.~\eqref{fukui_1}. Focusing on the simple case of $N=1$, 
	the discrete Wilson loop is found to assume the form (see Eq.~\eqref{discrete_wilson}),
	\begin{eqnarray}
		W_L(dS)&=&F_{\mu\nu}a^2+2\pi\eta(dS),\nonumber\\
		F_{\mu\nu}&=&\delta_{\mu}A_{\nu}-\delta_{\nu}A_{\mu},
	\end{eqnarray}
	where $\delta_{\mu}$ are finite difference operators on the discrete lattice. Thus, on summing over all such plaquettes on the compact BZ, one finds,
	\begin{equation}
		\nu_L=\sum\limits_{dS\in BZ}\eta(dS).
	\end{equation} 
	
	The {\it lattice Chern number} $\nu_L$ is thus also integer quantized despite the parameter space being discrete in a finite-size system. Note that the lattice Chern number is physically meaningful only if it approaches the Chern number defined in the thermodynamic limit as the system size is increased. However, since $\nu_L$ is always an integer, it cannot approach $\nu$ continuously as $L\rightarrow\infty$ and hence, there must be a critical size $L_c$ for which the lattice chern number exactly reaches the continuum Chern number. To this end, we note that the lattice Chern number can only change if the condition in Eq.~\eqref{branch} is violated for at least one plaquette within the BZ. In other words, $\nu_L$ can only change when
\begin{equation}\label{eq_ch_order}
\left|F_{\mu\nu}a^2\right|=\left|\frac{4\pi^2F_{\mu\nu}}{L^2}\right|\sim\pi.
	\end{equation} 
	In this case, one needs to choose an altogether different integer field $\eta(dS)$ so as to restore the condition in Eq.~\eqref{branch}. 
	We can now perform an order of magnitude analysis of the dependence of the lattice Chern number on the spatial dimension $L$. From Eq.~\eqref{fukui_2} one might estimate,
	\begin{equation}
		F_{\mu\nu}\sim\frac{2\pi\nu_L}{4\pi^2},
	\end{equation}
	where $4\pi^2$ is the total area of the BZ. Substituting in Eq.\eqref{eq_ch_order}, we obtain an order of magnitude dependence of the lattice Chern number with the lattice dimension,
	\begin{equation}
		L\sim\sqrt{2|\nu_L|}.
	\end{equation}
	This shows that as the system size $L$ increases and the lattice becomes finer, the lattice Chern number monotonically transits between different quantized values until at $L=L_c$, it reaches the exact Chern number defined in the continuum limit. Any subsequent increase in $L$ does not alter the lattice Chern number and hence\ct{fukui05},
	\begin{equation}
		\nu_L=\nu~~\text{for}~~L\gtrsim L_c=\sqrt{2|\nu|}.
	\end{equation}
	
	Since $L_{C}$ is a finite length, this method of finding the Chern number for the continuum theory is far more advantageous numerically as one does not have to go to the thermodynamic limit of system size as required in Eq.~\eqref{chern_number}. Here, it is also worth noting that the discrete lattice arises in finite-size systems which remain gapped even at the QCP, which exists in the continuum limit of the system and is necessarily gapless. Finally, we recall that the topological phases of a thermodynamically large system are separated by gapless QCPs, at which the adiabatic conditions break down. However, given that no truly gapless point exists in finite-size systems, it seems possible to connect phases characterized by distinct values of the lattice Chern number without any breakdown of adiabaticity. This is particularly remarkable as the lattice Chern number mimics the continuum Chern number above the critical size $L_c$.\\
	
\subsubsection{Bott invariant}\label{sec:bott}
	
	So far we have focused on topological invariants which classify translationally invariant systems. However, it turns out that it is possible to define a topological invariant which does not depend on the translational invariance of the system, yet which remarkably reduces to the conventional Chern number with periodic boundary conditions. Consider a two-dimensional insulating lattice system with the lattice coordinates being $\{x_i,y_i\}$ and described by the Hamiltonian $H$.
	It is then straightforward to see that the Bott index defined in Eq.~\eqref{eq_bott_main} of Sec.~\ref{nogo1} (also see Ref.~\onlinecite{rigol15}),
	\begin{equation}\label{eq_bott}
		\nu_{B}=\frac{1}{2\pi}{\rm Im}\left[{\rm Tr}\ln\left(\tilde{T}_X\tilde{T}_Y\tilde{T}_X^{\dagger}\tilde{T}_Y^{\dagger} \right)\right]
	\end{equation}
	is a real number which is independent of the boundary conditions imposed on the lattice. This quantity is known as the {\it Bott index}. Using $C^*$ algebras, it can be rigorously shown that the quantity $\nu_B$ is indeed an integer quantized topological invariant which can also topologically classify systems with broken translational symmetry as well as disordered quantum systems. Nevertheless, a simpler understanding of the Bott index can be made by inspecting it with periodic boundary conditions, which simplifies the analysis.
	
	As in preceding discussions, we describe the lattice in the momentum space $\{k_x,k_y\}$ such that in one BZ, $k_x,k_y\in[0,2\pi)$ with periodic boundary conditions. In the momentum space, the action of the operator $T_{X(Y)}$ is simply to generate translations in momentum direction $k_{x(y)}$ by an amount $\Delta k_{x(y)}=2\pi/L$. To see this, we note that the position operators $X$ and $Y$ have the representations $i\partial_{k_x}$ and $i\partial_{k_y}$, respectively, in reciprocal space. Consequently, the action of the operators $T_X,T_Y$ when expanded in a Taylor series gives,
\begin{eqnarray}
T_{X(Y)}\ket{\psi_n(k)} &=& \ket{\psi(k)}-\partial_{k_{x(y)}}\ket{\psi_n(k)}
\Delta k_{x(y)} \non \\
&& +\frac{1}{2}\partial_{k_{x(y)}}^2\ket{\psi_n(k)}\Delta k_{x(y)}^2-\dots
\non \\
&=& \ket{\psi_n(k-\Delta k_{x(y)})}, \end{eqnarray}
	Hence, it follows that
	\begin{eqnarray}
		\braket{\psi_m({k})|T_{X(Y)}|\psi_n({k}^{\prime})}&=&\braket{\psi_m({k})|\psi_n({k}^{\prime})}\delta_{{k^\prime},k+\Delta k_{x(y)}}\nonumber\\
		&=&\delta_{m,n}\delta_{{k^\prime},k+\Delta k_{x(y)}}.
	\end{eqnarray}
	From Eq.\eqref{eq_bott}, it is then easy to see that the Bott index with periodic boundary conditions reduces to,
	\begin{equation}\label{bott_wilson}
		\nu_B^{n}=\\
		\frac{1}{2\pi}\sum_{k\in BZ}{\rm Im}\left[\ln\Big(\left(T_{X}\Big)^{nn}\Big({T}_{Y}\Big)^{nn}\left({T}_{X}^{\dagger}\right)^{nn}\left({T}_{Y}^{\dagger}\right)^{nn}\right)\right].
	\end{equation}
	It is thus clear that for each $k$, the argument of the logarithms in the above equations traces out a Wilson loop (see Eq.~\eqref{fukui_1}) around a plaquette with $k$ residing at one corner of the plaquette. Thus, the summation is thus carried out over all such plaquettes covering the complete BZ. We therefore conclude that the Bott index is indeed a gauge invariant and quantized topological invariant which is equivalent to the lattice Chern number $\nu_L$ with periodic boundary conditions. We can then expect that, similar to the lattice Chern number, one does not need to cross a gapless QCP in the thermodynamic limit to connect two regions separated by inequivalent Bott index or lattice Chern numbers. The is significant in the context of the dynamical preparation of topologically non-trivial quantum states in real physical systems as discussed in Sec.~\ref{nogo1}.

\section{A perturbative expansion of the steady state Majorana correlations}\label{sec_perturb}

In Sec.~\ref{slow_defect}, we discussed the dynamical defects generated in the two-point Majorana correlations of one-dimensional
Kitaev-like systems when driven adiabatically in the presence of a Markovian bath. While the lowering of defects with increasing ramp duration $\tau$ (in the small $\tau$ limit) is a result of approach towards the adiabatic limit of ramping protocol, the linear rise following the optimal $\tau_0$ needs further scrutiny.
	
	{To this end, we perturbatively expand (see Ref.~\onlinecite{souvik20}) the solution of the dynamical equations of motion for the two-point Majorana correlations where we make use of the condition that $\tau\ll\tau_B$ or $\kappa\tau\ll 1$ ($\kappa\sim1/\tau_B$) to identify $\kappa\tau$ as the small parameter in the perturbation.}
	
	{The dynamical equations of motion for the two-point Majorana correlations can be expressed in terms of a $2L\times2L$ dimensional covariance matrix $C(t)$ defined as $C_{i,j}(t)=\Tr(a_ia_j\rho(t))-\delta_{i,j}$, which satisfies,
		\begin{equation}\label{correlation}
		\dot{C}(t)=-X(t)C(t)-C(t)X^T(t)+iY(t),
		\end{equation}
		where $X(t)=4(i\mathcal{H}(t)+\mathrm{Re}[M(t)])$, $Y(t)=4(\mathrm{Im}[M(t)]-\mathrm{Im}[M^T(t)])$ and $M(t)=\sum_il_i\otimes l_i^*$. The matrix $\mathcal{H}(t)$ corresponds to single particle Hamiltonian in Majorana basis while $M(t)$ encodes all the bath information and is therefore time independent in our case. Substituting the ansatz $C(t)=Q(t)C(0)Q^T(t)-iP(t)Q^T$ in the above equation, where $P(T)$ and $Q(t)$ are two real matrices, results in two simpler equations \cite{prosen11}
		\begin{subequations}
			\begin{align}
			\dot{Q}(t)=-X(t)Q(t),
\end{align}
\begin{align}
			\dot{P}(t)=-X(t)P(t)-Y(t)Q^{-T}(t),
			\end{align}
		\end{subequations}
		with $Q(0)=\mathcal{I}$ and $P(0)=\mathbb{O}$.
	}
	
	{The uniform and time independent coupling of the bath with the system in our case ($\kappa_i=\kappa$) allows us to rewrite $M(t)=M=\kappa\widetilde{M}$ and $Y(t)=Y=\kappa \widetilde{Y}$ where all elements of matrix $\widetilde{M}$ and $ \widetilde{Y}$, are dimensionless. Assuming natural units, we substitute $t=\tilde{t}\tau$, $\widetilde{\mathcal{H}}=\mathcal{H}\tau$, and $\tilde{\kappa}=\kappa\tau$ to arrive at the non-dimensionalized version of the above equations,
		\begin{subequations}
			\begin{align}\label{eq_Q}
			\dot{Q}(\ttil)=-4\left(i\Htil(\ttil)+\ktil Re[\Mtil]\right)Q(\ttil),
			\end{align}
			\begin{align}\label{eq_P}
			\dot{P}(\ttil)=-4\left(i\Htil+\ktil Re[\Mtil]\right)P(\ttil)-\ktil \Ytil Q^{-T}(\ttil).
			\end{align}
		\end{subequations}
		Solving the above pair of matrix equations perturbatively to first order in $\ktil$, we obtain}
	{\begin{widetext}
			\begin{align}\label{eq_Corr}
			C(\ttil)=V(\ttil)C(0)V^T(\ttil)-\ktil V(\ttil)\big[C(0)\Lambda^T(\ttil)+\Lambda(\ttil)C(0)-\Gamma(\ttil)\big]V^T(\ttil)+\mathcal{O}(\ktil^2),
			\end{align} 
		\end{widetext}
		where
		\begin{subequations}
			\begin{align}
			V(\ttil)=\mathcal{T}e^{-4i\int_0^{\ttil}\Htil(\ttil')d\ttil'},
			\end{align}	
			\begin{align}
			\Lambda(\ttil)=4\int_{0}^{\ttil}V^T(\ttil')Re[\Mtil]V(\ttil')d\ttil',
			\end{align}
			\begin{align}
			\Gamma(\ttil)=i\int_{0}^{\ttil}V^T(\ttil')\Ytil V(\ttil')d\ttil'.
			\end{align}
		\end{subequations}
	}
	{The end of the ramp protocol corresponds to $\ttil=1$ in the rescaled units; the covariance matrix at the end of the ramp is therefore obtained as
		\begin{align}
		C(1) = V(1)C(0)V^T(1)-\ktil K+\mathcal{O}(\ktil^2),
		\end{align}
		where $K=V(1)\big[C(0)\Lambda^T(1)+\Lambda(1)C(0)-\Gamma(1)\big]V^T(1)$ is a constant matrix with dimensionless elements. Next, we note that the first term in the above equation $V(1)C(0)V^T(1)$ correspond to an unitary time-evolution of the covariance matrix generated by the Hamiltonian $\Htil$. Further, this term captures the full evolution of the covariance matrix in the absence of bath ($\ktil=0$). We, therefore identify this term as the time-evolved covariance matrix $C^U(1)$ in the absence of any dissipative channels. Returning to the original unscaled units, we finally obtain,
		\begin{align}\label{corr}
		C(\tau)=C^U(\tau)-\kappa\tau K+\mathcal{O}(\kappa^2\tau^2).
		\end{align}
	}
	
	{Let us now consider the Majorana edge correlation $\mathrm{Tr}(\rho(\tau)\theta)=iC_{1,2L}(\tau)$ at the end of the ramp. As we are interested in the adiabatic limit $L\ll\tau$, we assume that the unitary ramp (in absence of dissipation) induces negligible excitations on its own, i.e., $iC^U_{1,2L}\simeq\bra{\psi_f^0}\theta\ket{\psi_f^0}$. Using Eq.~\eqref{corr}, the defect $\chi(\tau)$ is obtained as,
		\begin{align}
		\chi(\tau)=\kappa\tau K_{1,2L}+\mathcal{O}(\kappa^2\tau^2).
		\end{align}
		Hence, we conclude that in the limit $L\ll\tau\ll\tau_B$, the defect at the end of the ramp scales as $\chi(\tau)\sim\kappa\tau$ in the leading order of perturbation. Further, we note that the above scaling holds true for all pairwise Majorana correlations or elements of the covariance matrix.}
	
	{Finally, we note that the perturbation theory, in essence, extracts the effect of a weak environmental effect (perturbation). The defect generated because of non-adiabatic effects arising from a unitary finite ramp duration can not be captured within this framework.}\\

\end{document}